\documentclass[lettersize,journal]{IEEEtran}

\RequirePackage{xcolor}
\PassOptionsToPackage{binary-units, detect-all, per-mode=repeated-symbol, range-units=single, range-phrase=~to~}{siunitx}
\RequirePackage{siunitx}
\PassOptionsToPackage{acronyms,nohypertypes={acronym},nomain}{glossaries}
\RequirePackage{glossaries}
\RequirePackage[pscoord]{eso-pic}
\RequirePackage{multirow}
\RequirePackage{makecell}
\RequirePackage{threeparttable}
\RequirePackage{booktabs}
\RequirePackage{colortbl}
\RequirePackage[verbose]{placeins}
\RequirePackage{fancyhdr}
\RequirePackage[hidelinks]{hyperref}
\RequirePackage{MnSymbol}
\RequirePackage{wasysym}
\RequirePackage{xspace}
\RequirePackage{lipsum}
\RequirePackage{soul}
\RequirePackage{calc}

\RequirePackage{amsmath,amsfonts}
\RequirePackage{algorithmicx}
\RequirePackage{algorithm}
\RequirePackage{algpseudocode}
\RequirePackage{array}
\RequirePackage[caption=false,font=normalsize,labelfont=sf,textfont=sf]{subfig}
\RequirePackage{url}
\RequirePackage{verbatim}
\RequirePackage{graphicx}
\RequirePackage{pifont}

\RequirePackage{orcidlink}
\RequirePackage[noabbrev,capitalise]{cleveref}

\RequirePackage{cite}

\RequirePackage{arydshln}

\newcommand{\x}{$\times$}

\renewcommand{\subsubsection}[1]{\paragraph*{\textbf{#1}}}

\DeclareSIUnit{\x}{\!\ensuremath{\times}}
\DeclareSIUnit\bit{b}
\DeclareSIUnit\GE{GE}
\DeclareSIUnit\kGE{\kilo\GE}
\DeclareSIUnit\MGE{\mega\GE}
\DeclareSIUnit\flop{FLOP}
\DeclareSIUnit\flops{FLOPS}
\newlength\myheight
\newlength\mydepth
\settototalheight\myheight{Xygp}
\newcommand*\circnumsmall[1]{\tikz[baseline=(char.base)]{%
            \node[white,shape=circle,fill=ieee-dark-black-100,draw,inner sep=1pt] (char) {\small\color{ieee-bright-white-100}\sffamily #1};}}

\definecolor{ieee-bright-dblue-100}{rgb}{0.0, 0.3828, 0.6055}
\definecolor{ieee-bright-dblue-80}{rgb}{0.0, 0.4883, 0.6797}
\definecolor{ieee-bright-dblue-60}{rgb}{0.3633, 0.6094, 0.7617}
\definecolor{ieee-bright-dblue-40}{rgb}{0.5898, 0.7383, 0.8398}
\definecolor{ieee-bright-dblue-20}{rgb}{0.8906, 0.8984, 0.9219}
\definecolor{ieee-bright-red-100}{rgb}{0.7266, 0.0469, 0.1836}
\definecolor{ieee-bright-red-80}{rgb}{0.832, 0.3164, 0.3281}
\definecolor{ieee-bright-red-60}{rgb}{0.8906, 0.4922, 0.4805}
\definecolor{ieee-bright-red-40}{rgb}{0.9336, 0.6562, 0.6406}
\definecolor{ieee-bright-red-20}{rgb}{0.9688, 0.8203, 0.8125}
\definecolor{ieee-bright-orange-100}{rgb}{0.9961, 0.6367, 0.0}
\definecolor{ieee-bright-orange-80}{rgb}{0.9844, 0.6953, 0.3125}
\definecolor{ieee-bright-orange-60}{rgb}{0.9883, 0.7695, 0.4844}
\definecolor{ieee-bright-orange-40}{rgb}{0.9922, 0.8359, 0.6562}
\definecolor{ieee-bright-orange-20}{rgb}{0.9961, 0.9219, 0.8164}
\definecolor{ieee-bright-yellow-100}{rgb}{0.9961, 0.8164, 0.0}
\definecolor{ieee-bright-yellow-80}{rgb}{0.9961, 0.8477, 0.2148}
\definecolor{ieee-bright-yellow-60}{rgb}{0.9961, 0.875, 0.4492}
\definecolor{ieee-bright-yellow-40}{rgb}{0.9961, 0.9062, 0.6328}
\definecolor{ieee-bright-yellow-20}{rgb}{0.9961, 0.9531, 0.8125}
\definecolor{ieee-bright-lgreen-100}{rgb}{0.4688, 0.7422, 0.125}
\definecolor{ieee-bright-lgreen-80}{rgb}{0.5742, 0.7852, 0.332}
\definecolor{ieee-bright-lgreen-60}{rgb}{0.6875, 0.8398, 0.5039}
\definecolor{ieee-bright-lgreen-40}{rgb}{0.793, 0.8906, 0.6641}
\definecolor{ieee-bright-lgreen-20}{rgb}{0.8945, 0.9414, 0.8281}
\definecolor{ieee-bright-dgreen-100}{rgb}{0.0, 0.5156, 0.2383}
\definecolor{ieee-bright-dgreen-80}{rgb}{0.1641, 0.6055, 0.3867}
\definecolor{ieee-bright-dgreen-60}{rgb}{0.3906, 0.6953, 0.5234}
\definecolor{ieee-bright-dgreen-40}{rgb}{0.6094, 0.8008, 0.6719}
\definecolor{ieee-bright-dgreen-20}{rgb}{0.8047, 0.8945, 0.8359}
\definecolor{ieee-bright-purple-100}{rgb}{0.5938, 0.1133, 0.5898}
\definecolor{ieee-bright-purple-80}{rgb}{0.6992, 0.3281, 0.668}
\definecolor{ieee-bright-purple-60}{rgb}{0.7812, 0.4961, 0.7461}
\definecolor{ieee-bright-purple-40}{rgb}{0.8555, 0.6602, 0.8281}
\definecolor{ieee-bright-purple-20}{rgb}{0.9219, 0.8281, 0.9023}
\definecolor{ieee-bright-lblue-100}{rgb}{0.0, 0.6094, 0.6484}
\definecolor{ieee-bright-lblue-80}{rgb}{0.0, 0.6797, 0.7188}
\definecolor{ieee-bright-lblue-60}{rgb}{0.2109, 0.75, 0.7812}
\definecolor{ieee-bright-lblue-40}{rgb}{0.5469, 0.8242, 0.8438}
\definecolor{ieee-bright-lblue-20}{rgb}{0.7695, 0.918, 0.9219}
\definecolor{ieee-bright-cyan-100}{rgb}{0.0, 0.707, 0.8828}
\definecolor{ieee-bright-cyan-80}{rgb}{0.0, 0.7227, 0.9453}
\definecolor{ieee-bright-cyan-60}{rgb}{0.2656, 0.7812, 0.957}
\definecolor{ieee-bright-cyan-40}{rgb}{0.5547, 0.8438, 0.9688}
\definecolor{ieee-bright-cyan-20}{rgb}{0.7773, 0.9141, 0.9805}
\definecolor{ieee-bright-white-100}{rgb}{0.9961, 0.9961, 0.9961}
\definecolor{ieee-bright-white-80}{rgb}{0.9961, 0.9961, 0.9961}
\definecolor{ieee-bright-white-60}{rgb}{0.9961, 0.9961, 0.9961}
\definecolor{ieee-bright-white-40}{rgb}{0.9961, 0.9961, 0.9961}
\definecolor{ieee-bright-white-20}{rgb}{0.9961, 0.9961, 0.9961}
\definecolor{ieee-dark-red-100}{rgb}{0.5234, 0.1211, 0.2539}
\definecolor{ieee-dark-red-80}{rgb}{0.6445, 0.2812, 0.3828}
\definecolor{ieee-dark-red-60}{rgb}{0.7422, 0.4727, 0.5234}
\definecolor{ieee-dark-red-40}{rgb}{0.832, 0.6445, 0.6758}
\definecolor{ieee-dark-red-20}{rgb}{0.918, 0.8203, 0.832}
\definecolor{ieee-dark-orange-100}{rgb}{0.9062, 0.4648, 0.1328}
\definecolor{ieee-dark-orange-80}{rgb}{0.9648, 0.5664, 0.3164}
\definecolor{ieee-dark-orange-60}{rgb}{0.9766, 0.6758, 0.4805}
\definecolor{ieee-dark-orange-40}{rgb}{0.9844, 0.7773, 0.6523}
\definecolor{ieee-dark-orange-20}{rgb}{0.9922, 0.8789, 0.8125}
\definecolor{ieee-dark-yellow-100}{rgb}{0.9961, 0.7773, 0.1719}
\definecolor{ieee-dark-yellow-80}{rgb}{0.9961, 0.8086, 0.375}
\definecolor{ieee-dark-yellow-60}{rgb}{0.9961, 0.875, 0.4492}
\definecolor{ieee-dark-yellow-40}{rgb}{0.9961, 0.8984, 0.6875}
\definecolor{ieee-dark-yellow-20}{rgb}{0.9961, 0.9453, 0.8438}
\definecolor{ieee-dark-lgreen-100}{rgb}{0.3945, 0.5508, 0.0938}
\definecolor{ieee-dark-lgreen-80}{rgb}{0.5078, 0.6289, 0.293}
\definecolor{ieee-dark-lgreen-60}{rgb}{0.6367, 0.7188, 0.4688}
\definecolor{ieee-dark-lgreen-40}{rgb}{0.7539, 0.8047, 0.6367}
\definecolor{ieee-dark-lgreen-20}{rgb}{0.875, 0.9023, 0.8125}
\definecolor{ieee-dark-dgreen-100}{rgb}{0.0, 0.3867, 0.2539}
\definecolor{ieee-dark-dgreen-80}{rgb}{0.1836, 0.5, 0.3906}
\definecolor{ieee-dark-dgreen-60}{rgb}{0.3984, 0.6172, 0.5273}
\definecolor{ieee-dark-dgreen-40}{rgb}{0.5938, 0.7422, 0.6758}
\definecolor{ieee-dark-dgreen-20}{rgb}{0.793, 0.8711, 0.8359}
\definecolor{ieee-dark-purple-100}{rgb}{0.4648, 0.1445, 0.5117}
\definecolor{ieee-dark-purple-80}{rgb}{0.5898, 0.3242, 0.6016}
\definecolor{ieee-dark-purple-60}{rgb}{0.6914, 0.4883, 0.6953}
\definecolor{ieee-dark-purple-40}{rgb}{0.7969, 0.6523, 0.793}
\definecolor{ieee-dark-purple-20}{rgb}{0.8945, 0.8203, 0.8945}
\definecolor{ieee-dark-cyan-100}{rgb}{0.0, 0.4492, 0.4648}
\definecolor{ieee-dark-cyan-80}{rgb}{0.0, 0.5469, 0.5664}
\definecolor{ieee-dark-cyan-60}{rgb}{0.3047, 0.6602, 0.668}
\definecolor{ieee-dark-cyan-40}{rgb}{0.5586, 0.7695, 0.7734}
\definecolor{ieee-dark-cyan-20}{rgb}{0.7734, 0.8789, 0.8789}
\definecolor{ieee-dark-dblue-100}{rgb}{0.0, 0.1562, 0.332}
\definecolor{ieee-dark-dblue-80}{rgb}{0.1797, 0.3008, 0.4609}
\definecolor{ieee-dark-dblue-60}{rgb}{0.3828, 0.4609, 0.5859}
\definecolor{ieee-dark-dblue-40}{rgb}{0.5781, 0.6289, 0.7188}
\definecolor{ieee-dark-dblue-20}{rgb}{0.7852, 0.8047, 0.8555}
\definecolor{ieee-dark-grey-100}{rgb}{0.457, 0.4688, 0.4805}
\definecolor{ieee-dark-grey-80}{rgb}{0.5625, 0.5625, 0.5742}
\definecolor{ieee-dark-grey-60}{rgb}{0.6641, 0.6641, 0.6758}
\definecolor{ieee-dark-grey-40}{rgb}{0.7734, 0.7695, 0.7773}
\definecolor{ieee-dark-grey-20}{rgb}{0.8789, 0.8828, 0.8828}
\definecolor{ieee-dark-black-100}{rgb}{0.0, 0.0, 0.0}
\definecolor{ieee-dark-black-80}{rgb}{0.3438, 0.3477, 0.3555}
\definecolor{ieee-dark-black-60}{rgb}{0.5, 0.5078, 0.5195}
\definecolor{ieee-dark-black-40}{rgb}{0.6523, 0.6602, 0.6719}
\definecolor{ieee-dark-black-20}{rgb}{0.8164, 0.8242, 0.8281}

\def\reviewpass{v0.0.6}

\newcommand{\includegraphicssafe}[2][]{%
    \IfFileExists{#2}{%
        \includegraphics[#1]{#2}%
    }{}%
}

\definecolor{donegreen}{rgb}{0,0.3,0}
\definecolor{overhaultext}{rgb}{128,0,128}

\def\thetitle{Co-designing a  Programmable RISC-V Accelerator for MPC-based Energy and Thermal Management of Many-Core HPC Processors}

\newacronym{dtm}{DTM}{dynamic thermal management}
\newacronym{dpm}{DPM}{dynamic power management}
\newacronym{dtpm}{DTPM}{dynamic thermal and power management}
\newacronym{stm}{STM}{static thermal management}
\newacronym{hw}{HW}{hardware}
\newacronym{sw}{SW}{software}
\newacronym{ca}{CA}{command/address}
\newacronym{ip}{IP}{intellectual property}
\newacronym{ddr}{DDR}{double data rate}
\newacronym{lpddr}{LPDDR}{low-power double data rate}
\newacronym{rpc}{RPC}{reduced pin count}
\newacronym{dma}{DMA}{direct memory access}
\newacronym{axi}{AXI}{Advanced eXtensible Interface}
\newacronym{dram}{DRAM}{dynamic random access memory}
\newacronym[firstplural=static random access memories (SRAMs)]{sram}{SRAM}{static random access memory}
\newacronym{edram}{eDRAM}{embedded DRAM}
\newacronym[firstplural=systems on chip (SoCs)]{soc}{SoC}{system on chip}
\newacronym{mpsoc}{MPSoC}{multi-processor system on chip}
\newacronym{hesoc}{HeSoC}{heterogeneous system on chip}
\newacronym{sip}{SiP}{system in package}
\newacronym{fpga}{FPGA}{field-programmable gate array}
\newacronym{asic}{ASIC}{application-specific integrated circuit}
\newacronym{phy}{PHY}{physical layer}
\newacronym{ml}{ML}{machine learning}
\newacronym{iot}{IoT}{internet of things}
\newacronym{foss}{FOSS}{free and open source}
\newacronym{cmos}{CMOS}{complementary metal-oxide-semiconductor}
\newacronym{sut}{SUT}{system under test}
\newacronym{isut}{ISUT}{integrated system under test}
\newacronym{rtl}{RTL}{register transfer level}
\newacronym{hil}{HIL}{hardware-in-the-loop}
\newacronym{pil}{PIL}{processor in the loop}
\newacronym{fil}{FIL}{FPGA in the loop}
\newacronym{mil}{MIL}{model in the loop}
\newacronym{sil}{SIL}{software in the loop}
\newacronym{hpc}{HPC}{high performance computing}
\newacronym{mcu}{MCU}{microcontroller unit}
\newacronym{fub}{FUB}{functional unit block}
\newacronym{ecu}{ECU}{electronic control unit}
\newacronym{dcu}{DCU}{domain control unit}
\newacronym{adas}{ADAS}{advanced driver-assistance system}
\newacronym{fame}{FAME}{FPGA Architecture Model Execution}
\newacronym{pl}{PL}{Programmable Logic}
\newacronym{ps}{PS}{Processing System}
\newacronym{apu}{APU}{Application Processing Unit}
\newacronym{ocm}{OCM}{on-chip memory}
\newacronym{pcs}{PCS}{power controller system}
\newacronym{pcf}{PCF}{power control firmware}
\newacronym{pmca}{PMCA}{programmable multi-core accelerator}
\newacronym{bram}{BRAM}{block RAM}
\newacronym{lut}{LUT}{look-up table}
\newacronym{ff}{FF}{flip-flop}
\newacronym{fsbl}{FSBL}{First Stage BootLoader}
\newacronym{pvt}{PVT}{process, voltage, and temperature}
\newacronym{hls}{HLS}{high-level synthesis}
\newacronym{mqtt}{MQTT}{Message Queuing Telemetry Transport}
\newacronym{cots}{COTS}{commercial off-the-shelf}
\newacronym{cpu}{CPU}{central processing unit}
\newacronym{gpu}{GPU}{graphic processing unit}
\newacronym{ibmocc}{IBM OCC}{IBM on-chip controller}
\newacronym{clic}{CLIC}{Core-Local Interrupt Controller}
\newacronym{clint}{CLINT}{Core-Local Interruptor}
\newacronym{scmi}{SCMI}{System Control and Management Interface}
\newacronym{os}{OS}{operating system}
\newacronym{ospm}{OSPM}{operating system-directed configuration and power management}
\newacronym{mimo}{MIMO}{multiple-input multiple-output}
\newacronym{siso}{SISO}{Single-Input Single-Output}
\newacronym{bmc}{BMC}{Baseboard Management Controller}
\newacronym{qos}{QoS}{quality of service}
\newacronym{tdp}{TPD}{thermal design power}
\newacronym{dvfs}{DVFS}{dynamic voltage and frequency scaling}
\newacronym{dfs}{DFS}{dynamic frequency scaling}
\newacronym{dvs}{DVS}{dynamic voltage scaling}
\newacronym{rtu}{RTU}{Real Time Unit}
\newacronym{pe}{PE}{processing element}
\newacronym{noc}{NoC}{network on chips}
\newacronym{pid}{PID}{proportional integral derivative}
\newacronym{sota}{SoA}{state-of-the-art}
\newacronym{fpu}{FPU}{floating point unit}
\newacronym{pcu}{PCU}{Power Control Unit}
\newacronym{scp}{SCP}{System Control Processor}
\newacronym{mcp}{MCP}{Manageability Control Processor}
\newacronym{occ}{OCC}{On-Chip Controller}
\newacronym{smu}{SMU}{System Management Unit}
\newacronym{ap}{AP}{application-class processor}
\newacronym{vrm}{VRM}{voltage regulator module}
\newacronym{pfct}{PFCT}{periodic frequency control task}
\newacronym{pvct}{PVCT}{periodic voltage control task}
\newacronym{ipc}{IPC}{instructions per cycle}
\newacronym{simd}{SIMD}{single instruction, multiple data}
\newacronym{mctp}{MCTP}{Management Component Transport Protocol}
\newacronym{pldm}{PLDM}{Platform Level Data Model}
\newacronym{rtos}{RTOS}{real-time OS}
\newacronym{hlc}{HLC}{high-level controller}
\newacronym{llc}{LLC}{low-level controller}
\newacronym{acpi}{ACPI}{Advanced Configuration and Power Interface}
\newacronym{pdn}{PDN}{Power Delivery Network}
\newacronym{ewma}{EWMA}{Exponential Weight Moving Average}
\newacronym{ppa}{PPA}{power, performance and area}
\newacronym{pcb}{PCB}{printed circuit board}
\newacronym{dsa}{DSA}{domain-specific accelerator}
\newacronym{ha}{HA}{Hardware Accelerator}

\newacronym[longplural={high-bandwidth memories}]{hbm}{HBM}{high-bandwidth memory}
\newacronym{rapl}{RAPL}{Running Average Power Limit}
\newacronym{tsv}{TSV}{through silicon via}
\newacronym{fet}{FET}{field effect transistor}
\newacronym{fll}{FLL}{frequency locked loop}
\newacronym{pll}{PLL}{phase locked loop}

\newacronym{oca}{EBA}{Enhanced Baseline Algorithm}
\newacronym{fca}{FCA}{Fuzzy-ispired Iterative Control Algorithm}
\newacronym{vba}{VBA}{Voting Box Algorithm}

\newacronym{pm}{PM}{power management}
\newacronym{pmi}{PMI}{power management interface}
\newacronym{fw}{FW}{firmware}
\newacronym{opal}{OPAL}{OpenPower abstraction layer}
\newacronym{pmbus}{PMBUS}{Power Management Bus}
\newacronym{avsbus}{AVSBUS}{Adaptive Voltage Scaling}
\newacronym{psci}{PSCI}{Power State Coordination Interface}
\newacronym{uefi}{UEFI}{Unified Extensible Firmware Interface}
\newacronym{asl}{ASL}{ACPI source language}
\newacronym{aml}{AML}{ACPI machine language}
\newacronym{msr}{MSR}{model-specific register}
\newacronym{mpc}{MPC}{model predictive control}
\newacronym{qp}{QP}{quadratic programming}
\newacronym{fp32}{\texttt{FP32}}{\texttt{float32}}
\newacronym{fp16}{\texttt{FP16}}{\texttt{float16}}
\newacronym{fp64}{\texttt{FP64}}{\texttt{float64}}
\newacronym{fp8}{\texttt{FP8}}{\texttt{float8}}
\newacronym{gp}{GP}{general-purpose}
\newacronym{ds}{DS}{domain-specific}
\newacronym{spm}{SPM}{scratchpad memory}
\newacronym{flops}{FLOPS}{floating-point operations per second}
\newacronym{rac}{RAC}{runtime active control}
\newacronym{rl}{RL}{reinforcement learning}
\newacronym{etm}{ETM}{energy and thermal management}
\newacronym{uav}{UAV}{unmanned aerial vehicles}
\newacronym{pulp}{PULP}{parallel ultra-low power}
\newacronym{osqp}{OSQP}{operator-splitting quadratic programming}
\newacronym{admm}{ADMM}{alternating direction method of multipliers}
\newacronym{isa}{ISA}{instruction set architecture}
\newacronym{vlsi}{VLSI}{very large scale integration}
\newacronym{deepc}{DeePC}{data-enabled predictive control}
\newacronym{ssr}{SSR}{stream semantic register}
\newacronym{sssr}{SSSR}{sparse stream semantic register}
\newacronym{dsp}{DSP}{digital signal processing}
\newacronym{kkt}{KKT}{Karush-Kuhn-Tucker}
\newacronym{amd}{AMD}{approximate minimum degree}
\newacronym{lqr}{LQR}{linear quadratic regulator}
\newacronym{lti}{LTI}{linear time-invariant}
\newacronym{fg}{FG}{fast gradient}
\newacronym{pcg}{PCG}{preconditioned conjugate gradient}
\newacronym{lns}{LNS}{logarithmic number system}
\newacronym{dmp}{DMP}{discrete model pruning}
\newacronym{ge}{GE}{gate equivalent}
\newacronym{aot}{AOT}{ahead-of-time}
\newacronym{oom}{OoM}{order of magnitude}
\newacronym{csc}{CSC}{compressed sparse column}
\newacronym{fe}{FE}{forward elimination}
\newacronym{bs}{BS}{backward substitution}
\newacronym{sptrsv}{SpTRSV}{sparse triangular linear system solver}
\newacronym{hypt}{ParSPL}{parallel sparsity-pattern-leveraging triangular linear system solver}
\newacronym{sl}{SL}{streaming length}
\newacronym{dag}{DAG}{directed acyclic graph}
\newacronym{alap}{ALAP}{as-late-as possible}
\newacronym{asap}{ASAP}{as soon as possible}
\newacronym{mac}{MAC}{multiply-and-accumulate}
\newacronym{rhs}{RHS}{right-hand-side}
\newacronym{rmse}{RMSE}{root mean square error}

\newcommand{\mpcstep}{\gls{mpc} step}

\newcommand{\gfs}{{GlobalFoundries'}}
\newcommand{\gftech}{{GF12LP+}}

\newcommand{\cmark}{\ding{51}}%
\newcommand{\xmark}{\ding{55}}%

\begin{document}

%
%

\title{\thetitle}

\ifx\showrevision\undefined
    \newcommand{\todo}[1]{{#1}}
\else
    \newcommand{\todo}[1]{{\textcolor{red}{#1}}}
    \AddToShipoutPictureFG{%
        \put(%
            8mm,%
            \paperheight-1.5cm%
            ){\vtop{{\null}\makebox[0pt][c]{%
                \rotatebox[origin=c]{90}{%
                    \huge\textcolor{red!75}{\reviewpass}%
                }%
            }}%
        }%
    }
    \AddToShipoutPictureFG{%
        \put(%
            \paperwidth-6mm,%
            \paperheight-1.5cm%
            ){\vtop{{\null}\makebox[0pt][c]{%
                \rotatebox[origin=c]{90}{%
                    \huge\textcolor{red!30}{ETH Zurich - Unpublished - Draft}%
                }%
            }}%
        }%
    }
\fi

\ifx\showrebuttal\undefined
    \newcommand{\rev}[1]{#1}
\else
    \newcommand{\rev}[1]{{\textcolor{ieee-bright-lblue-100}{#1}}}
\fi

\ifx\showrebuttal\undefined
    \newcommand{\revdel}[1]{}
\else
    \newcommand{\revdel}[1]{\textcolor{ieee-bright-red-100}{\st{#1}}}
\fi

\ifx\showrebuttal\undefined
    \newcommand{\revrep}[2]{#2}
\else
    \newcommand{\revrep}[2]{\revdel{#1} \rev{#2}}
\fi

\ifx\showrebuttal\undefined
    \newcommand{\revprg}[1]{}
\else
    \newcommand{\revprg}[1]{\hspace{-0.5ex}\textcolor{red}{\scalebox{.2}[1.5]{$\blacksquare$}}\hspace{-0.5ex}}
\fi

\author{
    Alessandro~Ottaviano\orcidlink{0009-0000-9924-3536}~\IEEEmembership{Graduate Student Member, IEEE},
    Andrino~Meli\orcidlink{0009-0002-6423-0967}, 
    Paul~Scheffler\orcidlink{0000-0003-4230-1381},~\IEEEmembership{Graduate Student Member, IEEE},
    Giovanni~Bambini\orcidlink{0000-0000-0000-0000}~\IEEEmembership{Graduate Student Member, IEEE},
    Robert~Balas\orcidlink{0000-0002-7231-9315},~\IEEEmembership{Graduate Student Member, IEEE},
    Davide~Rossi\orcidlink{0000-0002-0651-5393},~\IEEEmembership{Senior Member,~IEEE},
    Andrea~Bartolini\orcidlink{0000-0000-0000-0000},~\IEEEmembership{Senior Member, IEEE},
    and Luca~Benini\orcidlink{0000-0001-8068-3806},~\IEEEmembership{Fellow,~IEEE}
    \IEEEcompsocitemizethanks{%
    \IEEEcompsocthanksitem A.~Ottaviano, A.~Meli, P.~Scheffler, R.~Balas, and L.~Benini are with the Integrated Systems Laboratory (IIS), ETH Zurich, Switzerland\protect\\
    E-mail: \{aottaviano,moroa,paulsc,balasr,lbenini\}@ethz.ch
    \IEEEcompsocthanksitem G.~Bambini, L.~Benini, D.~Rossi, and A.~Bartolini are with Department of Electrical, Electronic and Information Engineering (DEI), University of Bologna, Bologna, Italy\protect\\
    E-mail: \{giovanni.bambini2,luca.benini,davide.rossi,a.bartolini\}@unibo.it.
    }%
}

\markboth{Journal of \LaTeX\ Class Files,~Vol.~14, No.~8, August~2021}%
{Shell \MakeLowercase{\textit{et al.}}: A Sample Article Using IEEEtran.cls for IEEE Journals}

\maketitle


\begin{abstract}
Managing energy and thermal profiles is critical for many-core HPC processors with hundreds of application-class processing elements (PEs). 
Advanced model predictive control (MPC) delivers state-of-the-art performance but requires solving an online optimization problem over a thousand times per second (1kHz control bandwidth), with computational and memory demands scaling with PE count.
Traditional MPC approaches execute the controller on the PEs, but operating system overheads create jitter, and limit control bandwidth.
Running MPC on dedicated on-chip controllers enables fast, deterministic control but raises concerns about area and power overhead.
In this work, we tackle these challenges by proposing a hardware-software codesign of a lightweight MPC controller, based on an operator splitting quadratic programming solver, and an embedded multi-core RISC-V controller.
Key innovations include pruning weak thermal couplings to reduce model memory and ahead-of-time scheduling for efficient parallel execution of sparse triangular systems arising from the optimization problem.
The proposed controller achieves sub-millisecond latency when controlling 144 PEs at 500 MHz, delivering 33$\times$ lower latency and 7.9$\times$ higher energy efficiency than a single-core baseline. 
Operating within a compact $<$ 1 MiB memory footprint, it consumes as little as 325 mW while occupying less than 1.5\% of a typical \gls{hpc} processor's die area.
\end{abstract}


\glsresetall

\section{Introduction}
\label{sec:intro}

\IEEEPARstart{W}{ith} the growing computational demands of \gls{hpc}, processors have evolved into complex heterogeneous systems, integrating \gls{gp} and \gls{ds} sub-domains to efficiently tackle diverse compute-intensive tasks ranging from artificial intelligence (training and inference)~\cite{10707191}, to quantum and molecular simulations~\cite{MOLECULAR_DYNAMICS_UMBRELLA_III}, and computational biology~\cite{MOLECULAR_DYNAMICS_UMBRELLA_II}.

\Gls{hpc} processors must deliver high floating-point operations per second (\si{\flop\per\second}), and energy efficiency (\si{\flop\per\second\per\watt}), to address growing operational costs and sustainability challenges~\cite{green500}. 
Achieving these objectives requires real-time management of the processors' thermal and power profiles to implement effective thermal and power capping policies and optimize the energy efficiency of the whole computing system. 
One approach involves advanced cooling strategies, to improve heat dissipation~\cite{thermal_runaway,TILLI2022105099}. 
A complementary approach leverages dynamic thermal and power management techniques, or \gls{rac}, to mitigate the negative effects of increased power density in modern technology nodes --- including reduced component lifespan, electromigration, and dielectric breakdown --- which are often triggered by thermal hotspots and steep thermal gradients during workload execution.

\begin{figure}[t]
    \centering
    \includegraphics[width=\columnwidth]{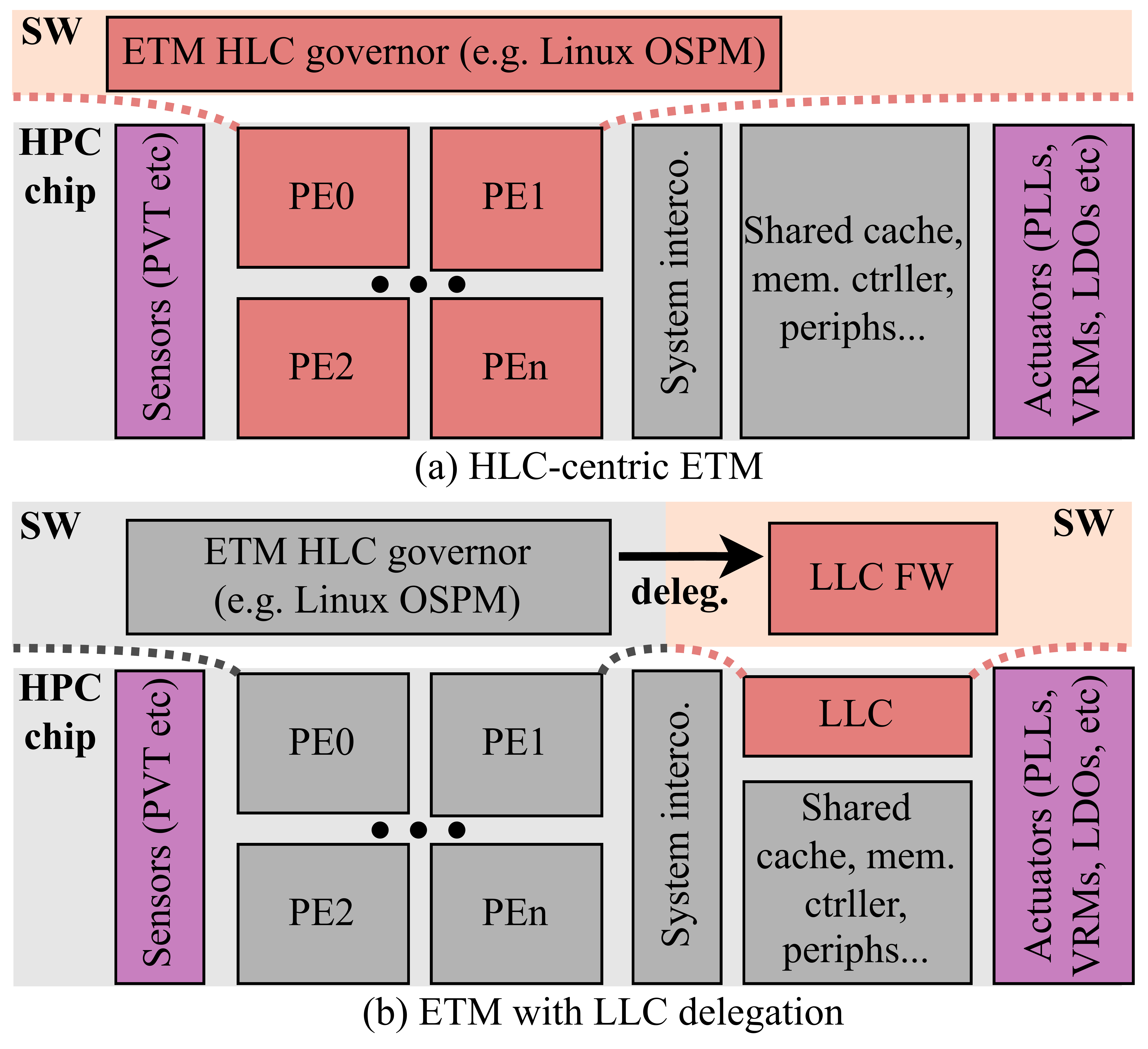}
    \caption{%
        Illustration of \gls{hlc} (a) and \gls{llc} (b) centric paradigms for \gls{etm} of \gls{hpc} chips. 
        Blocks in red indicate the hardware/software entity responsible for low-level interaction with sensors and actuators (in purple) during \gls{rac}.
    }
    \glsreset{hlc}\glsreset{llc}\glsreset{etm}
    \label{fig:etm-paradigms}
\end{figure}

\Gls{rac} uses \gls{pvt} data from on-chip sensors and online workload information to choose optimal \gls{dvfs} operating points for the chip within imposed power and thermal limits. 
These settings are applied at runtime to chip actuators like \glspl{pll} and \glspl{vrm} connected to the \glspl{pe}, which are commonly large, out-of-order application-class processors with vector engines.
The combination of static and dynamic methods is collectively called the \gls{etm} policy of an \gls{hpc} processor.
As shown in~\cref{fig:etm-paradigms}, the \gls{etm} landscape has shifted over the last decade from \textit{\gls{hlc} centric} policies --- software routines like the Linux \gls{ospm} governor running on the same \glspl{pe} to be controlled --- to a \textit{delegation-based} paradigm, where a hardware \gls{llc} collaborates with the \gls{hlc} to enforce power and thermal policies~\cite{ARM_PCSA, Ottaviano2024}.

The delegation-based approach brings several benefits to \gls{etm} from a real-time scheduling perspective.
For example, processor thermal conditions can change so quickly that \gls{hlc} software policies cannot deal with them in a timely manner~\cite{GROVER_ACPI_2003}. 
In contrast, \glspl{llc} can provide faster and more deterministic response times through (i) low-level and autonomous access to the controlled system, (ii) a streamlined and domain-specific hardware-software stack with fast and deterministic access to hardware sensors and actuators like a lightweight \gls{rtos}, and (iii) a microarchitecture tuned to meet real-time constraints.
Furthermore, the \glspl{pe} may be forced to remain active or wake up to perform low-level control functions without other primary tasks.
Decoupled \glspl{llc} provide an always-on intelligence that is low-power, flexible, timely, and able to act autonomously or in collaboration with the \gls{hlc}~\cite{ARM_PCSA}.
In its simplest form, a \gls{llc} is a low-end single-core \gls{mcu} with limited compute capabilities.
However, continuous advancements in transistor scaling have enabled the integration of more performant, heterogeneous, mid-end embedded \glspl{llc} on \gls{hpc} dies.
Examples include \emph{ControlPULP}~\cite{Ottaviano2024}, which combines a single-core manager domain with a \gls{pmca}, and IBM's on-chip controller~\cite{ibm_occ}, which integrates four microcode engines.

On the algorithmic side, established 
\gls{etm} control policies are closed-loop \gls{mimo} algorithms ranging from simple hardware-triggered capping without \gls{dvfs} to classic \gls{pid} controllers~\cite{ibm_occ} for reactive setpoint tracking and \gls{dvfs} enforcement.
However, these methods are not sufficiently flexible and powerful to capture and control the dynamic behavior of many-core processors~\cite{6178247,9163012}.
For this reason, in the last decade, predictive policies such as \gls{mpc} algorithms have been investigated as more advanced alternatives beyond reactive \gls{rac}~\cite{Wang2011AdaptivePC,6178247,TILLI2022105099,9163012}.
Notably, one of the main challenges of \gls{mpc} is the high computational cost of solving an optimization problem on every iteration. 
In \gls{etm} applications, this cost grows with the number of controlled \glspl{pe}, nowadays ranging from under 100~\cite{SIPEARL,AMDGENOA} to over 200~\cite{NVIDIAGRACE} on the same die.
For such complex systems, the large resulting model heavily impacts the memory footprint and computational work of the underlying optimization algorithm~\cite{JEREZ_2}.

These factors present challenges and opportunities for implementing predictive \gls{etm} policies within the modern delegation-based framework.
Challenges arise from an \gls{llc}'s limited computational and memory resources, even when enhanced with dedicated accelerators.
On the other hand, opportunities emerge from the growing interest in \gls{mpc} on embedded devices, driving the development of lightweight, optimized, and fast \gls{mpc} algorithms for resource-constrained \glspl{mcu} particularly in commodity and miniaturized robotics applications~\cite{Salzmann_2023, alavilli2023tinympc}.

In this paper, we present a fast, lightweight, hardware-software \gls{llc} capable of supporting \gls{mpc} for online energy and thermal management of \gls{hpc} processors. 
We leverage a low-cost RISC-V \gls{llc}, which integrates an energy-efficient eight-core \gls{pmca} with custom extensions maximizing compute utilization~\cite{Scheffler_2023}.
We base our approach on the \gls{sota} \gls{osqp} solver and the \gls{admm} solver~\cite{BoydAdmm2011} to minimize memory usage by exploiting the sparsity of the \gls{qp} formulation.
We further reduce model complexity by introducing a novel pruning technique that eliminates weak thermal couplings among \glspl{pe} far apart in the chip.
To enhance parallel execution on the \gls{llc}'s \gls{pmca}, we propose a novel \gls{aot} scheduling and code generation algorithm atop \gls{osqp} that extracts inherent parallelism from the problem’s sparsity pattern.
%
%
The proposed optimization framework applies to any \gls{qp} problem beyond the specific use case of \gls{etm}.
The hardware and software is released under a permissive license\footnote{\url{https://github.com/pulp-platform/control-pulp.git} and \url{https://github.com/andrino-meli/ParSPL}}. 

In more detail, this work makes the following contributions: 
\begin{itemize}

    \item Design of a model-predictive \gls{etm} control policy for many-core \gls{hpc} processors (\cref{sec:etm_mpc}), tailored for online execution on the multi-core embedded RISC-V \gls{llc} ControlPULP~\cite{Ottaviano2024} (\cref{sec:background:llc}). 
    Unlike \gls{sota} approaches~\cite{6178247}, our \gls{mpc} model incorporates both power and temperature dynamics. 
    The implementation utilizes \gls{osqp}'s native QDLDL direct solver (\cref{sec:background:osqp}), which solves a \gls{sptrsv} during the \gls{fe} and \gls{bs} passes. 
    The controller design links hardware-agnostic \gls{mil} optimization with hardware-aware deployment on the \gls{llc} through \gls{osqp}'s native code generation feature (\cref{subsec:meth:framework}).

    \item Development of a threshold-based pruning algorithm to optimize the model memory usage. 
    The algorithm removes low-magnitude couplings among \gls{pe} in the state matrices. 
    This optimization reduces the \gls{mpc} model complexity from quadratic to linear with respect to the number of \glspl{pe}, significantly enhancing scalability for centralized control in larger many-core systems (\cref{subsec:etm_mpc:dmp}).

    \item Design of an \gls{aot} scheduling algorithm, named \gls{hypt}, to accelerate QDLDL's \gls{sptrsv} on the multi-core \gls{llc}.
    \Gls{hypt} uses \gls{aot} partial level scheduling and tiling in the \gls{sptrsv} to extract more parallelism from the problem's sparsity pattern and schedule the execution in advance; this dramatically reduces the synchronization steps among the \gls{llc}'s compute cores compared to a naive parallelization by column/row (\cref{subsec:meth:codegen}).
    Furthermore, \gls{hypt} leverages the chosen \gls{llc}'s extensions for sparse workloads and floating-point hardware loops~\cite{Scheffler_2023} to eliminate memory and control overheads and maximize compute utilization (\cref{subsec:meth:sssr}).
    
    \item Deployment and assessment of the algorithm on the embedded multi-core \gls{llc} in cycle-accurate simulations.
    For one \gls{sptrsv} iteration on a large \gls{etm} problem controlling 144 \glspl{pe}, our methodology is 33$\times$ faster and 7.9$\times$ more energy efficient than vanilla single-core \gls{osqp} while fitting in $<$\SI{1}{\mebi\byte} of on-chip \gls{spm}.
    The controller consumes at most~\SI{325}{\milli\watt} while occupying under \SI{1.5}{\percent} of a typical \gls{hpc} processor die.
    Our approach consumes minimal power and maximizes energy efficiency, unlike \gls{hlc}-centric methods that incur high overhead from power-hungry, \gls{os}-based software stacks for thermal and power capping.
\end{itemize}
The paper is structured as follows:
\Cref{sec:background} covers fundamental concepts of \gls{qp} and its connection to \gls{mpc}.
\Cref{sec:background:llc} details the multi-core \gls{llc} hardware, including extensions for sparse workloads and memory hierarchy.
\Cref{sec:etm_mpc} introduces the thermal and power model for an \gls{hpc} processor, the \gls{mpc} controller architecture, and the threshold-based pruning algorithm, highlighting the application's demanding memory and computational needs.
\Cref{sec:methodology} describes the solver optimization framework, emphasizing numerical precision and the ParSPL approach.
Finally, \cref{sec:evaluation} assesses execution latency, memory footprint, and energy and area efficiency of the optimized solver on the \gls{llc} across varying problem sizes.

\section{Background}
\label{sec:background}

\glsreset{qp}\glsreset{osqp}\glsreset{admm}

In this section, we review the fundamentals of \gls{qp}, the \gls{osqp} solver, and its \gls{admm}-based solution approach.
Finally, we discuss \gls{mpc} and its formulation as a \gls{qp}.

\subsection{Convex quadratic program formulation}\label{sec:background:qp}


Solving a convex \gls{qp} optimization problem~\cite{NoceWrig06} means finding a vector of decision variables $x \in \mathbb{R}^{n}$ that:

\begin{equation}\label{eq:qp}
    \begin{aligned}
        & \textnormal{minimize} \quad (1/2)x^{T}Px + q^{T}x \\
        & \textnormal{subject to} \quad l \leq Ax \leq u
    \end{aligned}
\end{equation}

The objective function is defined by a positive semi-definite matrix $P \in \mathbb{R}_{+}^{n}$ and a vector $q \in \mathbb{R}^{n}$; the constraints function is defined by a matrix $A \in \mathbb{R}^{m \times n}$.
The constraints $l$ and $u$ belong to a set $\mathcal{C}$ s.t.

\begin{equation}\label{eq:lin_constr}
    \begin{aligned}
        \mathcal{C} = [l,u] = \{z \in \mathbb{R}^{m} \; | \; l_i \leq z_i \leq u_i, i = 1, ... m \}
    \end{aligned}
\end{equation}

If $l_i = u_i$ for some or all elements in $[u,l]$, the problem is equality-constrained. 
$n$ and $m$ represent the number of decision variables and constraints, respectively~\cite{Stellato_2020}.

The \emph{size} of the problem in \cref{eq:qp} is defined by a tuple $(n,m,N_{nnz})$. $N_{nnz}$ is the number of nonzero entries in the objective and constraint matrices $P$ and $A$, respectively:

\begin{equation}\label{eq:num_tot_nnz}
    \begin{aligned}
        N_{nnz} = nnz(P) + nnz(A); \; P \in \mathbb{R}_{+}^{n}, \; A \in \mathbb{R}^{m \times n}
    \end{aligned}
\end{equation}

where $nnz(\cdot)$ returns the number of non-zeros (\emph{fill-ins}) of a matrix.

\subsection{\Gls{qp} solution: the resurgence of first-order methods}\label{sec:background:fom}






Well-known \gls{qp} solution methods include \emph{gradient}, \emph{active-set}, \emph{interior-point}, and \emph{first-order} approaches~\cite{Stellato_2020}.

Gradient methods iteratively solve unconstrained \gls{qp} problems and project the solution onto the feasible set, with variants like the \gls{fg} being suitable for input-constrained problems~\cite{FerrauABB_1, JEREZ_2}.
Active-set methods iteratively modify the active set of constraints to converge to the optimal solution, while interior-point methods rely on barrier functions to handle constraints but face scalability issues, making them less practical for embedded platforms~\cite{activeset_intpoint_comp, NoceWrig06, Stellato_2020}.

In contrast, first-order methods only use information about the objective function's first derivative (gradient), are amenable to low-precision number formats, and can be efficiently parallelized~\cite{JEREZ_2}.
Despite providing low- and medium-accuracy solutions, high-quality control can still be achieved without solving the \gls{qp} in \cref{eq:qp} to full accuracy, particularly for real-time embedded optimization~\cite{fast_mpc_boyd, operator_splitting_boyd}; if high-accuracy is desired, techniques like \emph{solution polishing}~\cite{Stellato_2020} can enhance accuracy and robustness if necessary at the cost of additional computation. 
For these reasons, these methods have seen renewed interest in recent years.

In this work, we focus on a class of first-order methods called \emph{operator splitting}. 
They have been proven effective for problems that need to be solved in real-time under tight sampling periods, e.g., embedded control applications~\cite{operator_splitting_book_1}.
Decomposition schemes split the problem into two parts: a quadratic optimal control problem and a set of single-period optimization problems. An iteration alternating these two steps then converges to a solution~\cite{operator_splitting_boyd}.

\Gls{admm} is a particular splitting technique that applies the method of augmented Lagrange multipliers.
By introducing the splitting variables $\tilde{x} \in \mathbb{R}^{n}$ and $\tilde{z} \in \mathbb{R}^{m}$, the problem in \cref{eq:qp} can be rewritten in consensus form~\cite{operator_splitting_boyd, Stellato_2020}:

\begin{equation}\label{eq:qp_cons}
    \begin{aligned}
        & \textnormal{minimize} \quad (1/2)x^{T}Px + q^{T}x + \mathcal{I}_{Ax=z}(\tilde{x},\tilde{z}) + \mathcal{I}_{\mathcal{C}}(z) \\
        & \textnormal{subject to} \quad (\tilde{x},\tilde{z}) = (x,z)
    \end{aligned}
\end{equation}

where $\mathcal{I}_{Ax=z}(\tilde{x},\tilde{z})$ and $\mathcal{I}_{\mathcal{C}}(z)$ are the indicator functions that equal $0$ when $Ax=z$ and $z \in \mathcal{C}$, respectively, or are $+\infty$ otherwise. 
The $k$ \gls{admm} iteration of \cref{eq:qp_cons} is obtained from alternating the minimization of its augmented Lagrangian over $\tilde{x}$ and $\tilde{z}$ as follows:

\begin{equation}\label{eq:admm_xztilde}
    \begin{aligned}
        (\tilde{x}^{k+1},\tilde{z}^{k+1}) \leftarrow & \operatorname {argmin}_{(\tilde{x},\tilde{z}):A\tilde{x}=\tilde{z}} (1/2)\tilde{x}^{T}P\tilde{x} + q^{T}\tilde{x} \\
        & + (\sigma/2) \lVert\tilde{x}-x^k+\sigma^{-1}w^k\rVert^{2}_{2} \\
        & + (\rho/2) \lVert\tilde{z}-z^k+\rho^{-1}y^k\rVert^{2}_{2}
    \end{aligned}
\end{equation}
\begin{equation}\label{eq:admm_x}
    x^{k+1} \leftarrow \alpha\tilde{x}^{k+1} + (1-\alpha)x^k +\sigma^{-1}w^k
\end{equation}
\vspace{-\baselineskip} 
\begin{equation}\label{eq:admm_z}
    z^{k+1} \leftarrow \Pi \left( \alpha\tilde{z}^{k+1} + (1-\alpha)z^k +\rho^{-1}y^k \right)
\end{equation}
\vspace{-\baselineskip} 
\begin{equation}\label{eq:admm_y}
    y^{k+1} \leftarrow y^k + \rho \left( \alpha\tilde{z}^{k+1} + (1-\alpha)z^k -z^{k+1} \right)
\end{equation}

$x$ and $z$ are the primal variables, $\sigma > 0$ and $\rho > 0$ are the step-size parameters, $\alpha \in (0,2)$ is the relaxation parameter, and $\Pi = \max(\min(z,u),l)$ is the Euclidean projection onto $\mathcal{C}$~\cite{Stellato_2020, Stellato_osqp_conference}. 
$y^k$ is the dual variable associated with the equality constraint $\tilde{z}=z$.
Note that $x$ and $z$ are updated in an alternating fashion. Hence the term \emph{alternating direction}~\cite{BoydAdmm2011}.

\subsection{The \gls{osqp} general-purpose solver}\label{sec:background:osqp}

\begin{algorithm}[t]
\caption{\Gls{osqp} solver}\label{alg:osqp}
\begin{algorithmic}[1]
\State \textbf{given} init. vals. $x^0$, $z^0$, $y^0$, params. $\rho>0$, $\sigma>0$, $\alpha \in (0,2)$
\Repeat
    \State $ \textnormal{solve}    
    \left[ 
    \begin{array}{@{}cc@{}}
        P+\sigma I & A^T \\
        A & -\rho^-1I
    \end{array}
    \right]
    \left[ 
    \begin{array}{@{}c@{}}
        \tilde{x}^{k+1} \\
        \nu^{k+1}
    \end{array}
    \right]
    = 
    \left[ 
    \begin{array}{@{}c@{}}
        \sigma x^k -q \\
        z^k - \rho^-1 y^k
    \end{array}
    \right]    
    $ \label{alg:osqp:linsys_xnu}
    \State $z^{k+1} \leftarrow z^k + \rho^{-1}(\nu^{k+1}-y^k)$ \label{alg:osqp:linsys_z}
    \State $x^{k+1} \leftarrow \alpha\tilde{x}^{k+1} + (1-\alpha)x^k +\sigma^{-1}w^k$\label{alg:osqp:x}
    \State $z^{k+1} \leftarrow \Pi \left( \alpha\tilde{z}^{k+1} + (1-\alpha)z^k +\rho^{-1}y^k \right)$\label{alg:osqp:z}
    \State $y^{k+1} \leftarrow y^k + \rho \left( \alpha\tilde{z}^{k+1} + (1-\alpha)z^k -z^{k+1} \right)$\label{alg:osqp:y}
\Until termination criteria or \texttt{max\_iter} reached
\end{algorithmic}
\end{algorithm}

\Gls{osqp} is an \gls{sota} general-purpose solver for constrained \gls{qp} problems based on \gls{admm}. 
The solver pseudocode is shown in \cref{alg:osqp}. 
For a complete overview of how to derive the optimality conditions in \cref{alg:osqp:linsys_xnu} and \cref{alg:osqp:linsys_z} from the $\tilde{x}$ and $\tilde{z}$ updates in \cref{eq:admm_xztilde}, refer to~\cite{Stellato_2020} and~\cite{BoydAdmm2011},  Section 4.2.5.
We call the coefficient matrix in \cref{alg:osqp:linsys_xnu} of~\cref{alg:osqp} the \gls{kkt} matrix, and label it $K$.
%
\Gls{osqp} combines the advantages inherited from first-order operator splitting methods with a streamlined, open-source\footnote{\url{https://github.com/osqp/osqp}} software package to produce, among other outputs, embedded C code.
In the following, we summarize its main features.

\subsubsection{Precomputation, caching, and sparsity}
The default QDLDL \emph{direct solver} used to solve the linear system in \cref{alg:osqp:linsys_xnu} of \cref{alg:osqp} comprises two steps: 
(i) Cholesky factorization of the \gls{kkt} matrix $K$ so that $K = LDL^T$, where $L$ is a lower triangular matrix and $D$ a diagonal matrix with nonzero diagonal elements, and 
(ii) lower and upper \gls{sptrsv} through \gls{fe} and \gls{bs}, respectively. 
To make the algorithm division-free, thereby reducing the execution time of the \gls{qp} problem, one can divide all the rows of the triangular matrix $L$ by the diagonal element of that row and move the divider into the $D$ matrix as compensation. 
Then, the $D^{-1}$ can be stored offline.

To improve memory efficiency, \gls{osqp} leverages problem sparsity and allows for offline pre-computation of the coefficient matrix $K$, caching it before \gls{admm} is run online.
The matrix $K$ in~\cref{alg:osqp:linsys_xnu} is block-sparse and quasi-definite.
\Gls{osqp} uses \gls{amd}~\cite{AMD_SPARSE} to compute the $LDL^T$ 
factorization of sparse matrices.
\Gls{amd} reorders the matrix $K$ with a permutation matrix $P$ so that $PKP^T = LDL^T$, reducing the fill-ins introduced in $L$ during factorization.
The factorization of the permuted matrix then comprises two steps: 
(i) \emph{symbolic factorization}, where the sparsity pattern for $L$ is computed based only on the nonzero pattern of $K$, and 
(ii) \emph{numerical factorization}, which computes the numerical values of $L$.
The symbolic and numerical factorizations can be computed only once, then stored and reused if only the vectors $q$, $l$, and $u$ in \cref{eq:qp} change for each \gls{admm} iteration, a typical scenario for linear \gls{mpc}~\cite{Stellato_2020} and also this work.
%

\Cref{subsec:etm_mpc:osqp_param,subsec:meth:codegen} detail how these features are leveraged for optimal configuration of the \gls{osqp} library and \gls{aot} scheduling of the sparse QDLDL solver, respectively.

\subsubsection{Termination}
In each iteration \( k \), \cref{alg:osqp} produces the triplet \( (x^k, z^k, y^k) \). The problem in \cref{eq:qp} is considered solvable if the primal and dual residuals converge to zero:

\[
\lim_{k \to \infty} r_{\text{prim}}^k = 0, \quad \lim_{k \to \infty} r_{\text{dual}}^k = 0
\]

where the primal residual \( r_{\text{prim}} = Gx - z \) represents how well the current solution satisfies the problem constraints, and the dual residual \( r_{\text{dual}} = Px + q + G^T y \) reflects how far the current solution is from minimizing the objective function (optimality). 
Given termination tolerances \( \epsilon_{\text{prim}} > 0 \) and \( \epsilon_{\text{dual}} > 0 \), \cref{alg:osqp} terminates in iteration \( k \) if:

\[
\lVert r_{\text{prim}}^k \rVert_2^2 \leq \epsilon_{\text{prim}}, \quad \lVert r_{\text{dual}}^k \rVert_2^2 \leq \epsilon_{\text{dual}}
\]

Frequent checks for termination can slow down algorithms, particularly on embedded platforms with real-time constraints and limited computational resources. 
To mitigate this, \gls{osqp} allows termination after a fixed number of iterations, controlled by \texttt{max\_iter} in~\cref{alg:osqp}. 
Proper selection of \texttt{max\_iter} is critical to maintaining control performance, as we will discuss in~\cref{subsec:etm_mpc:osqp_param}.

\subsubsection{Warm starting}
By providing an initial guess for the primal and dual solutions and setting the solution of a prior iteration as the initial value in the following iteration, \emph{warm starting} can improve the execution time on embedded systems and is ideal when the solution to \cref{alg:osqp} does not vary significantly between iterations.

\subsection{Model Predictive Control}\label{sec:background:mpc}

\Gls{mpc} is a predictive control technique that uses a dynamic model to forecast the future evolution of a system by solving an optimization problem at each discrete time step, called the \emph{\mpcstep} or \emph{sample time}~\cite{JEREZ_2}.
Controlling a constrained, linear time-invariant dynamical system evaluates to~\cite{Stellato_2020}:

\begin{equation}\label{eq:mpc}
    \begin{aligned}
        & \textnormal{minimize} \quad && x_{H_p}^{T}Q_{H_p}x_{H_p} + \sum_{h=0}^{H_p-1} x_{h}^TQx_h + u_{h}^TRu_h \\
        & \textnormal{subject to} \quad 
        && x_{h+1} = Dx_h + Eu_h \\
        &&& x_h \in \mathcal{X}, \; u_h \in \mathcal{U} \\
        &&& x_0 = x_{init}
    \end{aligned}
\end{equation}

Here, $x_h \in \mathbb{R}^{n_x}$ and $u_h \in \mathbb{R}^{n_u}$ represent the state and control input vectors, respectively; $\mathcal{X}$ and $\mathcal{U}$ denote feasible sets (constraints); $H_p$ is the prediction horizon; and $x_{init} \in \mathbb{R}^{n_x}$ is the initial system state. Cost matrices $Q$, $R$, and $Q_{H_p}$ penalize state and input deviations from their respective target values over the prediction horizon.
\Cref{eq:mpc} can be recast as a standard \gls{qp} from \cref{eq:qp} by defining a compound variable $z = (x_0, ..., x_{H_p}, u_0, ..., u_{H_p-1})$. This allows us to apply first-order solvers like \gls{osqp}, described in \cref{sec:background:fom} and \cref{sec:background:osqp}. The corresponding \gls{qp} problem size is $n = \textnormal{dim(D)} = n_x(H_p + 1)$ and $m = \textnormal{dim(E)} = n_u H_p$.
At each \mpcstep, the controller:
\textbf{(1)} measures the current system state ($x_{init}$);
\textbf{(2)} solves the \gls{qp} in~\cref{eq:mpc} for $u_h$;
\textbf{(3)} applies only the first control input $u_0$ from the optimized control sequence $\{u_0, \ldots, u_{H_p-1}\}$ to the system's actuators (i.e., control knobs), while the remaining values $\{u_1, \ldots, u_{H_p-1}\}$ are discarded.
As detailed in~\cref{subsec:etm_mpc:archi}, in our energy and thermal management use case, the state vector $x_h$ contains temperature readings for each \gls{pe}, while the control input vector $u_h$ specifies the power allocated to each \gls{pe} at time step $h$. Once the \gls{qp} is solved, the first input $u_0$ determines the power to apply via \glspl{vrm} and \glspl{pll}, which sets the corresponding frequency-voltage pairs through an inverse power model. This process repeats at each {\mpcstep} with updated sensor data.

This online computation of the first control input defines the \emph{implicit} \gls{mpc} scheme, targeted in this work, which contrasts with \emph{explicit} \gls{mpc}, where control actions are precomputed and stored for runtime lookup~\cite{fast_mpc_boyd}. While explicit approaches have been explored for next-generation 3D chips because of their speed~\cite{MPC_3D_1}, their exponential memory growth makes them unsuitable for embedded controllers managing large-scale systems~\cite{fast_mpc_boyd}.

\section{Embedded RISC-V \gls{llc}}\label{sec:background:llc}

\begin{figure}[t]
    \centering
    \includegraphics[width=\columnwidth]{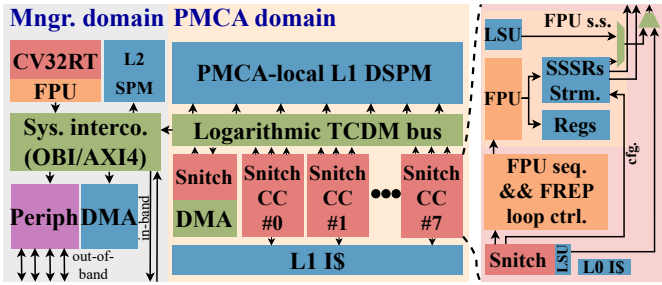}
    \caption{%
        ControlPULP microarchitecture. The Snitch \gls{pmca} and core complex with \glspl{sssr} and \texttt{frep} extensions are highlighted.
    }
    \label{fig:cpulp-archi}
\end{figure}

The embedded \gls{llc} employed in this work is based on \emph{ControlPULP}~\cite{Ottaviano2024}, an open-source, heterogeneous, 32-bit RISC-V platform managed by an \gls{rtos}-based software stack (FreeRTOS).
It comprises a single-core manager domain and an eight-core \gls{pmca}.
\Cref{fig:cpulp-archi} shows its architecture. 

\subsection{ControlPULP single-core domain}

ControlPULP's manager domain leverages the \gls{rtos} task scheduling mechanism to dispatch computationally heavy tasks to the \gls{pmca}, thereby accelerating the \gls{etm} control algorithm.
It oversees the on-chip AXI4-based \textit{in-band} communication channel, encompassing \gls{dma}-facilitated readouts from \gls{pvt} sensors, and regulates the allocation of frequencies to the \gls{hlc}'s \glspl{pe} according to the control policy.
Additionally, it controls the off-chip \textit{out-of-band} communication channel through dedicated power management peripherals for voltage control.

\subsection{Snitch \gls{pmca}}

The Snitch cluster~\cite{Scheffler_2023} is a RISC-V-based multicore \gls{pmca} specialized for energy-efficient floating-point computation. It serves as the \gls{llc}'s main compute unit.
The cluster features eight RISC-V compute cores, each extended with a hardware loop (\texttt{frep}) and sparsity-capable memory streaming units (\glspl{sssr}), described below, to maximize the utilization of its 
multi-precision \gls{fpu}. 
The compute cores share a 32-bank \gls{spm} accessed through a single-cycle logarithmic interconnect, as well as an L1 instruction cache.
An additional \gls{dma} core controls a tightly-coupled \gls{dma} engine that facilitates bulk data transfer between the \gls{pmca}'s \gls{spm} and the manager domain.
While the compute cores use a 32-bit instruction set, the \glspl{fpu} and memory system can be be configured as either 32 or \SI{64}{\bit} wide, depending on the requirements of the target application.

\subsection{\Glspl{sssr} and hardware loops}

Each of the \gls{pmca}'s compute cores features three \glsunset{ssr}\emph{stream semantic registers}~(SSRs)~\cite{Scheffler_2023}, which map streaming accesses to the shared \gls{spm} directly to floating-point registers.
\glspl{ssr} generate stream addresses using dedicated hardware units; this frees the compute core of issuing the loads, stores, and address computations required for streaming memory accesses and enables the near-continuous issuing of useful \gls{fpu} compute instructions.
All three \glspl{ssr} are capable of up to 4-dimensional strided streams to accelerate compute kernels with regular memory access patterns.
We further use the \emph{sparse SSR}~(SSSR) extension so that two of the three \glspl{ssr} are additionally capable of \emph{indirect} streams, significantly accelerating the irregular access sequences of sparse workloads like \gls{qp}.
These indexed \glspl{sssr} support reading 8-, 16-, 32-, and 64-bit index arrays from \gls{spm} to compute addresses for indirect read (gather) or write (scatter) streams accessing \gls{spm}.

To autonomously execute repeated sequences of floating-point operations, such as those in iterative loops, the \gls{pmca} additionally incorporates the \texttt{frep} (floating-point repetition) extension.
\texttt{frep} allows floating-point instruction sequences to be offloaded into a dedicated loop buffer, effectively decoupling \gls{fpu} instruction issuing from the integer core and thereby enhancing energy efficiency and floating-point utilization.

\subsection{Memory system hierarchy}
The \gls{llc} requires more than the few tens of \si{\kibi\byte} of memory typically found on low-end microcontrollers.
Instead, they are powerful mid-end embedded devices leveraging specialized libraries to handle \gls{mimo} interactions and process large volumes of data. 
This results in memory requirements on the order of \SI{1}{\mebi\byte}.

ControlPULP's memory system uses on-chip \glspl{spm}, which is still compatible wth on-chip integration at an acceptable cost, thereby reducing access latency, improving predictability, and increasesing the controller's autonomy within the integrated system.
To improve data locality in memory-intensive control algorithms, the \gls{pmca}'s local L1 \gls{spm} can be scaled up, leaving a smaller shared L2 memory for instruction and data storage in the manager domain.
Alternatively, if implementation constraints dictate a smaller \gls{spm} size for the \gls{pmca}, double-buffering can hide \gls{spm} refill latency from the manager domain thanks to the \gls{pmca}'s dedicated \gls{dma} core (\cref{fig:cpulp-archi}).


\subsection{Real-time control constraints}\label{sec:background:rt}

\Glspl{llc} for \gls{etm} are subject to \emph{soft} real-time constraints. Misbehavior of the control policy could negatively affect the chip's performance in case of deadline misses, causing temperature hot spots and overshoots.
For these reasons, ControlPULP features streamlined interrupt processing and fast context switch capabilities~\cite{Ottaviano2024}.
A well-designed controller should be fast enough to meet its soft deadlines and minimize response variation (jitter) through hardware-software cooperation.
An \gls{llc} typically measures the controlled system's state and applies operating points periodically.
In ControlPULP, the FreeRTOS-based \gls{etm} application layer is organized in \emph{tasks} with different priorities and periods responsible for the overall power and thermal policy~\cite{Ottaviano2024}.
Assuming a task set \(\Theta\) with periods \(\rho_\theta\) for \(\theta \in \{0, 1, ..., |\Theta|-1\}\), the least common multiple (LCM) of these periods, denoted as \(LCM(\{\rho_0, \rho_1, ..., \rho_{|\Theta|-1}\})\), is called the \emph{hyperperiod}~\cite{HYPERPERIOD}.

In this work, the \gls{mpc} controller is encapsulated as a FreeRTOS task $\theta_{MPC}$ with periodicity $\rho_{MPC}$. Hence, the \gls{mpc} step introduced in~\cref{sec:background:mpc} equals $\rho_{MPC}$. 
The time to run \cref{alg:osqp} must be less than $\rho_{MPC}$. We call \emph{slack} the free time left after task execution ends and before a new \gls{mpc} step starts. 

\section{\Gls{mpc} controller design and motivation}\label{sec:etm_mpc}

\glsreset{mil}

We first present the power and thermal model for \gls{mpc} design~(\cref{sec:background:tpm}).
Then, we review the \gls{mpc} architecture in \cref{subsec:etm_mpc:archi}, and discuss the parameterization of the \gls{osqp} solver used throughout the work in~\cref{subsec:etm_mpc:osqp_param}.
Finally, in~\cref{subsec:etm_mpc:dmp}, we propose an optimization procedure, termed \gls{dmp}, to increase the sparsity of the problem, thereby reducing its computational complexity and storage requirements.
We use the \gls{osqp} solver MATLAB interface with the YALMIP framework to conduct the \gls{mil} optimization of the control algorithm. 
The end-to-end flow down to hardware-aware optimization is presented in~\cref{subsec:meth:framework}.

\subsection{Thermal and power model for multi-core chiplets}\label{sec:background:tpm}

\begin{figure}[t]
    \centering
    \includegraphics[width=\columnwidth]{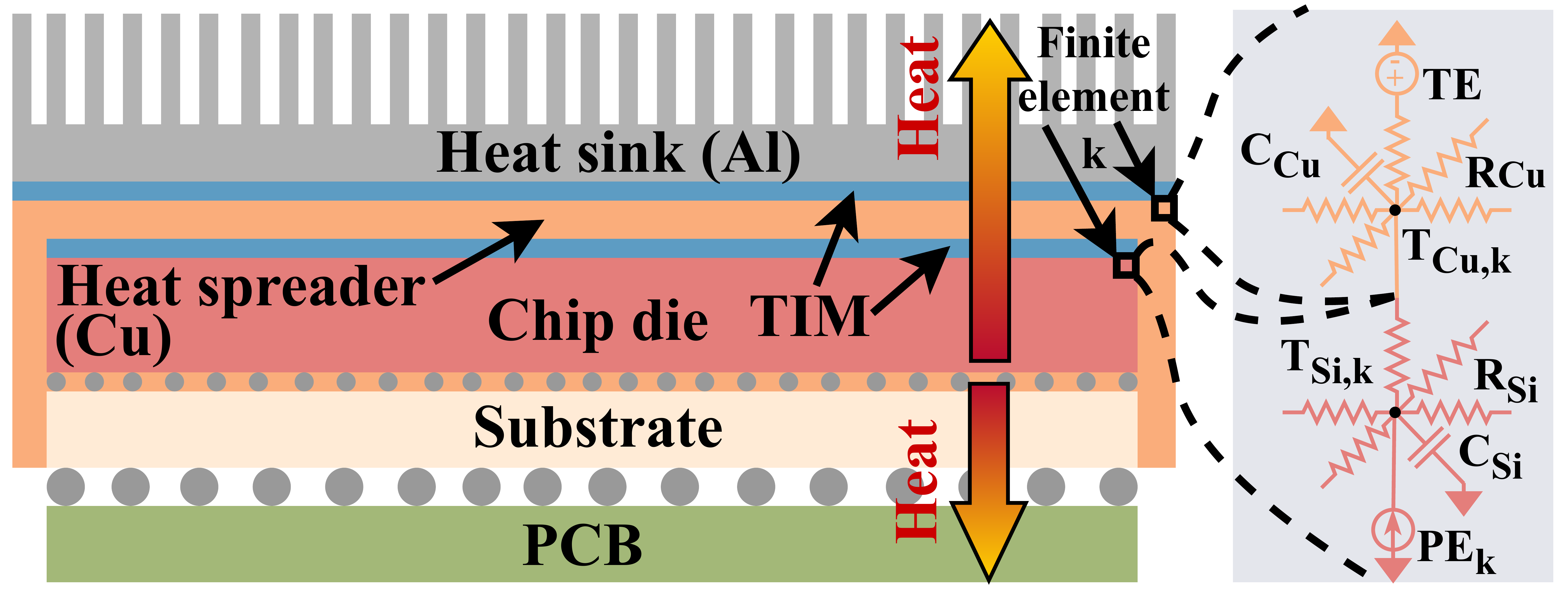}
    \caption{%
        Thermal components with heat propagation flow of an \gls{hpc} chiplet (left) and lumped parameter circuit of two finite elements from the silicon die and copper heat spreader, respectively (right).
    }
    \label{fig:chiplet-model}
\end{figure}

In this section, we introduce the thermal and power model of a multi-core chiplet, which is used to define the optimization problem described in the following \cref{subsec:etm_mpc:archi}.
For a detailed treatment, we refer the reader to~\cite{bambini2024arxiv,thermal_survey}.

The thermal structure is shown~\cref{fig:chiplet-model}.
It comprises, from top to bottom, a silicon die integrating a grid of $N_c = N_w \times N_h$ \glspl{pe}, a substrate layer, and the carrier \gls{pcb}. The grid is chosen to be square ($N_w = N_h$).
The main heat dissipation path, shown on the right side of~\cref{fig:chiplet-model}, passes through a copper heat spreader placed over the active silicon devices and an aluminum heat sink.
Thermal interface materials (TIMs) facilitate thermal conductance across layers.

The thermal and power model is derived from the finite element spatial discretization of the partial differential equations (PDEs) describing the silicon and copper heat dissipation.
The continuous model is discretized into $h_C$ finite elements associated with the $N_c$ \glspl{pe}.
An element is modeled with a lumped-parameter circuit, encapsulating the thermal capacitance and resistance of the neighboring materials. Each \gls{pe} is interpreted as an independent power source $P_i$ in the lumped representation.
An element has two thermal state variables for the (local) silicon die and metallic heat spreader temperatures.
Spatial discretization allows one to obtain a set of ordinary differential equations (ODEs) tractable for control design purposes.

\subsubsection{Power model dynamics}\label{sec:background:tpm:pm}

The power model dynamics are non-linear. The power source $P_i$ associated with \gls{pe} $i$ is:

\begin{equation}\label{eq:act_pow_map}
    \begin{aligned}
        P_{i} &= P_{i,\text{stat}} + P_{i,\text{dyn}} \\
        &= k_{s_0} + (I_{cc,i} V_i) \cdot \mathcal{K}(T_{Si,i}, V_i) + C_{eff,i} F_i V_{i}^2
    \end{aligned}
\end{equation}

where $P_{i,\text{stat}}$, $P_{i,\text{dyn}}$ are the static and dynamic component of the \gls{pe}'s power consumption. 
The effective capacitance $C_{\text{eff,i}}$ is correlated to the runtime workload, i.e., the type of instructions $\omega_i(t)$ executed by the \gls{pe}.
$\mathcal{K}(T_{Si,i}, V_i)$ is a non-linear mapping that encapsulates the dependency of the static leakage power on the voltage and temperature of the component. In this work, we use an exponential relation based on~\cite{bartolini2019advances}

\begin{equation}\label{eq:exp_leakage}
    \mathcal{K}(T_{Si,i}, V_i) = e^{
    k_v V_i(t) + k_T T_i(t) + k_{T_0}
    }
\end{equation}

where the $k$ parameters are constant and computed on the critical values of $V_{\text{MAX}}$ and $T_{\text{MAX}}$.

\subsubsection{Thermal model dynamics}\label{sec:background:tpm:tm}

Collecting the temperatures of all the components shown in \cref{fig:chiplet-model} for all the $h_C$ elements obtained after discretization in a unique vector $T=(T_{Si,1}, T_{Cu,1}, \ldots, T_{Si,n_s}, T_{Cu,n_s}, T_{{Al}})^T$, the ODEs for silicon and copper can be written in compact form:

\begin{equation}\label{eq:lumped_par_mod_comp}
    \begin{cases}
        \dot{T}(t)=A_T \cdot T(t) + B_T \cdot P(t)  \\
        T_{Si}=C_T \cdot T(t)
    \end{cases}
\end{equation}

where $P = (P_1, \ldots, P_{n_c})^T$ is the vector of power sources associated with each discrete element, and $A_T$, $B_T$, $C_T$ encapsulate the thermal constants of the lumped representation~\cite{bambini2024arxiv}. 

The dynamics in \cref{eq:lumped_par_mod_comp} and the algebraic power model in \cref{eq:act_pow_map} jointly characterize the system's thermal and power behavior.

\subsection{\Glsentrytext{mpc} Controller architecture}\label{subsec:etm_mpc:archi}

\begin{figure}[t]
    \centering
    \includegraphics[width=\columnwidth]{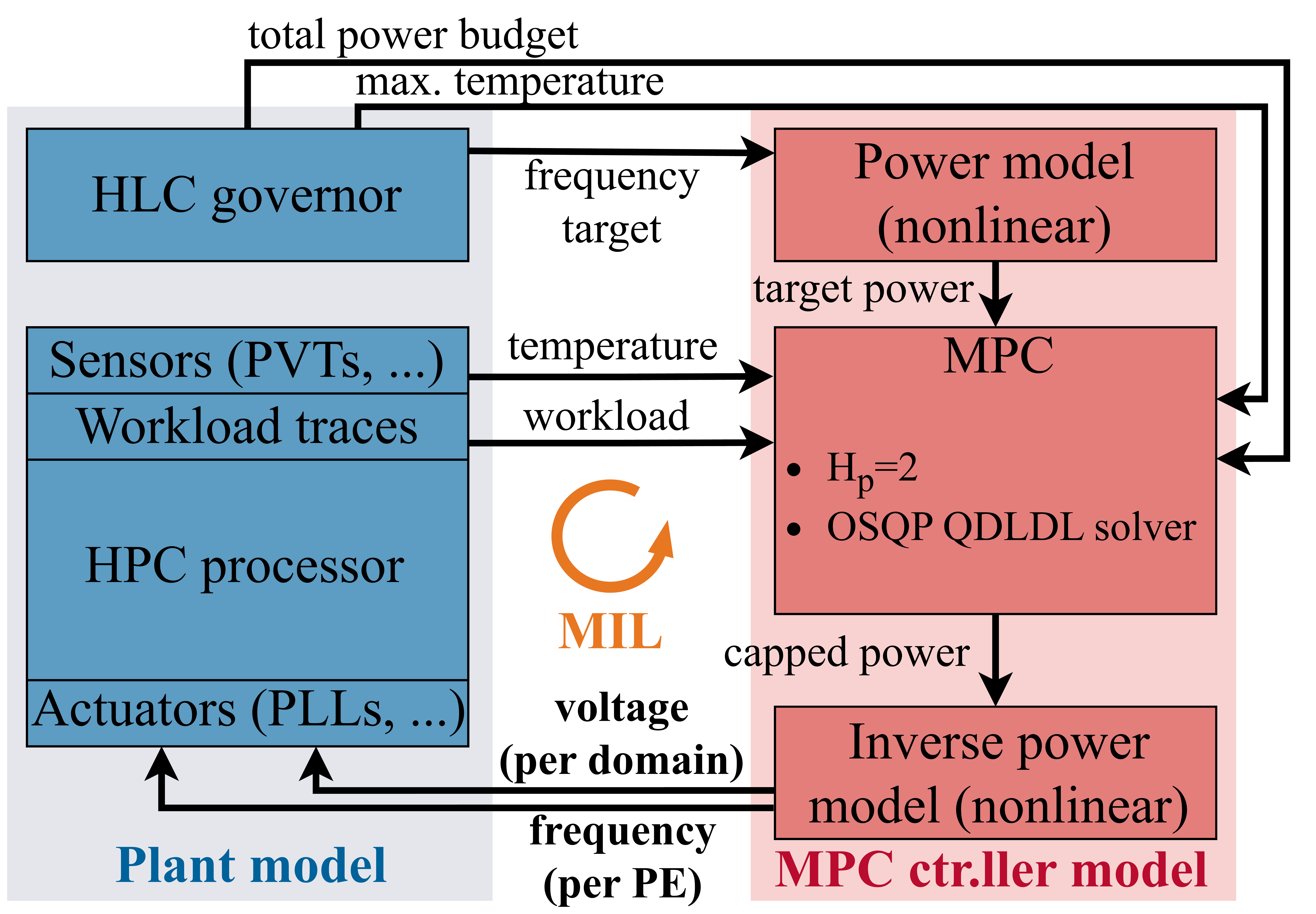}
    \caption{%
        Architectures of \gls{etm} \gls{mpc} controller (right) and plant (left). Both are modeled in MATLAB; the controller architecture is progressively refined using \gls{mil}.
    }
    \label{fig:etm-controller-archi}
\end{figure}

\Cref{fig:etm-controller-archi} illustrates the control loop between the controller (right) and the controlled \gls{hpc} processor (left) models during \gls{mil} design. 
From a control perspective, the controlled system is referred to as the \emph{plant}.
As described in~\cref{sec:background:tpm}, the plant is assumed structured as a grid of $N_c$ \glspl{pe}.
The control problem for managing the power and temperature of each \gls{pe} in the grid with a prediction horizon $H_p$ is denoted as $\mathtt{P N_w \times N_h\_{Hp}}$. 
For instance, $\mathtt{P5\times5\_H2}$ refers to controlling a $5 \times 5$ grid ($N_c = 25$) with $H_p = 2$. 
Since we gather sensor information and dispatch control action to all the \glspl{pe} through a single controller, this approach is \emph{centralized}.

The controller operates in three stages, separating non-linearity handling from a centralized linear constrained optimization. 
First, a power model computes the target power $P_i^*$ for each \gls{pe} based on the set-point \gls{dvfs} commands from the \gls{hlc}, as well as the measured workload and temperature from the previous interval using on-chip sensors and counters.
In the second stage, a linear \gls{mpc} controller aims to achieve the target power while respecting thermal and power constraints.
If constraints are exceeded, the \gls{mpc} reduces the target power $P_i^*$ (\emph{power-capping}).
In the third stage, the output powers $P_i$ are converted into per-domain voltage and per-\gls{pe} frequency pairs using an inverse power model and dispatched to \glspl{pll} and \glspl{vrm} actuators. \Glspl{pe} are grouped into power domains $\mathcal{D}_j$, each associated to a \gls{vrm} and its relative power limit~\cite{bambini2024arxiv}.

The chosen design for the \gls{mpc} controller not only aims at capping the temperature of each single \gls{pe}, as most \gls{sota} controllers do~\cite{6178247,TILLI2022105099}, but enforces shared power constraints and distributes power among \glspl{pe} too, representing an advancement over \gls{sota}. 
With this structure, power and thermal constraints are enforced simultaneously, obtaining a globally optimal operating point.
%
%
An additional global power budget constraint $P_{\text{B}}$ is enforced for the plant.
Being shared among the \glspl{pe}, these power constraints require either a centralized \gls{mpc} structure or the relaxation of these constraints in a distributed design.
The centralized approach thus avoids approximations and delays in effectively enforcing shared power constraints. However, this design choice increases model complexity and computational demands, seen in the dimensional growth of $P$, $A$, $l$, and $u$ in~\cref{eq:qp}.

The \gls{mpc} constraints can be summed up as follows:
\begin{equation}\label{eq:mpc-constr}
	\begin{cases}
		T_i \leq T_{\text{L}}, \qquad i=1,\dots ,  N_c, \\
        P_{\text{min}} \leq P_i \leq P_{\text{max}}, \\
		\sum_{i=1}^{N_c} P_i \leq P_{\text{B}}[t], \\
		\sum_{i \in \mathcal{D}_j} P_i \leq P_{\mathcal{D}_j}[t], \quad j=1,\dots ,  n_d,
	\end{cases}
\end{equation}
where $T_{\text{L}}$ is the temperature limit and $P_{\text{min}}$ and $P_{\text{max}}$ represent the plant physical limits. 
In addition to these constraints, the linear thermal model~\cref{eq:lumped_par_mod_comp} is used as a model constraint.

The optimization function $J$ penalizes deviations from the target power:
\begin{equation}\label{eq:mpc-costj}
	J = ( {P^*} - {P})^{T} D ({P^*} - {P}),
\end{equation}
where $P$ and $P^*$ are the vectors of dispatched and target power for all \glspl{pe}, and $D$ is a diagonal matrix.

The choice of the prediction horizon $H_p$ significantly impacts computational feasibility and controller performance. 
Longer horizons would better anticipate future dynamics, but in this application, high-amplitude noise~\cite{bambini2024arxiv} causes significant prediction divergence, invalidating the effectiveness of high $H_p$ values. 
An appropriate $H_p$ must balance thermal dynamics timing and computational cost. For a control interval $t_s \in [1,3] \ \text{ms}$, a range of $H_p \in [2,4]$ is a suitable choice based on the fastest thermal time constant~\cite{bambini2024arxiv}.
In this setting, high-frequency power spikes can be handled by a parallel, faster control loop reacting to oscillations.

As recalled in~\cref{sec:background:osqp}, a favorable case for fast online \gls{mpc} arises when the dynamical system and constraints remain constant at each \mpcstep.
In such cases, the matrices $P$ and $A$ in the \gls{qp} formulation are constant, and any time-varying dynamics can be modeled as updates to the lower and upper bounds $l$ and $u$.
This makes the \gls{kkt} matrix $K$ in \cref{alg:osqp} time-invariant, allowing its symbolic and numerical factorizations to be precomputed and stored only once.
%
%
Thus, the initial condition in~\cref{eq:mpc} is treated as an equality constraint instead of integrating it into the dynamic equation.\looseness=-1

\subsection{\Gls{osqp} parameterization}\label{subsec:etm_mpc:osqp_param}

We use the default \gls{osqp} library configuration, selecting the direct QDLD solver with Cholesky decomposition. 
This is preferred to indirect methods as it is more suitable for very large problems ($N_{nnz} > 10^6$)~\cite{Stellato_2020,SCHUBIGER202055}.
\Gls{amd} reordering during factorization ensures that fill-ins are minimized, preserving the system's sparsity and memory efficiency.

For termination, we set the primal and dual tolerances to $\epsilon_{prim} = \epsilon_{dual} = 0.01$, achieving a balance between accuracy and computational cost.
The convergence-critical \gls{admm} step-size parameters $\rho$ and $\sigma$ are fixed at $\rho = 0.1$ and $\sigma = 10^{-6}$. 
The relaxation parameter is set to $\alpha = 1.5$.

To further enhance performance on embedded platforms~(\cref{subsec:eval:func}), we fine-tune the \gls{osqp} parameterization during \gls{mil}. 
We combine warm-starting, using the previous iteration's solution as the initial value~(\cref{sec:background:osqp}), with a fixed \texttt{max\_iter}\;$=15$ to reduce computation time while ensuring constraint satisfaction. 
Convergence behavior with this iteration limit is discussed in~\cref{subsec:meth:precision}.
However, in cases with sharp reference changes, additional iterations may be required for convergence.

Moreover, we disable \gls{osqp}'s adaptive step-size scheme, which would otherwise adjust $\rho$ based on the primal-to-dual residual ratio at runtime, introducing an expensive online division operation.
Lastly, solution polishing is disabled to avoid resolving a linear system with only the active constraints~\cite{Stellato_2020}, prioritizing execution speed.
%

\subsection{Discrete model pruning}\label{subsec:etm_mpc:dmp}

\begin{figure}[t]
    \centering
    \includegraphics[width=\columnwidth]{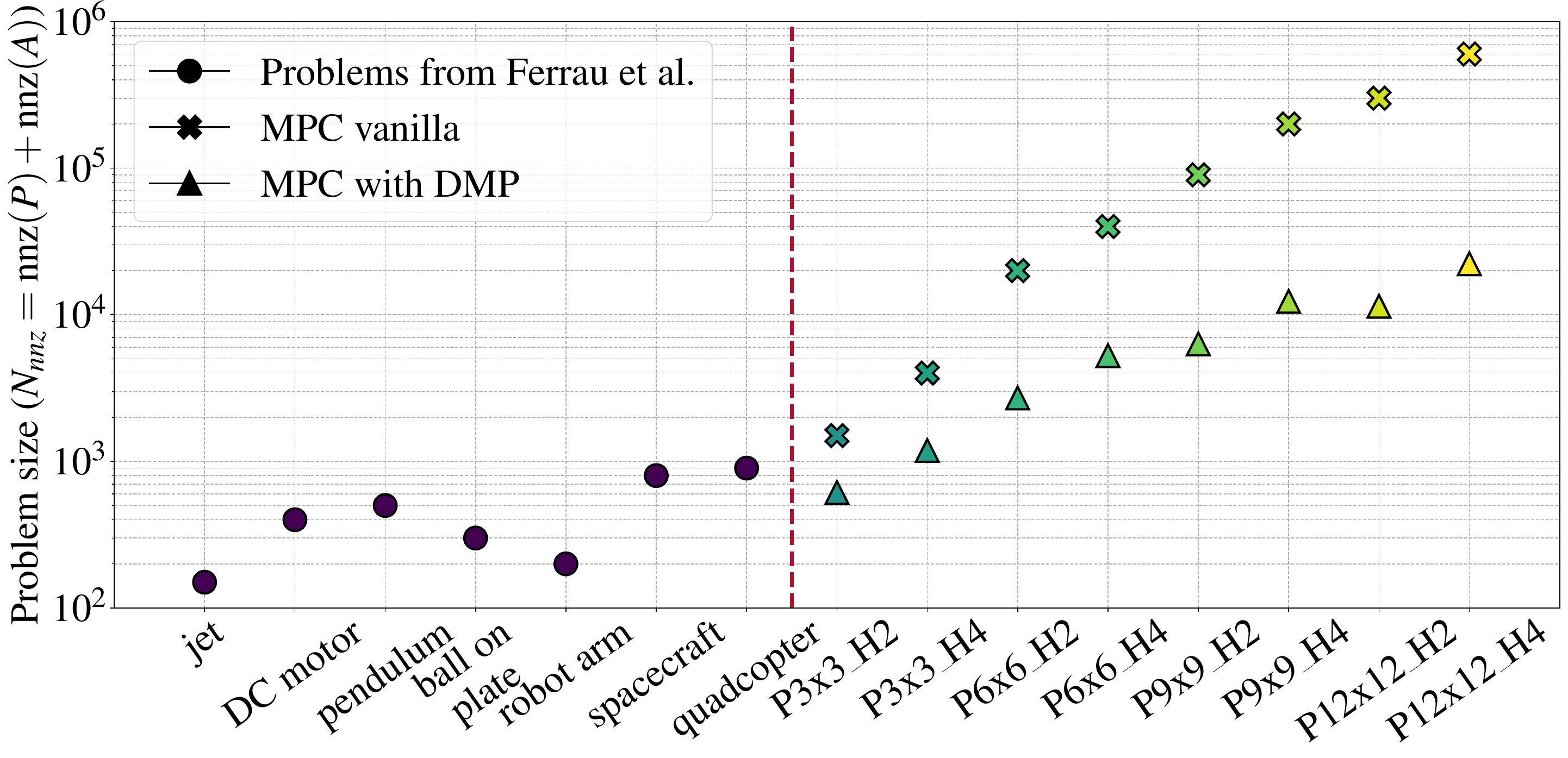}
    \caption{%
        Problem size (\cref{eq:num_tot_nnz}) of vanilla and pruned \gls{mpc} for \gls{etm} at a varying number of controlled \glspl{pe} and fixed horizon $H_p$=2. \Gls{dtm} uses a cutoff of $0.005$. 
        For comparison, the figure reports the problem size of other \gls{mpc} problems, reproduced from~\cite{FerrauABB_1}.
    }
    \label{fig:mpc-vs-others-mem}
\end{figure}

\begin{figure}[t]
    \centering
    \includegraphics[width=\columnwidth]{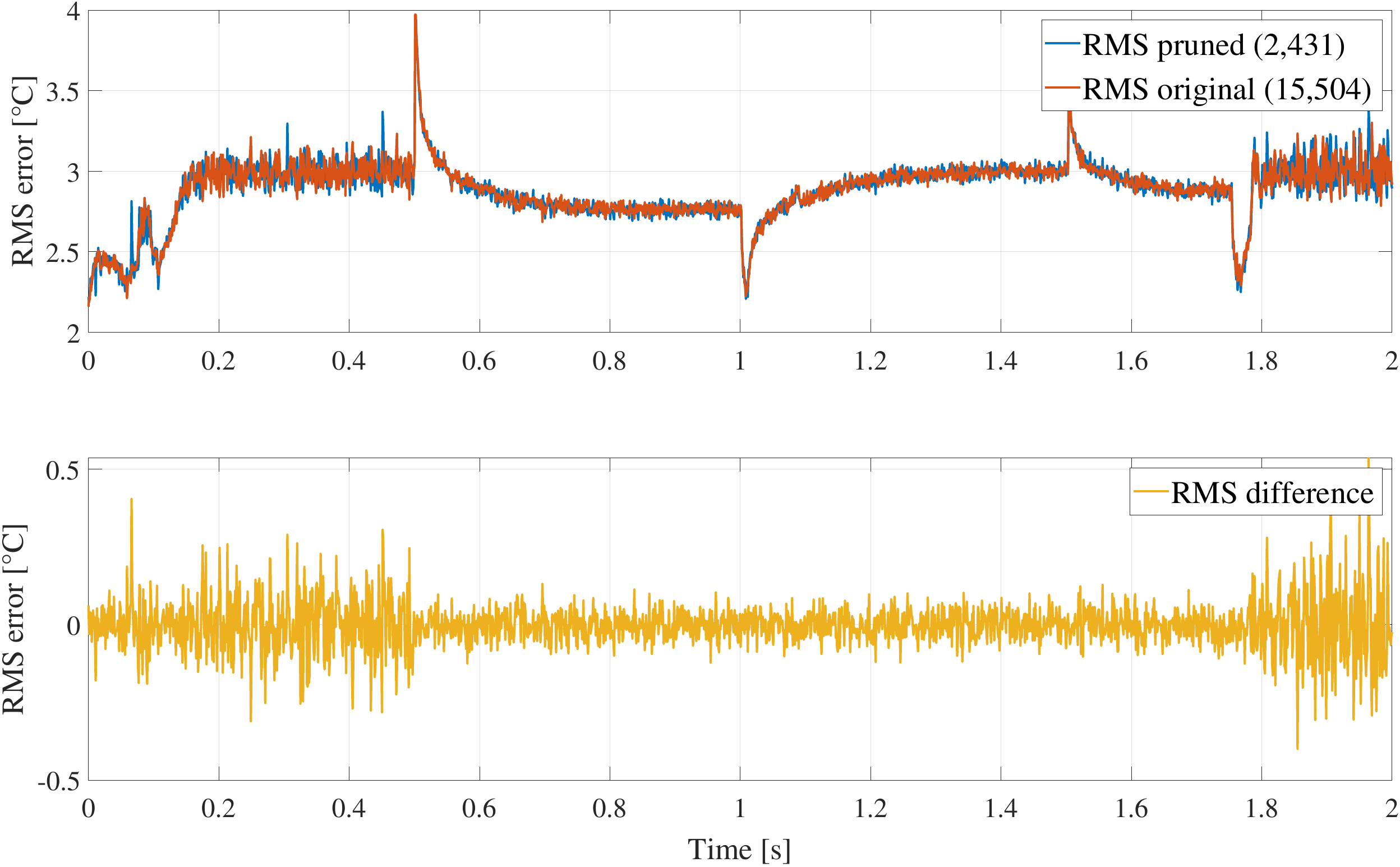}
    \caption{%
        Comparison of vanilla \gls{mpc} and \gls{mpc} after \gls{dmp} on the \texttt{P9$\times$9\_H2} problem using a MATLAB-based \gls{mil} simulation running for $2s$. The top chart shows the \gls{rmse} between predicted and measured \gls{mpc} temperatures over time, while the bottom chart highlights the difference between the two \glspl{rmse}. Spikes and offsets in the top chart are due to power budget changes and power model uncertainties, respectively, and are independent of the \gls{mpc}, hence the \gls{rmse} difference is unaffected.
    }
    \label{fig:mpc-og-vs-prun}
\end{figure}

\begin{figure}[t]
    \centering
    \includegraphics[width=\columnwidth]{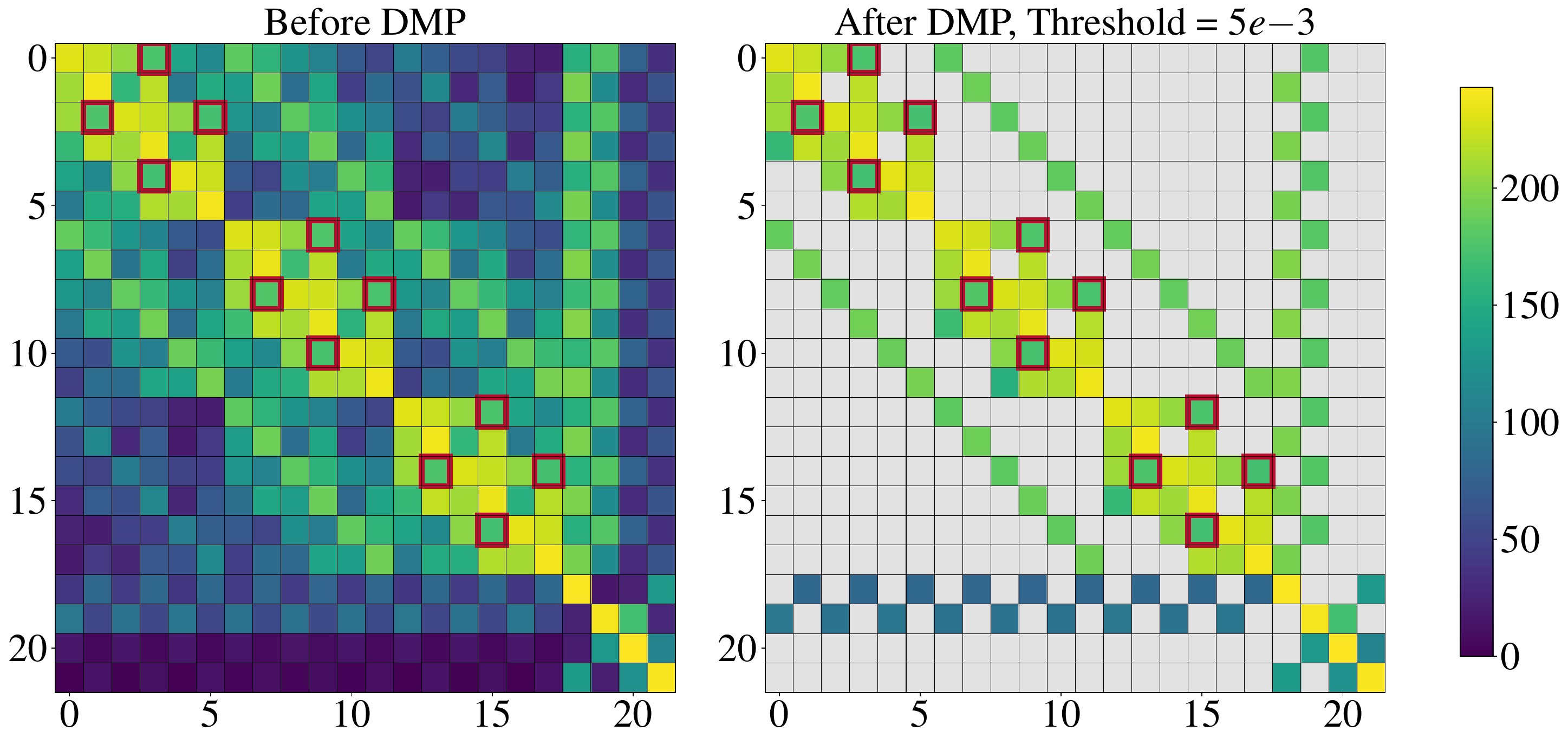}
    \caption{%
        The rationale behind \gls{dmp}. (Left) vanilla, (right) pruned state matrix $A_T$ in~\cref{eq:lumped_par_mod_comp}. We show the problem \texttt{P3$\times$3\_H2} for readability purposes. 
        Pruned entries are represented in grey. Entries highlighted in red indicate the non-zero values emerged after discretization.
    }
    \label{fig:dmp}
\end{figure}


The right-hand side of~\cref{fig:mpc-vs-others-mem} shows the size of various \gls{mpc} problems, quantified by the number of non-zero elements ($N_{nnz}$) in the corresponding \gls{qp} from \cref{eq:num_tot_nnz}. 
$N_{nnz}$, along with the number format, significantly influences the memory footprint; data precision is thoroughly discussed in~\cref{sec:methodology}. 
The solid cross markers in~\cref{fig:mpc-vs-others-mem} depict a standard problem formulation using the discretized thermal model recalled in~\cref{sec:background:tpm}.
We consider prediction horizons of 2 and 4 to demonstrate the influence of the horizon length on the problem size.

From a thermal perspective, \glspl{pe} primarily influence their immediate neighbors~\cite{TM-Gray-Beneventi}. 
Through these direct couplings, their impact propagates across all other \glspl{pe}, albeit with diminishing amplitude.
When discretizing the continuous system, depending on the discretization method and the chosen sampling interval $t_s$, additional couplings among \glspl{pe} in $A_T$ in~\cref{eq:lumped_par_mod_comp} may appear compared to the continuous model structure.
By selecting a sufficiently small $t_s$ relative to the thermal time constants, these additional couplings exhibit minimal amplitudes and negligible influence.
This is because the iterative time-step computation of the thermal model itself effectively captures the thermal influence propagation~\cite{TM-Gray-Beneventi}.

We propose a heuristic threshold-based pruning mechanism to exploit these weak, near-zero thermal couplings, thereby reducing the number of non-zero elements in the state matrix and, consequently, the memory footprint. 
Connections with values above the threshold remain unaltered.
Additional measures are taken to preserve the continuous-time structure of the thermal model matrices $A_T$ and $B_T$, ensuring that the plant dynamics are not truncated.
The pruning process is solely intended for memory and computational optimization, without introducing complex approximation criteria.
%
%
We call the method \gls{dmp}, as visually illustrated in~\cref{fig:dmp}.
%
%
Given a candidate cutoff threshold, the pruning algorithm computes the \glspl{rmse} between the predicted and measured \gls{mpc} temperature for all system states in the original and pruned model. 
It then selects the cutoff threshold that minimizes the difference in time between the \glspl{rmse} within an allowed range that we set to $\pm$\SI{0.5}{\degreeCelsius}.
In this work, we identify a cutoff threshold of $0.005$ for all problems. 
\Cref{fig:mpc-og-vs-prun} shows the \gls{rmse} evolution for both models of the \texttt{P9$\times$9\_H2} problem and their difference over time. 
The pruned thermal model maintains accurate temperature predictions, with the \gls{rmse} remaining within the allowed deviation range.

It is important to stress that, despite pruning, \gls{mpc} remains a more stable and safer alternative to classic \gls{pid} control. 
\Gls{mpc} integrates all control variables and constraints into a unified optimization framework by design, whereas \gls{pid}, as a traditionally single-input single-output method, addresses them separately~\cite{6178247}.
Furthermore, it has been shown~\cite{bambini_phd_2025} that the vanilla (non-pruned) \gls{mpc} outperforms the traditional voting-box \gls{pid}~\cite{ibm_occ}, achieving a worst-case temperature excedance relative to the temperature limit of less than \SI{1.7}{\degreeCelsius}, compared to up to \SI{40}{\degreeCelsius} for \gls{pid}.
Similar trends are observed in the average power exceedance relative to the power budget, as well as in the total time exceeding the temperature and power limits~\cite{bambini_phd_2025}.
Therefore, the approximation of \gls{dmp}, bounded within \SI{0.5}{\degreeCelsius}, slightly penalizes the original \gls{mpc} performance but preserves its significant advantages over \gls{pid}.

\Gls{dmp} reduces the model complexity (and size) from $O(N_c^2)$ to $O(N_c)$.
This reduction is crucial for scaling centralized control as the number of \glspl{pe} on a silicon die increases.
We show this behavior in~\cref{fig:mpc-vs-others-mem}, where the solid triangular markers represent the problem sizes after \gls{dmp}.
As anticipated, the reduction is more pronounced for larger $N_c$, resulting in up to 27$\times$ decrease in the number of non-zeros for \texttt{P12$\times$12\_H4}, compared to a 2.5$\times$ decrease for \texttt{P3$\times$3\_H2}.

To contextualize these values, the left-hand side of~\cref{fig:mpc-vs-others-mem} shows typical sizes of embedded \gls{qp} problems for various applications.
Ferrau~et~al.~\cite{FerrauABB_1} propose an open-source benchmarking framework for comparing the numerical performance of embedded \gls{qp} algorithms based on a general \gls{mpc} formulation. 
We replicate their results, extracting the problem sizes defined in~\cref{eq:num_tot_nnz}.
For the largest evaluated example, \texttt{P12$\times$12\_H2}, the vanilla \gls{mpc} formulation is approximately 335$\times$ larger than the quadcopter problem from~\cite{FerrauABB_1}\footnote{We note that the quadcopter reference given in~\cite{FerrauABB_1} lacks the pitch, yaw, and roll variable decoupling commonly applied in practice for the embedded domain, which would further reduce its complexity.}.
Despite the complexity reduction achieved by \gls{dmp}, the pruned \gls{mpc} model still remains over 13$\times$ larger.

This section highlighted the significant memory challenges of the analyzed problem. The next \cref{sec:etm_mpc:resources} addresses these challenges from an execution time perspective.

\subsection{Impact of MPC on execution time}\label{sec:etm_mpc:resources}

\begin{figure}[t]
    \centering
    \includegraphics[width=\columnwidth]{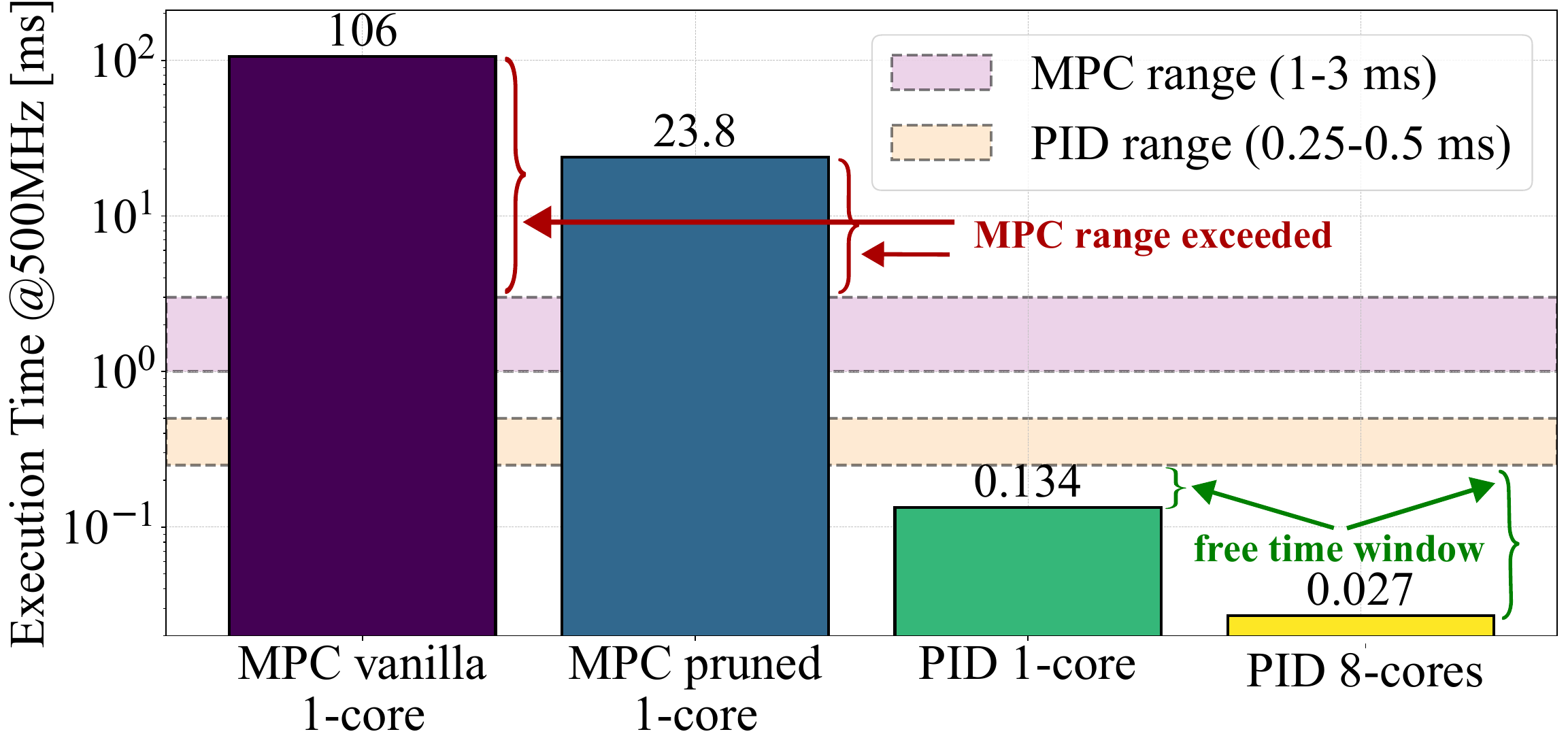}
    \caption{%
        Execution time of vanilla and pruned \gls{mpc} and \gls{pid}'s \gls{etm} policies on ControlPULP at \SI{500}{\mega\hertz}.
        Data for \gls{pid} are taken from~\cite{Ottaviano2024} when controlling 72 \glspl{pe}.
        \Gls{mpc} is executed for 15 iterations until convergence using vanilla \gls{osqp} for \texttt{P9$\times$9\_H2} problem, i.e., 81 \glspl{pe}.
    }
    \label{fig:mpc-vs-others-speed}
\end{figure}

We contextualize the high computational demands of \gls{mpc} for \gls{etm} by providing a quantitative comparison of the execution time compared to a \gls{pid} control policy, which represents the current \gls{sota} in this application domain~\cite{Ottaviano2024}.
The comparison is shown in~\cref{fig:mpc-vs-others-speed}. All algorithms are executed on ControlPULP in different configurations. For the \gls{mpc}, the \gls{qp} solver runs 15 iterations. \Gls{pid} results are taken from~\cite{Ottaviano2024}.
The light yellow and orange areas indicate the permissible execution ranges for \gls{pid} and \gls{mpc} algorithms. 
\Gls{pid} is computationally lightweight and generally faster than \gls{mpc}. \gls{sota} controllers for power and thermal management can execute one control policy iteration in \SIrange{250}{500}{\micro\second} (hyperperiod range, \cref{sec:background:rt}).  Its optimization on the \gls{pmca} further increases available slack (green arrows in~\cref{fig:mpc-vs-others-speed}). 
In contrast, both vanilla and \gls{dmp} \gls{mpc} fail to meet their deadline (red arrows in~\cref{fig:mpc-vs-others-speed}) by a large margin, being 35.3$\times$ and 7.9$\times$ slower than the \SI{3}{\milli\second} upper bound of the permissible range discussed in~\cref{subsec:etm_mpc:archi}.
Together with the problem size comparison in~\cref{fig:mpc-vs-others-mem}, this further motivates the need for hardware-software co-optimization to streamline and accelerate the execution of expensive \gls{etm} predictive control schemes on resource-constrained systems. 
In the following section, we detail our methodology to achieve this goal.
\section{Hardware-software co-optimization and acceleration methods}
\label{sec:methodology}

We first outline the optimization metrics and framework considered for efficient deployment of the \gls{mpc} solver algorithm on the target multi-core embedded platform (\cref{,subsec:meth:metrics,subsec:meth:framework}).
We then present and discuss the proposed hardware-aware optimizations to accelerate the algorithm and maximize resource utilization.
The optimization scope includes numerical precision considerations (\cref{subsec:meth:precision}), \gls{hypt} for multi-core acceleration and code generation of the \gls{sptrsv} (\cref{subsec:meth:codegen}), and leveraging the \gls{llc}'s \glspl{sssr} and hardware loop extensions (\cref{subsec:meth:sssr}).

\subsection{Optimization metrics}\label{subsec:meth:metrics}

The following optimization metrics are essential for efficient deployment on a resource-constrained system:

\paragraph{\textbf{Memory Footprint}} The goal is twofold: (i) to keep data close to the compute resources in low-latency on-chip \gls{spm} memories, as accessing off-chip \gls{dram} increases latency and reduces the autonomy of the \gls{llc}, and (ii) to ensure that the compiled algorithm size remains below \SI{1}{\mebi\byte} as discussed in~\cref{sec:background:llc}. 
Preserving sparsity in the state matrices is crucial to achieving this goal. This motivates the use of \gls{dmp} during \gls{mil} and \gls{amd} reordering during Cholesky factorization.

\paragraph{\textbf{Execution speed}} To meet the real-time constraints described in~\cref{sec:background:rt,subsec:etm_mpc:dmp} and illustrated in~\cref{fig:mpc-vs-others-speed}, optimizations must focus on solving the \gls{qp} problem within or below the target \gls{mpc} step time.

%

\paragraph{\textbf{Hardware utilization}} Factors such as synchronization overhead, parallelization inefficiencies, and data non-locality prevent the full utilization of compute units, wasting execution cycles on stalls or non-compute operations and thus reducing execution speed and energy efficiency.
We aim to maximize \gls{fpu} utilization of the eight \gls{pmca} cores through \gls{aot} scheduling and the use of \glspl{sssr} (\cref{sec:background:llc}). 

\paragraph{\textbf{Energy-efficiency}} The power manager's energy profile is critical, as its overhead can impact the overall HPC die performance~\cite{FORGOTTEN_UNCORE}. Maximizing compute utilization of the control algorithm  enhances its energy efficiency (\si{\flop\per\second\per\watt}), thereby reducing the LLC's chip-level impact.

\subsection{Optimization framework}\label{subsec:meth:framework}

\begin{figure}[t]
    \centering
    \includegraphics[width=\columnwidth]{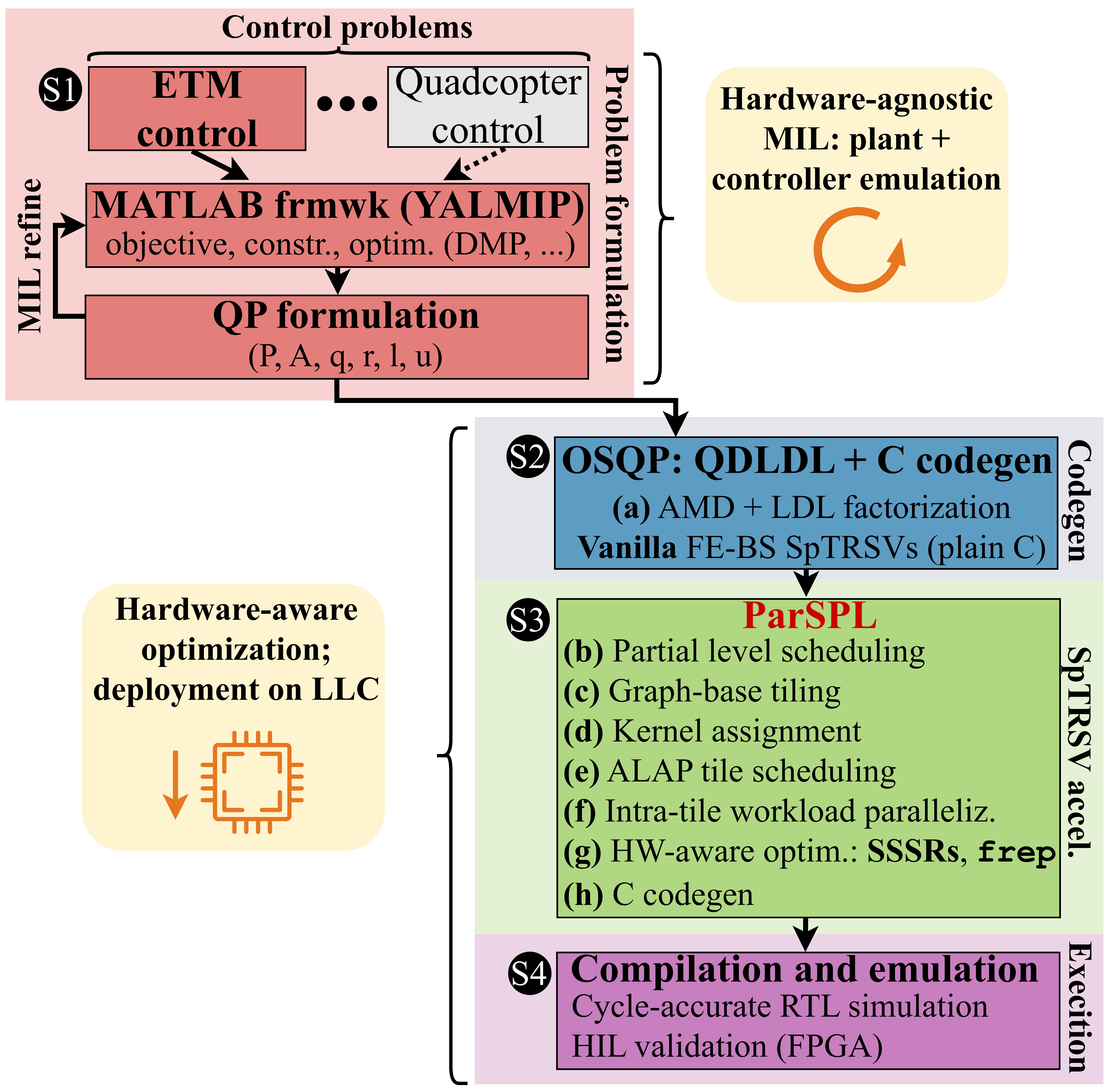}
    \caption{%
        Optimization framework stages highlighted from hardware-agnostic \gls{mil} to hardware-aware multi-core scheduling, optimization, and deployment of the \gls{mpc} control algorithm on the target \gls{llc}.
    }
    \label{fig:optim-framework}
\end{figure}

\cref{fig:optim-framework} shows the progression from hardware-agnostic \gls{mil} to hardware-aware deployment and \gls{hil} emulation.

In the \gls{mil} simulation phase~\circnumsmall{S1}, the \gls{qp} solver is validated independently of the hardware platform. For instance, \gls{dmp} (\Cref{subsec:etm_mpc:dmp}) is evaluated and applied at this stage. This level of abstraction relies on MATLAB and YALMIP, which has native support for the \gls{osqp} solver. 

Then, the deployment stage progresses toward a hardware-aware design space that considers \gls{llc}-specific \gls{isa} extensions, memory, and computing resources.
First, \gls{osqp}'s native code generation capabilities translate the high-level MATLAB representation of the \gls{qp} solver into embedded C \cite{Stellato_osqp_conference}, as shown in~\circnumsmall{S2}.
At this stage, \gls{amd} reordering and LDL factorization are performed by \gls{osqp}; the resulting triangular $L$ matrix is generated in \gls{csc} format.
Second, the \gls{hypt} scheduler further preprocesses the matrix $L$ to achieve extremely fast and efficient multi-core execution of the \gls{sptrsv} in the \gls{fe} and \gls{bs} phases of QDLDL (stage~\circnumsmall{S3}). We detail \gls{hypt} in~\cref{subsec:meth:codegen}.
%
%
Finally, to further increase resource utilization on the embedded platform, the deployed code harnesses the hardware-specific \gls{sssr} extensions available on the \gls{llc}'s \gls{pmca} (\cref{sec:background:llc}).

This framework automatically generates the optimized solver code for~\cref{alg:osqp}. 
Before compilation and deployment on the target platform, the solver is encapsulated as a FreeRTOS task with periodicity larger than the minimum \gls{mpc} step size achievable after optimization (\cref{sec:background:rt}) and integrated within the control firmware.
%

\subsection{Data precision format}\label{subsec:meth:precision}

\begin{figure}[t]
    \centering
    \includegraphics[width=\columnwidth]{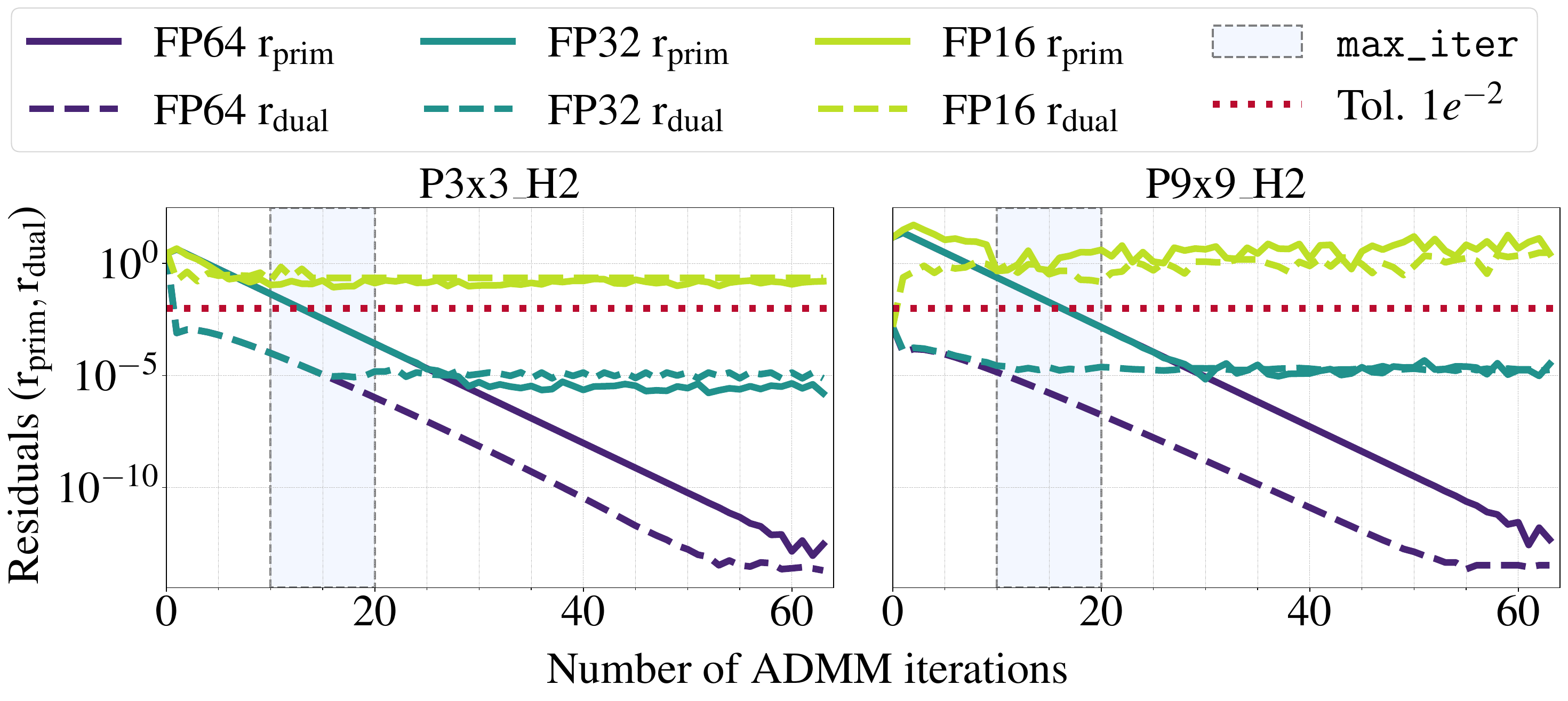}
    \caption{%
        Primal and dual residual evolution for \texttt{P3$\times$3\_H2} and \texttt{P9$\times$9\_H2} at different precision formats. We use \gls{osqp} Python's interface with cold starting on a desktop-grade x86 machine. The solver is executed for $\texttt{max\_iter}=64$ iterations. We highlight a feasible range for \texttt{max\_iter} in [10-25] (light-blue region).
    }
    \label{fig:mpc-convergence}
\end{figure}

As mentioned in~\cref{subsec:etm_mpc:dmp}, data precision impacts the memory footprint in embedded \gls{mpc} deployments.
%
%
On embedded platforms, lower precision formats such as \gls{fp32}, \gls{fp16}, and \gls{fp8} are commonly used to minimize storage consumption. 
However, lower precision can introduce numerical errors for iterative solvers like \gls{osqp}, affecting constraint satisfaction and solver convergence~\cite{BoydAdmm2011,JEREZ_1}.

\Cref{fig:mpc-convergence} illustrates the evolution of $r_{\text{prim}}$ and $r_{\text{dual}}$ at different data precisions for two problems of different sizes, \texttt{P3$\times$3\_H2} and \texttt{P9$\times$9\_H2}.
We leverage \gls{osqp}'s Python interface on a desktop-grade x86 machine.
\Cref{alg:osqp} is executed for \texttt{max\_iter}\;$=64$ iterations, with tolerances $\epsilon_{\text{prim}} = \epsilon_{\text{dual}} = 0.01$ serving as termination criteria.
%
%
We observe that \gls{fp64} delivers the best convergence behavior for primal and dual residuals. 
\Gls{fp32} performs slightly worse, while \gls{fp16} experiences numerical instability early in the solver progression. Because of the poor convergence behavior of \gls{fp16}, we did not explore \gls{fp8}.

In the following, we uniformly adopt \gls{fp32} across the entirety of~\cref{alg:osqp}. 
This choice provides a balanced tradeoff between computational precision and memory efficiency, mitigating the risks of numerical instability.
Other alternatives could be explored, such as a mixed-precision approach where only critical components of~\cref{alg:osqp} essential for numerical stability and convergence are computed in higher precision to reduce the memory footprint further.

\subsection{\Gls{hypt}: \gls{aot} scheduling and parallelism extraction}\label{subsec:meth:codegen}
\begin{figure*}[t]
    \centering
    \includegraphics[width=\linewidth]{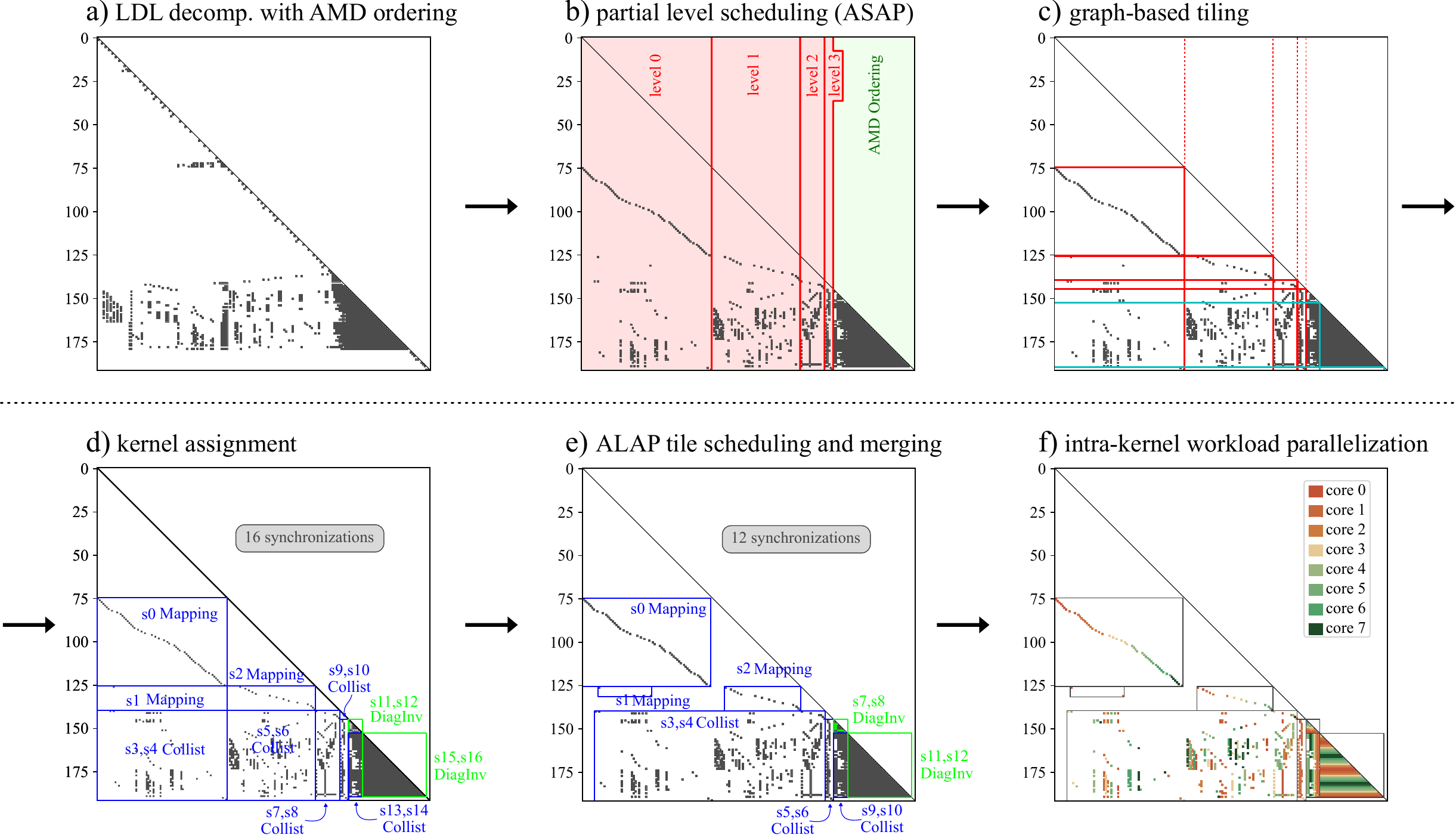}
    \caption{%
    \Gls{hypt} steps applied to the lower triangular matrix $L$ for the problem \texttt{P3$\times$3\_H2}. The starting point in \textbf{(a)} is the LDL decomposition.
    In \textbf{(b)}, partial level scheduling extracts inherent parallelism from the sparsity pattern.
    Strongly dependent components (i.e., denser regions) are retained in the original AMD ordering to preserve the beneficial structure, particularly along the diagonal.
    These strongly interconnected components are separated in \textbf{(c)} using a graph-based approach.
    Cuts from partial level scheduling are shown in red, while graph-based tiling cuts are in cyan.
    The resulting tiles in \textbf{(c)} are merged and assigned to one of three kernels in \textbf{(d)} using static rules in a black-box manner without considering specific data dependencies within and between tiles.  
    In step \textbf{(d)}, the tile region, kernel name, and synchronization steps are visualized.
    To further reduce synchronizations, tiles are dynamically scheduled using \gls{alap} and merged into shards in \textbf{(e)}.
    This exposes concurrency based on the actual sparsity pattern and the dependencies arising from the non-zero data locations.
    Finally, in \textbf{(f)}, intra-shard workloads are distributed across the 8 \gls{pmca} cores.
    }
    \label{fig:aot-scheduling-parspl}
\end{figure*}

The most computationally expensive step in~\cref{alg:osqp}, and the most challenging to parallelize on a multi-core system with limited resources, is the linear system solver described in~\cref{alg:osqp:linsys_xnu}.
In contrast, the updates in~\cref{alg:osqp:linsys_z,alg:osqp:x,alg:osqp:y,alg:osqp:z} of~\cref{alg:osqp} are dense, component-wise-separable vector operations, accounting for only \SI{11}{\percent} of the total solver pass execution time during single-core execution.
Since the triangular matrix $L$ after \gls{amd} reordering is constant, \cref{alg:osqp} repeatedly solves the linear system with a changing right-hand side vector using \gls{fe} and \gls{bs} until termination.
The need for multiple solver passes and the limited resources of the target platform justify the extensive precomputation and preprocessing adopted by \gls{hypt} to maximize the figures of merit discussed in~\cref{subsec:meth:metrics}.

\paragraph{\textbf{LDL decomposition with \gls{amd} ordering}}
Since any column-based approach in \gls{fe} is transposed to a row-based approach in \gls{bs}, we store the lower triangular matrix $L$ according to the \gls{csc} representation, and the following discussion will focus exclusively on $L$.
\Cref{fig:aot-scheduling-parspl}-a illustrates the structure of $L$ after \gls{amd} reordering for the \texttt{P3$\times$3\_H2} problem, chosen to improve readability.

Two key aspects of \gls{fe} and \gls{bs} pose challenges to achieving high execution performance on single- and multi-core systems.
First, the index arrays in the \gls{csc} representation of $L$ lead to indirect and irregular memory access patterns, negatively impacting compute utilization.
Second, both algorithms exhibit inherent dependencies: \gls{fe} follows a \emph{fan-out} pattern, where updates propagate to multiple entries, while \gls{bs} follows a \emph{fan-in} reduction structure. 
These dependencies hinder straightforward parallelization and necessitate extensive synchronization to manage shared data across the computing cores, significantly impacting performance as we will show in~\cref{subsec:eval:func}.

To efficiently parallelize the triangular solver, \gls{hypt} uses a hybrid method, combining partial level scheduling and tiling of $L$. 
Figures \ref{fig:aot-scheduling-parspl}b-f show the process steps.

\paragraph{\textbf{Partial level scheduling (ASAP)}}
Level scheduling is a common technique to extract parallelism in a \gls{sptrsv} by rearranging the matrix into a sequence of levels~\cite{saad2003iterative}.
A level is a set of independent columns (or equations) that can be processed in parallel. Levels are executed sequentially, as they depend on one another.
Sparser columns with fewer dependencies are prioritized and rearranged to be processed \gls{asap} to create the levels.

As a result, level scheduling removes all unnecessary synchronizations across the \gls{pmca} compute cores.
%
However, its effectiveness diminishes as independent columns thin out across levels, leading to two primary challenges. 
The first is the increased synchronization demands as the algorithm proceeds because deeper columns become denser and more likely to have blocking dependencies with others; this results in many small blocking levels with fewer members.
The second is the disruption of dense sub-triangular structures near the lower diagonal, created by \gls{amd} reordering, which are easier to parallelize when densely populated.

To address the limitations of full level scheduling, we use partial level scheduling, discarding levels with fewer columns than a set threshold similarly to~\cite{DIRECT_SPARSE_TRIANG_SOLV}.
Columns in the discarded levels retain the original \gls{amd} ordering.
The choice of the threshold is based on heuristics and derived by varying the problem size.
In the target application considered in this work, at most the first four levels typically satisfy the threshold constraint. 
\cref{fig:aot-scheduling-parspl}-b depicts $L$ after this step.

\paragraph{\textbf{Graph-based tiling}}

\begin{figure}[t]
    \centering
    \includegraphics[width=.65\columnwidth]{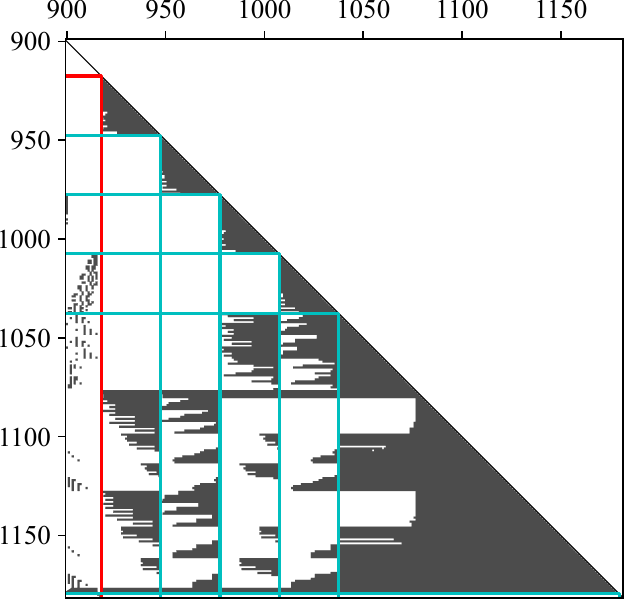}
    \caption{%
        Graph-based tiling for \texttt{P8$\times$8\_H2}. Red lines denote cuts derived from partial level scheduling, while cyan lines indicate cuts introduced by graph-based tiling.
    }
    \label{fig:autocut}
\end{figure}

Following partial level scheduling, $L$ is automatically partitioned into a few large, irregular tiles that minimize inter-tile synchronization while maximizing intra-tile parallelism.
Partial level scheduling naturally divides a portion of the $L$ matrix into tiles, as shown in~\cref{fig:aot-scheduling-parspl}-c with solid vertical and horizontal red lines.
For the remaining \gls{amd}-ordered part, we use graph-based partitioning to extract the dense triangular sub-matrices along the diagonal.
The remaining $L$ matrix is interpreted as an adjacency graph where nodes correspond to the columns/rows indices; an edge connects two nodes $i$ and $j$ if $l_{ij}$ is non-zero.
The edge \emph{weight} for an element $l_{ij}$ is defined as \(\frac{1}{(i-j)^2}\), assigning greater importance to elements near the diagonal and less to those farther away.
Using the Ford-Fulkerson algorithm and the Max-Flow Min-Cut theorem, the graph is partitioned into two subgraphs by cutting along the weakest links.
The obtained cuts are shown in cyan in~\cref{fig:aot-scheduling-parspl}-c.
This process identifies dense sub-diagonal triangular regions and is applied recursively until a heuristic termination criterion.
The collection of all cuts defines the tiling boundaries.
To highlight the importance of graph-based partitioning for larger problems, \cref{fig:autocut} illustrates the process on the \texttt{P8$\times$8\_H2} problem, where dense sub-triangular matrices dominate.

\paragraph{\textbf{Kernel assignment}}
Tiles are assigned to specialized kernels.
Decomposing the vanilla QDLDL algorithm into heterogeneous kernels allows one to leverage the diverse sparsity structures and data representations of the tile topologies introduced by the previous steps.

We identify four kernel types, each with two variants for \gls{fe} and \gls{bs}. 
\texttt{Mapping} is used for regions where each row and each column at most contain one element. An element corresponds to a floating-point \emph{multiply and accumulate} operation that is independent both in \gls{fe} and \gls{bs}, so it can be trivially computed concurrently.
\texttt{Diaginv} is used for dense triangles along the diagonal. The insight behind this kernel is that inverting an (almost) dense triangular matrix results again in an (almost) dense triangular matrix.
The matrix inversion is computed \gls{aot} to eliminate all synchronizations at runtime.
The corresponding matrix-vector multiplication is well-known and parallelizable.
The diagonal inverse multiplication, i.e., the multiplication by $D^{-1}$ from the LDL decomposition after \gls{fe}, is a variant of \texttt{Diaginv}. 
Finally, \texttt{Collist}, or column list, deals with regular sparse data stored in \gls{csc} format and is used for all remaining rectangular tiles.
The rectangular shape ensures disjoint memory read and write locations, avoiding any synchronizations.
%

After the initial assignment, horizontally aligned tiles requiring the same kernel type are merged using static rules to minimize unnecessary inter-tile synchronizations (see~\cref{fig:aot-scheduling-parspl}-d).
Kernels with conflicting memory write during \gls{fe} or \gls{bs} are assigned a second synchronization step to accommodate a reduction.

\paragraph{\textbf{ALAP tile scheduling and merging}}
Stage (e) performs inter-tile scheduling and merging, assuming tiles are black boxes.
In~\cref{fig:aot-scheduling-parspl}-f, the actual data dependencies within and between tiles are considered based on the non-zero locations. 
For each tile, the data-dependent read and write accesses are determined. This information is then used to build a~\gls{dag} of tiles that are scheduled \gls{alap}.
The choice over \gls{asap} is illustrated next.
Looking at the $L$ matrix, \gls{asap} schedules vertical tiles as they share variables to be read before computation.
In contrast, horizontal tiles share variables to be written to a region in the \gls{rhs} vector after computation. 
As horizontal \emph{Collist} tiles were merged in~\cref{fig:aot-scheduling-parspl}-d using static rules, the \gls{asap} potential is largely exhausted, whereas \gls{alap} is found to significantly decrease the number of synchronizations.
An example is a reduction from 53 to 34 in \texttt{P9$\times$9\_H2}.
The tiles obtained by \gls{hypt} are denoted \emph{shards} as they can overlap, are non-rectangular, and do not have to cover the entire matrix.

\paragraph{\textbf{Intra-shard workload parallelization}}
While shards are processed sequentially, 
the data within a shard is inherently designed for full parallelization, enabled by prior optimizations.
Therefore, workload distribution to the \gls{pmca} compute cores is trivial and depicted in \cref{fig:aot-scheduling-parspl}-f, where each core represents a color. 
\Gls{hypt} automatically generates C code from the resulting scheduling (\cref{fig:optim-framework}), with appropriate data structures and data types that minimize memory footprint.

\subsection{\Glspl{sssr} and hardware loops}\label{subsec:meth:sssr}
As shown in~\cref{fig:optim-framework}, the kernels identified during the \gls{sptrsv} optimization are further optimized using dedicated \gls{pmca}'s hardware extensions to boost hardware utilization and performance of the compute cores, i.e., \glspl{sssr} and \texttt{frep} (\cref{sec:background:llc}).
%
%
Both memory streams and regular loops are programmed ahead of computing.
We introduce the average \gls{sl} as $\mathrm{SL} = \frac{nnz(L)}{\mathrm{\#synchr.}}$.
\Gls{sl} is the key metric in determining \gls{sssr} impact and resulting \gls{fpu} utilization and indicates the average number of non-zeros per shard. 
The lower $nnz(L)$ (the sparser the matrix), the lower \gls{sl} and the benefits of accelerated indirect memory streams. 
Hence, techniques oriented towards memory efficiency, like \gls{dmp}, negatively affect the inherent parallelism of the problem. 
Similarly, the fewer synchronizations required for parallel execution, the greater the performance improvement.

In the next section, we provide an exhaustive evaluation of the presented methodology~\cref{subsec:meth:metrics}.
\section{Deployment and Evaluation}
\label{sec:evaluation}


We evaluate the performance of the optimized solver using the metrics and methodology outlined in~\cref{sec:methodology}. 
To assess improvements in execution speed and hardware utilization on the PMCA, we conduct cycle-accurate RTL simulations (stage~\circnumsmall{S4} in~\cref{fig:optim-framework}). The memory footprint is analyzed using LLVM 15.0.0, targeting \emph{RV32IMAFCXsssrXfrep}.
Finally, we perform gate-level assessments of power consumption, energy, and area efficiency of the controller using {\gfs} {\gftech} node with a 13-metal-stack, 7.5-track standard cell library targeting the typical corner (\SI{25}{\celsius}, \SI{0.8}{\volt}).

\subsection{Functional Performance}\label{subsec:eval:func}

\begin{figure}[t]
    \centering
    \includegraphics[width=\columnwidth]{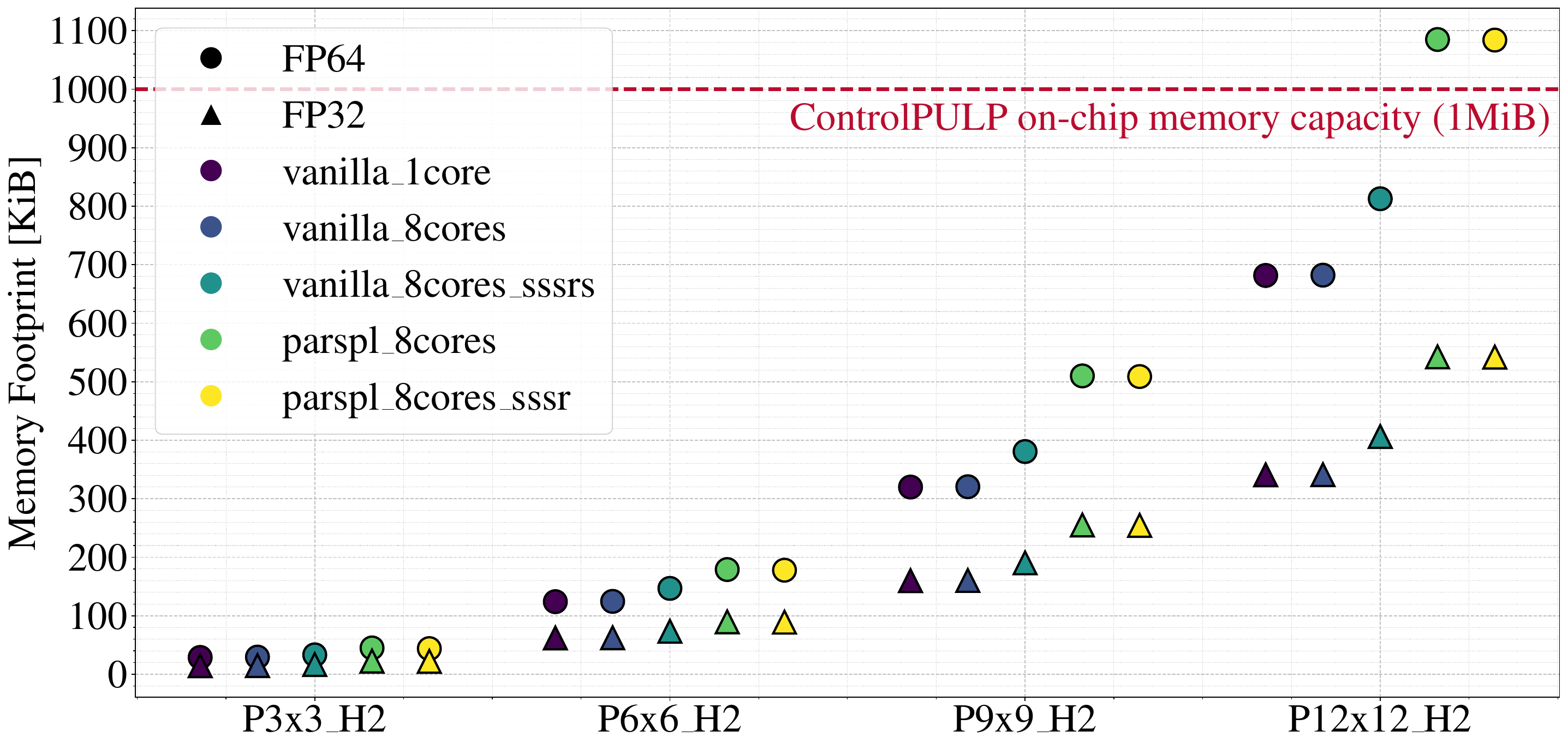}
    \caption{%
        Memory footprint of various \gls{mpc} problems with \gls{fp64} (circle) and \gls{fp32} (triangle) data precisions, optimized with different strategies (color-coded) after compilation for ControlPULP. All problems apply \gls{dmp} in \gls{mil}. Both data precisions allow to stay well within the \gls{llc} memory constraints discussed in~\cref{sec:background:llc}.
    }
    \label{fig:mpc-mem-footprint}
\end{figure}

\begin{figure}[t]
    \centering
    \includegraphics[width=\columnwidth]{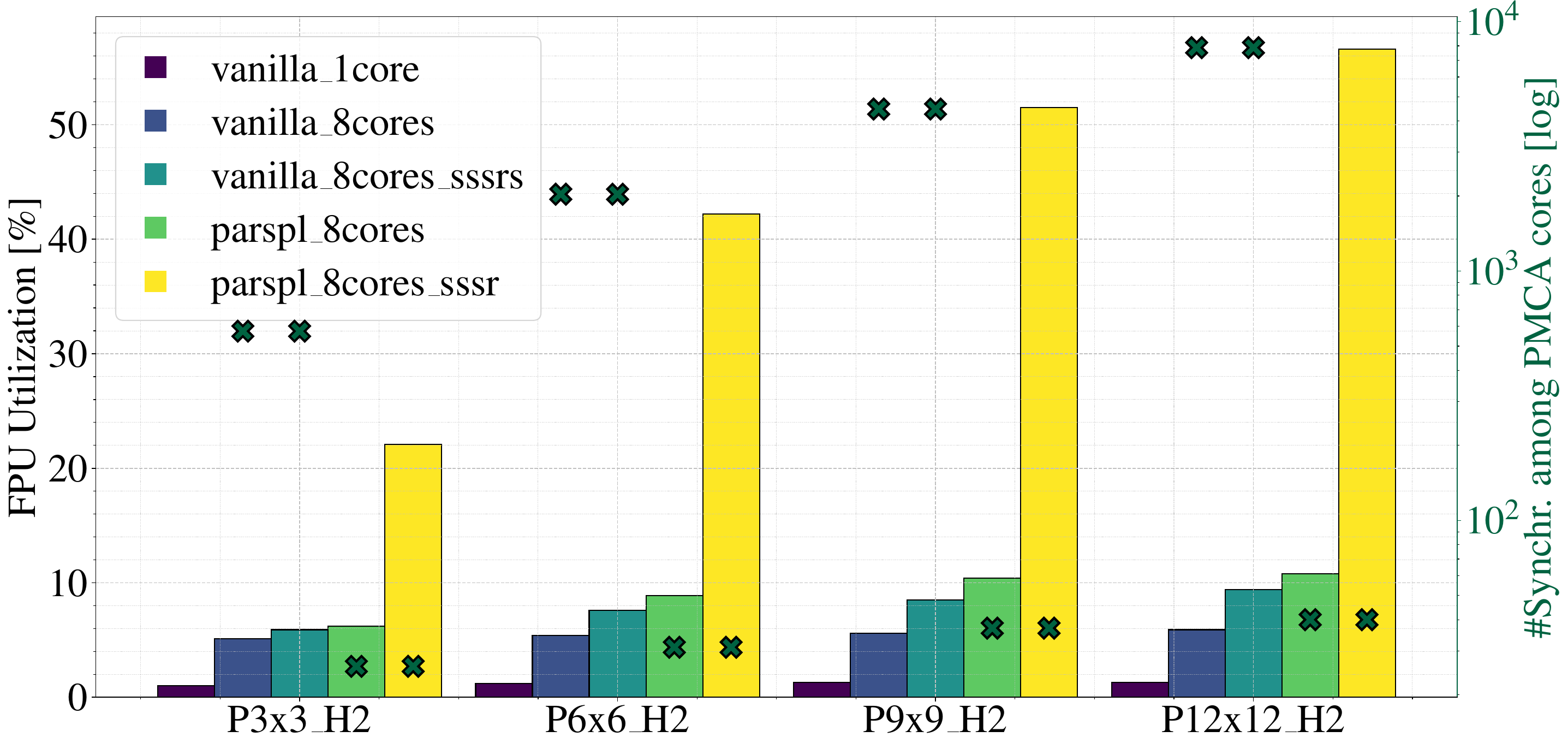}
    \caption{%
        (Bars) \Gls{fpu} utilization of \cref{alg:osqp:linsys_xnu} during \gls{mpc} execution. (Points) Number of synchronization steps required among the \gls{pmca} compute cores. Both metrics apply to a problem size with different optimization methods.
    }
    \label{fig:mpc-hw-util}
\end{figure}

\begin{figure}[t]
    \centering
    \includegraphics[width=\columnwidth]{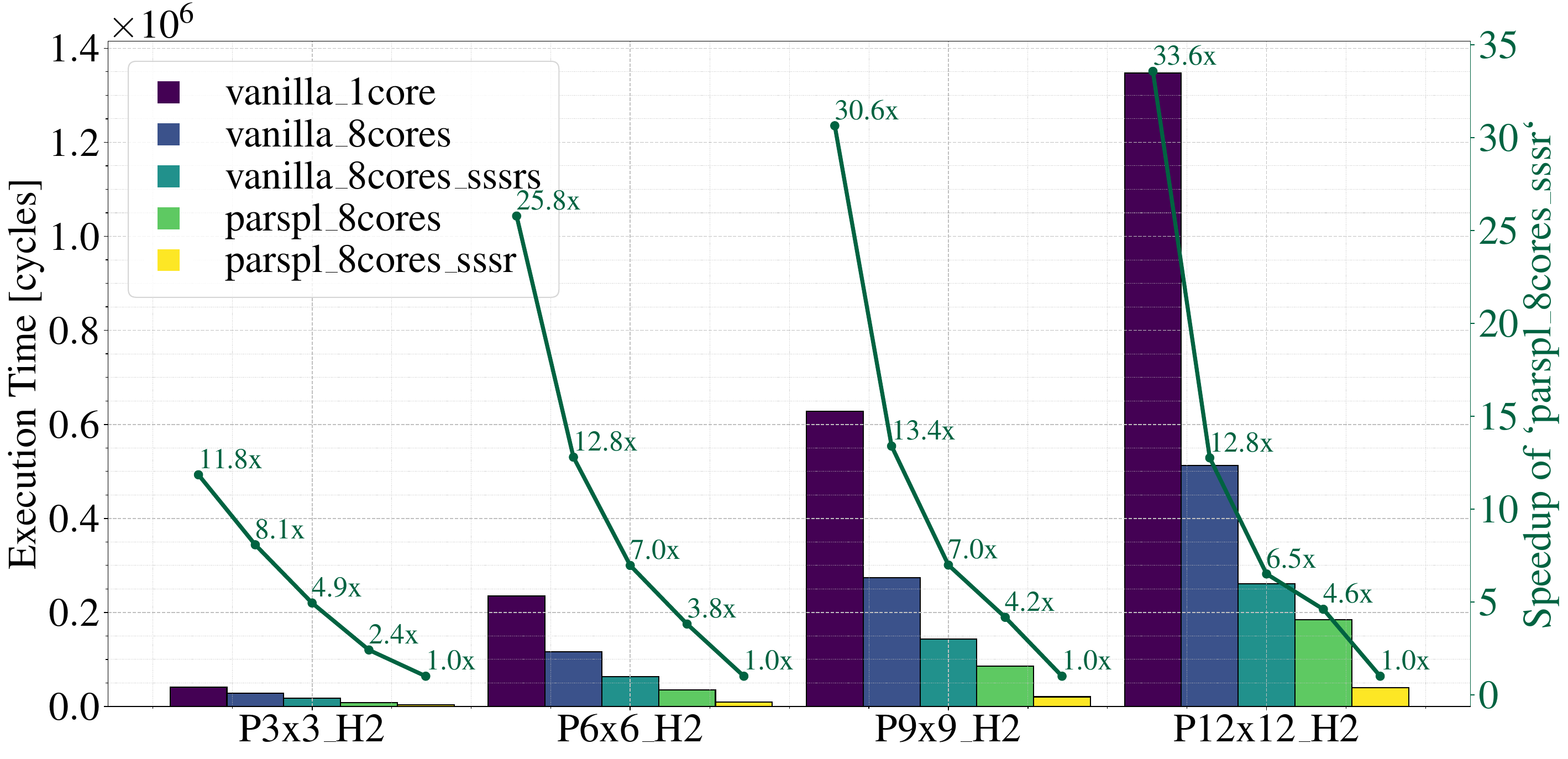}
    \caption{%
        (Left) \Gls{mpc} execution speed for different problems and optimization methods in clock cycles. 
        The chart considers one iteration of the linear system solver (\cref{alg:osqp:linsys_xnu} in~\cref{alg:osqp}). (Right) Speedup of the best method (\texttt{parspl\_8cores\_sssr}) compared to the others.
    }
    \label{fig:mpc-exec-speed}
\end{figure}

We evaluate problems of sizes $N_w=N_h = [3, 6, 9, 12]$, i.e, grids with up to 144 \glspl{pe} on the same silicon die.
For each problem, we assess various optimization methods: vanilla single-core (\texttt{vanilla\_1core}, used as a baseline reference), naive multi-core parallelization across the columns of $L$ (\texttt{vanilla\_8cores}), and \gls{hypt}.

\paragraph{\textbf{Memory Footprint}}

\Cref{fig:mpc-mem-footprint} illustrates the memory footprint in \si{\kibi\byte} for \gls{fp64} (circle) and \gls{fp32} (triangle) precisions.  
All problems apply \gls{dmp}. 
Optimization methods that utilize \glspl{sssr} employ a \SI{16}{\bit} index to reduce overhead.
The density of the triangular matrix $L$ ranges from \SIrange{1.8}{8}{\percent} for the smallest to largest evaluated problems.

For problem sizes \(N_w = N_h > 9\), \gls{hypt} introduces a memory overhead of approximately \SI{100}{\kibi\byte} to \SI{300}{\kibi\byte} compared to the baseline \gls{osqp}.
In the \gls{fe} phase, this overhead arises from the matrix inversions applied by \gls{hypt} to submatrices near the lower diagonal of $L$, which increase the number of fill-ins relative to the original $L$ (\cref{fig:aot-scheduling-parspl}).
The majority of the overhead, however, stems from the current implementation, which stores the inverted triangular dense matrices in a rectangular format, including the upper triangular zeros.
While storing only the triangular part could reduce memory usage, it would degrade the utilization of \glspl{sssr}. 
%

Overall, a problem at scale like \texttt{P12$\times$12\_H2} fits in less than \SI{550}{\kibi\byte} with \gls{fp32} precision. This provides a significant margin for controlling larger systems --- over 200 \glspl{pe} on a single silicon die --- without exceeding the storage constraint discussed in~\cref{sec:background:llc}.

\paragraph{\textbf{Hardware Utilization}}

The bars in \cref{fig:mpc-hw-util} depict the average \gls{fpu} utilization across the eight \gls{pmca} cores during one solver iteration.
We observe the beneficial impact of \gls{hypt} coupled with \glspl{sssr} and hardware loops, enabling efficient acceleration of both affine and indirect memory streams (\texttt{parspl\_8cores\_sssr}). 
This configuration achieves \SI{56.6}{\percent} \gls{fpu} utilization, approximately 43$\times$ higher than the single-core implementation using vanilla \gls{osqp} and 6$\times$ higher than the naive column-based parallelization (\texttt{vanilla\_8cores\_sssr}).

The utilization improvement in a multi-core setting mainly correlates with the reduced synchronization steps between the \gls{pmca} compute cores.
The points in~\cref{fig:mpc-hw-util} show the evolution of synchronization steps for each optimization method. 
%
To provide a more detailed analysis, we concentrate on \texttt{P12$\times$12\_H2}.
The single-core case incurs no synchronization overhead, as it is single-threaded. 
The plain vanilla parallelization method incurs around 7900 synchronizations, while \gls{hypt} reduces this to about 40.

When \glspl{sssr} and hardware loops are used, execution improves \gls{fpu} utilization for both parallelization methods but at different scales: 1.6$\times$ for the vanilla approach and 5.2$\times$ for \gls{hypt}. 
This gap arises due to the longer average \gls{sl} of $L$ introduced by \gls{hypt} (\cref{subsec:meth:sssr}). 
As the \gls{sl} grows with the inverse of the synchronization count, the benefits of \glspl{sssr} are significantly enhanced when fewer synchronizations are required. For instance,  on \texttt{P12$\times$12\_H2}, the \gls{sl}  is $7.9$ for \texttt{vanilla\_8cores} and $1538.2$ for \texttt{parspl\_8cores}.

\paragraph{\textbf{Execution Time (1 iteration)}}

We report the latency of~\cref{alg:osqp} in clock cycles. We first focus on the sparse linear system solver at~\cref{alg:osqp:linsys_xnu} of ~\cref{alg:osqp} as it is the bottleneck of the computation (\cref{subsec:meth:codegen}). 
The left-hand side of~\cref{fig:mpc-exec-speed} illustrates the execution time for one iteration.
The right-hand side shows the speedup of \texttt{parspl\_8cores\_sssr}, the most performant optimization method, against others.
Compared to the vanilla single-core \gls{osqp} case, \gls{hypt} is 11.8$\times$ faster for smaller problems like \texttt{P3$\times$3\_H2} and up to 33.6$\times$ faster for the largest evaluated problem, \texttt{P12$\times$12\_H2}.

\texttt{parspl\_8cores} achieves a high parallelization efficiency across the eight PMCA compute cores, delivering a 7.2$\times$ speedup over single-core performance, close to the theoretical maximum. In contrast, \texttt{vanilla\_8cores} reaches only a 5.2$\times$ speedup.
Similarly, \texttt{parspl\_8cores\_sssr} outperforms \texttt{vanilla\_8cores\_sssr} with nearly 5$\times$ faster execution.
These results arise from the superior parallelism extracted by \gls{hypt}, where the number of dependent sub-kernels requiring synchronization steps is minimized, while their inner workload is divided into independent chunks across the \gls{pmca}'s compute cores (\cref{fig:aot-scheduling-parspl}-f).

\paragraph{\textbf{End-to-end Performance}}

To demonstrate the end-to-end solution time improvement introduced by \gls{hypt}, we recall the chart in~\cref{fig:mpc-vs-others-speed}, which emphasizes the limitations of vanilla \gls{mpc} when controlling 81 \glspl{pe} (\texttt{P9$\times$9\_H2}) on an embedded single-core \gls{llc}. 
Using \texttt{parspl\_8cores\_sssr} and terminating the solver after $\texttt{max\_iter}=15$, the sparse linear system solver in~\cref{alg:osqp:linsys_xnu} takes approximately 307k clock cycles.
The dense vector operations involved in~\cref{alg:osqp} (\cref{alg:osqp:linsys_z,alg:osqp:x,alg:osqp:y,alg:osqp:z}) are accelerated through static distribution of equal chunks to each computing core, coupled with direct stream and loop acceleration in hardware, eliminating the overhead of memory and control operations.
This reduces the execution time of these operations from 2490k clock cycles of vanilla single-core \gls{osqp} to just 37k clock cycles, approximately 67$\times$ faster.
Assuming a conservative controller operating frequency of \SI{500}{\mega\hertz}, the total latency of a solver pass for this problem becomes \SI{0.69}{\milli\second}.
This result drastically reduces the time required to solve the \gls{qp} optimization problem online, enabling very short \gls{mpc} steps under \SI{1}{\milli\second}. 
Even allowing the solver to converge for more iterations would still leave a reasonable free time window within the \mpcstep.
This enables an online \gls{etm} control policy to manage a few hundred \glspl{pe} using a centralized embedded controller, and achieves more than \SI{1}{\kilo\hertz} control bandwidth.

\subsection{Area and Energy efficiency}

\begin{figure}[t]
    \centering
    \includegraphics[width=\columnwidth]{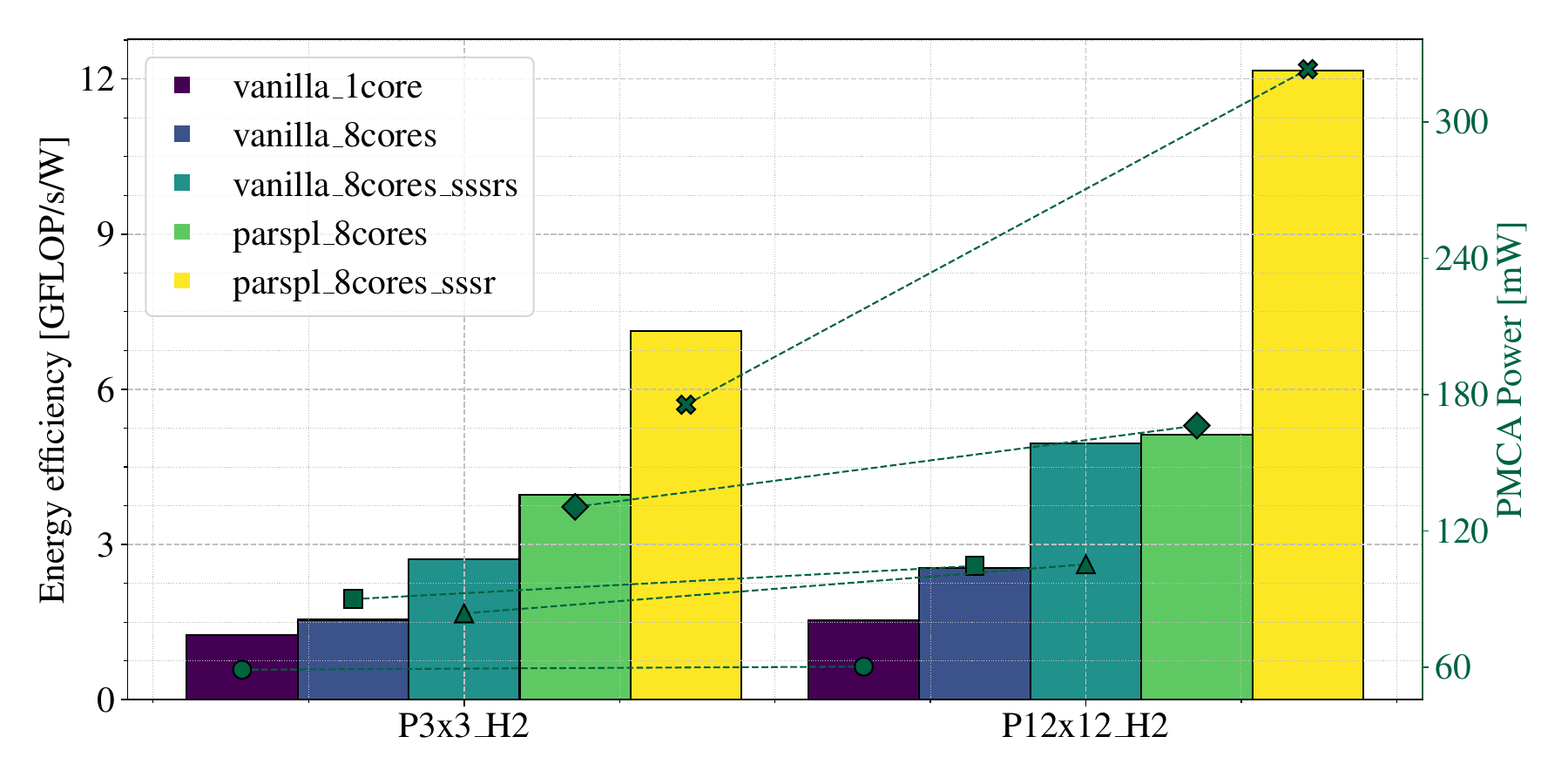}
    \caption{%
        The energy efficiency of the linear system solver in~\cref{alg:osqp:linsys_xnu},~\cref{alg:osqp} for the smallest and largest evaluated problems in different optimization scenarios.
    }
    \label{fig:mpc-energy-efficiency}
\end{figure}

\paragraph{\textbf{Area Efficiency}}

We implement the design using Synopsys Fusion Compiler at \SI{500}{\mega\hertz} under typical conditions (\SI{25}{\celsius}, \SI{0.8}{\volt}). We configure ControlPULP with \SI{128}{\kibi\byte} of L2 \gls{spm} and \SI{1}{\mebi\byte} of L1 \gls{spm} for the manager and \gls{pmca} domains, respectively.
%
%
%
A modern \gls{hpc} die, fabricated in a similar advanced technology node, typically has an area ranging from \SIrange{200}{800}{\milli\metre\squared}~\cite{ALDER_LAKE,NVIDIAGRACE}, depending on the packaging technology, number of \glspl{pe}, and their microarchitecture.
In such a large system, the area overhead of our controller, assuming a single instance for centralized control, is a negligible \SIrange{0.2}{1.5}{\percent}.
Given the growing number of \glspl{pe} integrated in each processor generation and the trend toward compute chiplets, even a scale-out approach using multiple controllers to manage different compute domains --- such as a multi-\gls{pmca} configuration --- remains justified due to its minimal area overhead.

\paragraph{\textbf{Energy Efficiency}}

\Cref{fig:mpc-energy-efficiency} shows the estimated energy efficiency (points) and power consumption (bars) of the \gls{llc}'s \gls{pmca} when solving, at \SI{32}{\bit} data precision, one iteration of the \gls{admm}'s linear system solver on the \texttt{P3$\times$3\_H2} and \texttt{P12$\times$12\_H2} problems. Results are shown across all optimization scenarios discussed in~\cref{subsec:eval:func}. Power estimates are obtained using Synopsys PrimeTime.
As expected, the median power consumption of the \gls{pmca} increases when using \glspl{sssr}. \Gls{hypt} introduces additional overhead due to higher \gls{fpu} and L1 \gls{spm} utilization. However, the \gls{sssr}-optimized version of \gls{hypt} achieves energy efficiency improvements of up to 2.4$\times$ compared to the vanilla \gls{hypt}, and up to 7.9$\times$, 2.4$\times$, and 4.8$\times$ compared to vanilla single-core and vanilla parallelization with and without \glspl{sssr}, respectively.
The largest problem instance, \texttt{P12$\times$12\_H2}, in its most optimized configuration, maintains a total power consumption below \SI{325}{\milli\watt}. This is nearly three orders of magnitude lower than that of traditional \gls{hlc}-centric approaches for executing \gls{mpc} policies, which typically consume power in the range of tens of Watts due to the overhead of the \gls{os} stack and the inherently power-hungry architectures of conventional application-class \glspl{pe}. These approaches are reviewed in the next~\cref{sec:relwrk}.

\section{Related Work}
\label{sec:relwrk}

We initially examine \gls{sota} \gls{mpc} schemes for \gls{etm} (\cref{sec:relwrk:mpc_etm}).
Then, in~\cref{sec:relwrk:mpc_transition}, we explore works targeting fast online \gls{mpc} in a wide range of control scenarios and problem scales, mostly relying on the \gls{admm} algorithm.
%
%
\Cref{tab:related_work} captures the main aspects discussed in this section.

\newcommand{\rot}[1]{\rotatebox[origin=c]{0}{#1}}
\newcommand{\tilt}[1]{\hspace{-1cm}\rotatebox[origin=c]{32}{#1}}
\newcommand{\noc}{\textcolor{ieee-dark-grey-40}{0}}
\newcommand{\dk}{\textcolor{ieee-bright-red-100}{?}}
\newcommand{\na}{\textcolor{ieee-dark-grey-40}{n.a.}}
\newcommand{\nad}{\textcolor{ieee-dark-grey-40}{-}}
\newcommand{\dl}[2]{\makecell[cc]{#1 \\ #2}}
\newcommand{\dll}[2]{\makecell[cl]{#1 \\ #2}}
\newcommand{\dllb}[2]{\makecell[cl]{\textbf{#1} \\ \textbf{#2}}}
\newcommand{\dlb}[2]{\makecell[cc]{\textbf{#1} \\ \textbf{#2}}}
\newcommand{\tl}[3]{\makecell[cc]{#1 \\ #2 \\ #3}}
\newcommand{\tll}[3]{\makecell[cl]{#1 \\ #2 \\ #3}}
\newcommand{\tlb}[3]{\makecell[cc]{\textbf{#1} \\ \textbf{#2} \\ \textbf{#3}}}

\begin{table*}[t]
    \caption{\glsentrytext{qp} acceleration solutions for various applications and control problems using \glsentrytext{mpc}.}
    \begin{center}
    \setlength{\tabcolsep}{1.5pt}
    \renewcommand{\arraystretch}{2.0}
    \resizebox{\linewidth}{!}{
    \begin{threeparttable}
    \begin{tabular}{lcc|cccc|cccccc|ccc}
    
    \arrayrulecolor{ieee-dark-black-100}\toprule &
    \multicolumn{2}{c}{\textbf{\glsentrytext{mpc} appl. profile}} &
    \multicolumn{4}{c}{\textbf{Algorithm (best referenced configuration)}} &
    \multicolumn{6}{c}{\textbf{Deployment Platform}} &
    \multicolumn{3}{c}{\textbf{\Gls{mpc} Performance}} \\

    \arrayrulecolor{ieee-dark-black-100}\midrule

    &
    \rot{\textbf{Class}} &
    \rot{\dlb{Referenced}{application}} &
    \rot{\dlb{\glsentrytext{qp} Sol.}{Method}} &
    \rot{\dlb{Solver}{ }} &
    \rot{\dlb{Exploits}{sparsity?}} &
    \rot{\dlb{Data}{precision}} &
    \rot{\dlb{Pred.}{horizon}} &
    \rot{\dlb{\glsentrytext{hw}}{topology}} &
    \rot{\dlb{\glsentrytext{hw}}{platform}} &
    \rot{\dlb{Mem.}{hierarchy}} &
    \rot{\dlb{Target freq.}{[\si{\mega\hertz}]}} &
    \rot{\dlb{Peak Power}{[\si{\watt}]}} &
    \rot{\dlb{Mem. footprint}{[\si{\kibi\byte}]}} &
    \rot{\dlb{Time to}{solve}} &
    \rot{\dlb{Best}{sample time}} \\

    \arrayrulecolor{ieee-dark-black-40}\cdashline{1-16}
    \multicolumn{16}{c}{\textbf{Online \gls{mpc} for various control applications}} \\
    \arrayrulecolor{ieee-dark-black-40}\cdashline{1-16}

    Alavilli \emph{et al.}~\cite{alavilli2023tinympc} &
    Emb. & 6-DOF quadr. & ADMM & tinyMPC & \textcolor{ieee-bright-red-100}{\xmark} & \texttt{\glsentrytext{fp32}} & 15-20 &
    MCU & STM32F405 &
    \dl{\SI{192}{\kibi\byte} DRAM +}{\SI{1}{\mebi\byte} flash} &
    168 &
    $<$0.1 &
    $<$192 & $<$2ms-\na & 2ms \\

    Bitjoka \emph{et al.}~\cite{NDJE202119} &
    Emb. & DC motor & FNLQDMC & \na & \textcolor{ieee-bright-red-100}{\xmark} &
    Custom & 2 & MCU & Arduino Due &
    \dl{\SI{96}{\kibi\byte} \glsentrytext{spm} +}{\SI{512}{\kibi\byte} flash} &
    84 &
    \na &
    83 & 0.8-2.9ms & 3-8ms \\

    Chaber \emph{et al.}~\cite{8611361} &
    Emb. & Servo & \dl{FGPC}{FDMC} & Custom & \textcolor{ieee-bright-red-100}{\xmark} &
    \glsentrytext{fp32} & \dl{50}{50} & MCU & STM32F746 &
    \dl{\SI{320}{\kibi\byte} \glsentrytext{spm} +}{\SI{1}{\mebi\byte} flash} &
    216 &
    $<$0.6 &
    \na & \dl{\SI{11}{\micro\second}}{\SI{584}{\micro\second}} & \dl{\SI{250}{\micro\second}}{1ms} \\

    Jerez \emph{et al.}~\cite{JEREZ_1, JEREZ_2} &
    Emb. & AFM control & \dl{\glsentrytext{fg}}{\glsentrytext{admm} (expl.)} & Custom &
    \textcolor{ieee-bright-red-100}{\xmark} & Fix. point & 16 & \glsentrytext{dsa} &
    \dl{Virtex6 \glsentrytext{fpga}\tnote{a}}{Spartan6 \glsentrytext{fpga}\tnote{a}} & \nad &
    \dl{400}{230} &
    \na &
    \na & \dl{4.9-8.9$\mu$s}{0.5-0.9$\mu$s} & 1-1.5$\mu$s \\

    Jerez \emph{et al.}~\cite{JEREZ_SPARSE} &
    Emb. & Spring-mass &
    \dl{Int.-point +}{Newton} & Custom & \textcolor{ieee-bright-dgreen-100}{\cmark} &
    Fix. point & \dl{4}{9} & \glsentrytext{dsa} & Virtex6 \glsentrytext{fpga} &
    \nad & 400 &
    \na &
    \na & 371ms & 388ms \\

    Malouche \emph{et al.}~\cite{7348173} &
    Emb. & Robot & GPC & Custom & \textcolor{ieee-bright-red-100}{\xmark} &
    \glsentrytext{fp32} & 20 & MCU & STM32F407 &
    \dl{\SI{192}{\kibi\byte} \glsentrytext{spm} +}{\SI{1}{\mebi\byte} flash} &
    168 &
    $<$0.07 &
    10.6 & \SI{4.9}{\micro\second} & 5.3ms \\

    Sabo \emph{et al.}~\cite{SABO} &
    Emb. & Double integrator & Primal-dual & Custom & \textcolor{ieee-bright-dgreen-100}{\cmark} &
    Fix. point & 9 & \glsentrytext{dsa} & Digil. Nexys 4 &
    \nad & 100 &
    \na &
    \na & \na & \na \\

    Vouzis \emph{et al.}~\cite{4814488} &
    Emb. & \dl{Rot. antenna}{Glucose regul.} & Newton & Custom & \textcolor{ieee-bright-red-100}{\xmark} &
    \glsentrytext{lns} & \dl{20}{5} & \glsentrytext{dsa} & Virtex4 \glsentrytext{fpga} &
    2 BlockRAMs & 50 &
    \na &
    \na & \dl{0.45ms}{0.68ms} & \dl{\na}{few min} \\

    Wills \emph{et al.}~\cite{5692131} &
    Emb. & 14th-order harmon. & Active-set & Custom &
    \textcolor{ieee-bright-red-100}{\xmark} &
    Custom\tnote{c} & 12 & \glsentrytext{dsa} & Stratix III \glsentrytext{fpga} &
    \nad & 70 &
    \na &
    \na & 30\si{\micro\second} & 200\si{\micro\second} \\

    Wang \emph{et al.}~\cite{fast_mpc_boyd} &
    HPC & \tl{Oscill. masses}{Supply chain}{Random} & Int.-point & Custom &
    \textcolor{ieee-bright-dgreen-100}{\cmark} & \na & 10/30 &
    Desk. CPU & Athlon & \na & 3000 &
    35 (TDP) &
    \na\tnote{e} & \dl{$<$1ms}{$<$12ms\tnote{b} \\ $<$25ms} &
    \dl{5ms}{12ms \\ 25ms} \\

    Schubiger \emph{et al.}~\cite{SCHUBIGER202055} &
    \glsentrytext{hpc} & Random & ADMM & \dl{cuOSQP}{(\glsentrytext{pcg})} &
    \textcolor{ieee-bright-dgreen-100}{\cmark} &
    \dl{\texttt{\glsentrytext{fp32}}}{\texttt{\glsentrytext{fp64}}} & \na & GP-GPU &
    \dl{i9-9900K +}{RTX2080Ti} & \dl{\SI{64}{\gibi\byte} DDR4}{\SI{11}{\gibi\byte} VRAM} &
    3600 &
    \dl{250}{(TDP of RTX)} &
    \na & \dl{100ms-90s}{200ms-100s} & \dl{200ms-100s}{ } \\

    \arrayrulecolor{ieee-dark-black-40}\cdashline{1-16}
    \multicolumn{16}{c}{\textbf{Online \gls{mpc} for energy and thermal management of \gls{hpc} processors}} \\
    \arrayrulecolor{ieee-dark-black-40}\cdashline{1-16}

    Bartolini \emph{et al.}~\cite{6178247} &
    \glsentrytext{hpc} & \dl{\glsentrytext{etm}}{distrib. 1 \gls{pe}} &
    Active-set & \na & \textcolor{ieee-bright-red-100}{\xmark} &
    \texttt{\glsentrytext{fp64}} & 2 & Desk. CPU & \na & \na\tnote{d} &
    2400 &
    \na &
    \na & 4.69\si{\micro\second}\tnote{f} & 1-10ms \\

    Tilli \emph{et al.}~\cite{TILLI2022105099} &
    \glsentrytext{hpc} & \dl{\glsentrytext{etm}}{distrib. 1 \gls{pe}} &
    \na & \na & \textcolor{ieee-bright-red-100}{\xmark} &
    \texttt{\glsentrytext{fp64}} & 2 & Desk. CPU & Intel i7 8th gen. & \na\tnote{e} &
    4600 &
    95 (TDP) &
    \na & 3\si{\micro\second}\tnote{f} & 1ms \\

    Wang \emph{et al.}~\cite{Wang2011AdaptivePC} &
    \glsentrytext{hpc} & \dl{\glsentrytext{etm}}{centr. 16 \glspl{pe}} &
    Least-square & \texttt{lsqlin} & \textcolor{ieee-bright-red-100}{\xmark} &
    \na & 8 & Desk. CPU & Xeon X5365 & \na\tnote{e} & 3000 &
    150 (TDP) &
    \na & \na & 40ms \\

    Maity \emph{et al.}~\cite{FUTURE_AWARE} &
    Emb. & \dl{\glsentrytext{etm}}{centr. 4 \glspl{pe}} &
    \dl{Heur.}{beam search} & \na & \textcolor{ieee-bright-red-100}{\xmark} &
    \texttt{\glsentrytext{fp32}} & \na & \glsentrytext{dsa} & Odroid XU4 &
    \na & 2000 &
    \na &
    \na & Few ms & \na \\

    \textbf{This work} &
    \textbf{Emb.} & \textbf{\dl{\glsentrytext{etm}}{centr. 81 \glspl{pe}~\tnote{g}}} & \textbf{ADMM} &
    \textbf{\dl{OSQP}{(QDLDL)}} & \textcolor{ieee-bright-dgreen-100}{\cmark} &
    \textbf{\dl{\texttt{\glsentrytext{fp32}}}{\texttt{\glsentrytext{fp64}}}} & \textbf{2} &
    \textbf{\dl{Spec.}{paral. MCU}} & \textbf{ControlPULP} &
    \textbf{$<$\SI{1}{\mebi\byte}~\glsentrytext{spm}~\tnote{h}} &
    \textbf{500} &
    \textbf{$<$0.325}~\tnote{i} &
    \textbf{600}~\tnote{i} & \textbf{0.69ms}~\tnote{i} & \textbf{$<$1ms}~\tnote{i} \\

    \arrayrulecolor{ieee-dark-black-100}\bottomrule
    \end{tabular}
    \begin{tablenotes}[para, flushleft]
        \fontsize{7.7pt}{7.7pt}\selectfont
        \item[a] The papers report only potential performance and resource usage without effective measurements.
        \item[b] The paper reports average time per iteration; no explicit sample time is given.
        \item[c] Custom floating-point format with 7 exponent bits and a mantissa in the range [7,15].
        \item[d] Data from Fig. 9 of~\cite{6178247} obtained on a dual-core desktop processor.
        \item[e] The paper lacks details on the exact memory hierarchy configuration, despite it being a commercial processor.
        \item[f] Time per single controlled \glsentrytext{pe}.
        \item[g] We report the largest problem evaluated in this work
        \item[h] Relies only on on-chip \glsentrytext{spm} hierarchy for manager domain and \glsentrytext{pmca}. See~\cref{sec:background:llc}.
        \item[i] Refers to controlling 81 \glspl{pe}, which fits the memory constraint of \SI{1}{\mebi\byte}.
    \end{tablenotes}
    \end{threeparttable}
    }
    \label{tab:related_work}
    \end{center}
\end{table*}

\subsection{\gls{mpc} for \gls{etm}}\label{sec:relwrk:mpc_etm}

%

Due to \gls{mpc}'s complexity, classical works leverage the computational capabilities of the controlled out-of-order, application-class \glspl{pe} to execute the controller, which becomes part of the \gls{hlc} governor (\cref{fig:etm-paradigms}-a).
Bartolini~et~al.~\cite{6178247} propose an \gls{mpc} architecture limited to thermal capping that leverages a distributed approach to address the complexity of centralized methods.
Taking advantage of the parallel architecture of the multi-core \gls{soc}, each \gls{pe} computes autonomously its future frequency in line with the incoming workload requirements. 
The deployment platform, a desktop \gls{cpu} operating at \SI{3}{\giga\hertz}, enables online solving of the \gls{qp} problem using active-set methods in approximately \SI{5}{\micro\second} per PE, achieving sampling intervals of \SIrange{1}{10}{\milli\second}.
Similarly, Tilli~et~al.~\cite{TILLI2022105099} propose a two-layer distributed \gls{mpc} framework combining a decentralized thermal capping layer for guaranteed feasibility and a distributed \gls{mpc} layer for performance optimization.
The approach is software-based, and evaluation is performed executing the controller on last-generation Intel processors.
Wang~et~al.~\cite{Wang2011AdaptivePC}, instead, adopts a centralized strategy.
It provides experimental data for controlling up to 16 \glspl{pe}, using a \SI{3}{\giga\hertz} Intel Xeon X5 and achieving an \gls{mpc} sample time below $5ms$.
In contrast, our centralized scheme controls 5$\times$ more \glspl{pe} with 10$\times$ lower execution time and running the controller's cores at 6$\times$ lower frequency, three orders of magnitude smaller power consumption, and at least two orders of magnitude smaller memory footprint.
Maity et al.~\cite{FUTURE_AWARE} propose a thermally optimizing scheduler for CPU/GPU mobile architectures like the Odroid-XU4. They use a lightweight heuristic beam search, executed by a quad-core Cortex-A7 \gls{llc}, to minimize peak temperature over a hyper-period. Scalability is not addressed, as their \gls{etm} policy manages a quad-core Cortex-A15 and a 6-shader Mali GPU in a fixed configuration.

These works focus on controller design, formulation, and reliability for the \gls{etm} problem, but omit details on the \gls{qp} solver or problem data structure (sparse/dense). Their \gls{hlc}-centric approach for thermal and power capping also incurs high overhead from power-hungry, \gls{os}-based software stacks, as shown in~\cref{tab:related_work}.

\subsection{Online \gls{mpc}: from high-end to embedded systems}\label{sec:relwrk:mpc_transition}

Early efforts to make online \gls{mpc} viable on embedded systems focus on accelerating sparse optimization methods.
For instance, the seminal work from Wang~et~al.~\cite{fast_mpc_boyd} fully exploits the problem sparsity with an interior-point method, enabling a linear growth of the number of operations per \mpcstep with $H_p$, rather than cubic.
This strategy allows for tackling fairly large problems --- a few hundred to thousands of state variables and constraints --- with sample times of \SIrange{5}{25}{\milli\second}, over 100$\times$ faster than a dense optimizer. 
While the approach inspired today's \gls{sota}, such as \gls{osqp}, it does not target embedded platforms directly and evaluates the control problems on a \SI{3}{\giga\hertz} desktop machine.

One way to tackle \gls{mpc}'s computational complexity is to design \glspl{dsa} to accelerate the \gls{qp} solver pass.
Vouzis~et~al.~\cite{4814488} propose an auxiliary co-processor with a microcode interface to accelerate the computationally intensive steps of Newton’s algorithm. 
To reduce memory footprint, a single 16-bit \gls{lns} unit iterates over the input data and stores intermediate results in local memories.
However, the evaluated problems are limited in size, with fewer than six decision variables ($n$) and constraints ($m$), making them 16$\times$ smaller than the smallest \texttt{P3$\times$3\_H2} problem analyzed in this work ($n=95 \; \mathrm{and} \; m=97$).
Wills~et~al.~\cite{5692131} employ a similar approach using an active-set method, which entails expensive matrix inversion computation.
In \cite{JEREZ_1,JEREZ_2}, Jerez~et~al. present hardware circuits to accelerate both \gls{fg} and \gls{admm} methods with dense dynamics, while in~\cite{JEREZ_SPARSE}, they exploit sparsity to achieve a \SI{75}{\percent} memory reduction and a 6.5$\times$ speedup over an equivalent software implementation running on a \gls{cpu}.
%
The accelerators proposed in~\cite{JEREZ_1,JEREZ_2} show that sample intervals in the range of \SIrange{1}{1.5}{\micro\second} using \gls{fg} and \SIrange{5}{25}{\micro\second} using \gls{admm} are feasible on \glspl{fpga} clocked at \SI{150}{\mega\hertz}, with a \gls{qp} size of $n=704$ and $m=544$ (smaller than \texttt{P3$\times$3\_H2} with $n=1310 \; \mathrm{and} \; m=1312$).

All discussed \glspl{dsa} have been implemented on \glspl{fpga}, reporting excellent execution times for fast \gls{mpc}. However, memory footprint is often not analyzed, performance on the deployed device is only estimated~\cite{JEREZ_2}, and metrics like power consumption and incurred area are omitted, as accurate evaluation would require ASIC implementation.
%

Recent research initiatives, notably those centered around the \gls{osqp} solver, have rejuvenated interest in \gls{admm} methods.
Alavilli~et~al.~\cite{alavilli2023tinympc} propose TinyMPC, a \gls{qp} solver that accelerates and compresses \gls{admm}, combined with \glspl{lqr}, for extremely resource-constrained systems.
TinyMPC's key approach is the pre-computation (caching) of the Riccati solution of the \gls{lqr} problem, equivalent to the primal update in \gls{admm}'s~\cref{eq:admm_xztilde}.
With this approach, it attains a speedup of up to 8.8$\times$ over vanilla \gls{osqp} on a single-core \gls{mcu} while adhering to stringent memory limitations ($<$~\SI{512}{\kibi\byte}) using random, small- and medium-scale problem instances (up to $n=696 \; \mathrm{and} \; m=490$).
TinyMPC could be a promising candidate for \gls{etm} as the memory usage and execution time exhibit more linear growth than vanilla \gls{osqp} at varying $n$, $m$, and $H_p$.
\Gls{lqr} policies for thermal management of \gls{hpc} processors have been previously studied~\cite{ZANINI1_TCM}.
They prioritize maintaining a uniform thermal profile rather than minimizing power consumption, and lack explicit threshold constraints on the chip maximum temperature in their optimization formulation. 
Instead, constraints like frequency bounds and thermal limits are addressed externally through post-optimization adjustments. 
This reliance on external corrections limits the effectiveness of \gls{lqr} under heavy workloads, as it cannot dynamically account for nonlinearities or spatial and temporal thermal gradients, which are essential for efficient thermal management in modern \gls{hpc} processors.

Schubiger~et~al.~\cite{SCHUBIGER202055} focus instead on accelerating \gls{osqp} on large-scale problems ($N_{nnz} \geq 10^{4}$). 
They propose cuOSQP, a CUDA-optimized solver for NVIDIA \glspl{gpu} based on the \gls{pcg} indirect method to solve the \gls{admm} linear system, achieving a peak speedup of $15\times$ and $5\times$ over single-threaded, vanilla QDLDL \gls{osqp} and multi-threaded, MKL-Pardiso \gls{osqp}, respectively, for the largest evaluated problem ($N_{nnz} = 10^{8}$) on a NVIDIA RTX2080Ti.
However, their assessment shows that vanilla QDLDL \gls{osqp} is roughly $10\times$ faster than \gls{pcg}-based cuOSQP for $N_{nnz} < 10^{6}$, the same problem range evaluated in this work.
While this supports our choice of a QDLDL-based approach, indirect methods remain promising candidates for very large \gls{etm} control problems involving thousands of \glspl{pe}, a scale not yet achieved by any major commercial vendor.

Several other works have addressed the mapping of fast \gls{mpc} algorithms on commercial \glspl{mcu}~\cite{8611361,7348173,NDJE202119}. 
Despite achieving sub-\si{\milli\second} solution times with a memory footprint in the tens of \si{\kibi\byte}, the tackled problems are small, with tens of decision variables and constraints. 
Furthermore, conventional \glspl{mcu} rely on memory hierarchies with an external flash alongside on-chip \glspl{spm} or caches.
While external memory is an option for embedded \glspl{llc}, a self-contained, on-chip memory hierarchy improves responsiveness and autonomy as we explain in~\cref{sec:background:llc}.

The comparative analysis shows that our hardware/software co-optimization framework (\cref{subsec:meth:framework}) outperforms \gls{sota} \gls{mpc} schemes for \gls{etm} in \gls{hpc} processors, requiring fewer computing resources and storage while delivering low-power execution and high energy efficiency at a small fraction of the \gls{hpc} processor die area. 
It also competes with fast \gls{mpc} methods deployed on various architectures like \glspl{mcu}, \glspl{dsa}, and \glspl{gpu} and validated on other control applications. 
This efficiency stems from leveraging \gls{aot} pre-computation and parallelization derived from the problem matrix's sparsity pattern.
Moreover, unlike fixed-function accelerators, our hardware is flexible and can be adapted to different algorithms and problems, helping avoid premature obsolescence that often affects hardwired accelerators.



\section{Conclusion}
\label{sec:conclusion}
In this work, we presented a comprehensive framework for implementing \gls{mpc} for energy and thermal management of \gls{hpc} processors using a resource-constrained, multi-core, embedded RISC-V \gls{llc}. 
The proposed method addressed the computational challenges of solving quadratic programming problems in real-time by leveraging the \gls{admm} algorithm and optimizing the \gls{sota} \gls{osqp} solver. 
To achieve this goal, we presented several key optimization techniques, including a threshold-based pruning algorithm to optimize the model memory usage and ahead-of-time scheduling with the \gls{hypt} framework to extract inherent parallelism from the sparse triangular system solved at runtime.
The optimized \gls{mpc} framework achieved significant reductions in execution time, memory footprint, and energy consumption, enabling the control of up to 144 processing elements on a single silicon die with sub-millisecond latency within a few hundred of \si{\kibi\byte}.
This solution outperforms \gls{sota} \gls{hlc}-centric methods that routinely require desktop-grade computing resources, demonstrating the feasibility of delegating advanced \gls{etm} policies to embedded, on-chip controllers. 
The approach was validated through extensive cycle-accurate simulations, paired with a physical implementation of the \gls{llc} using an advanced {\gfs} {\gftech} node for power estimation, achieving up to 33$\times$ speedup and 7.9$\times$ energy efficiency gains in solving \glspl{sptrsv} compared to single-core \gls{osqp}, with a memory footprint below \SI{1}{\mebi\byte}, well within the \gls{llc} constraints, and a power consumption below \SI{325}{\milli\watt}.


\renewcommand{\baselinestretch}{1.0}

\bibliographystyle{IEEEtran}
\bibliography{bib/references.bib}

\begin{thebibliography}{10}
\providecommand{\url}[1]{#1}
\csname url@samestyle\endcsname
\providecommand{\newblock}{\relax}
\providecommand{\bibinfo}[2]{#2}
\providecommand{\BIBentrySTDinterwordspacing}{\spaceskip=0pt\relax}
\providecommand{\BIBentryALTinterwordstretchfactor}{4}
\providecommand{\BIBentryALTinterwordspacing}{\spaceskip=\fontdimen2\font plus
\BIBentryALTinterwordstretchfactor\fontdimen3\font minus \fontdimen4\font\relax}
\providecommand{\BIBforeignlanguage}[2]{{%
\expandafter\ifx\csname l@#1\endcsname\relax
\typeout{** WARNING: IEEEtran.bst: No hyphenation pattern has been}%
\typeout{** loaded for the language `#1'. Using the pattern for}%
\typeout{** the default language instead.}%
\else
\language=\csname l@#1\endcsname
\fi
#2}}
\providecommand{\BIBdecl}{\relax}
\BIBdecl

\bibitem{10707191}
Z.~Guo, Y.~Tang, J.~Zhai, T.~Yuan, J.~Jin, L.~Wang, Y.~Zhao, and R.~Li, ``A survey on performance modeling and prediction for distributed dnn training,'' \emph{IEEE Trans. on Parallel and Distrib. Syst.}, vol.~35, no.~12, pp. 2463--2478, 2024.

\bibitem{MOLECULAR_DYNAMICS_UMBRELLA_III}
R.~Stocks, E.~Palethorpe, and G.~M.~J. Barca, ``Multi-gpu ri-hf energies and analytic gradients-toward high-throughput ab initio molecular dynamics,'' \emph{J. Chem. Theory Comput.}, vol.~20, no.~17, pp. 7503--7515, 2024.

\bibitem{MOLECULAR_DYNAMICS_UMBRELLA_II}
S.~Cheng, X.~Zhao, G.~Lu, J.~Fang, T.~Zheng, R.~Wu, X.~Zhang, J.~Peng, and Y.~You, ``Fastfold: Optimizing alphafold training and inference on gpu clusters,'' in \emph{Proc. of the 29th ACM SIGPLAN Annu. Symp. on Princ. and Pract. of Parallel Program.}, ser. PPoPP '24.\hskip 1em plus 0.5em minus 0.4em\relax New York, NY, USA: Association for Computing Machinery, 2024, p. 417–430.

\bibitem{green500}
W.-c. Feng and K.~Cameron, ``The green500 list: Encouraging sustainable supercomputing,'' \emph{Computer}, vol.~40, no.~12, pp. 50--55, 2007.

\bibitem{thermal_runaway}
A.~Vassighi and M.~Sachdev, ``Thermal runaway in integrated circuits,'' \emph{IEEE Trans. on Device and Mater. Rel.}, vol.~6, no.~2, pp. 300--305, 2006.

\bibitem{TILLI2022105099}
A.~Tilli, E.~Garone, C.~Conficoni, M.~Cacciari, A.~Bosso, and A.~Bartolini, ``A two-layer distributed mpc approach to thermal control of multiprocessor systems-on-chip,'' \emph{Control Eng. Pract.}, vol. 122, p. 105099, 2022.

\bibitem{ARM_PCSA}
\BIBentryALTinterwordspacing
Arm, ``Power control system architecture,'' 2023. [Online]. Available: \url{https://developer.arm.com/documentation/den0050/d/?lang=en}
\BIBentrySTDinterwordspacing

\bibitem{Ottaviano2024}
A.~Ottaviano, R.~Balas, G.~Bambini, A.~Del~Vecchio, M.~Ciani, D.~Rossi, L.~Benini, and A.~Bartolini, ``Controlpulp: A risc-v on-chip parallel power controller for many-core hpc processors with fpga-based hardware-in-the-loop power and thermal emulation,'' \emph{Int. J. of Parallel Program.}, vol.~52, no.~1, pp. 93--123, Apr 2024.

\bibitem{GROVER_ACPI_2003}
A.~Grover, ``Modern system power management: Increasing demands for more power and increased efficiency are pressuring software and hardware developers to ask questions and look for answers.'' \emph{Queue}, vol.~1, no.~7, p. 66–72, oct 2003.

\bibitem{ibm_occ}
T.~Rosedahl, M.~Broyles, C.~Lefurgy, B.~Christensen, and W.~Feng, ``{Power/Performance Controlling Techniques in OpenPOWER},'' in \emph{High Perform. Comput.}, J.~M. Kunkel, R.~Yokota, M.~Taufer, and J.~Shalf, Eds.\hskip 1em plus 0.5em minus 0.4em\relax Cham: Springer International Publishing, 2017, pp. 275--289.

\bibitem{6178247}
A.~Bartolini, M.~Cacciari, A.~Tilli, and L.~Benini, ``Thermal and energy management of high-performance multicores: Distributed and self-calibrating model-predictive controller,'' \emph{IEEE Trans. on Parallel and Distrib. Syst.}, vol.~24, no.~1, pp. 170--183, 2013.

\bibitem{9163012}
X.~Sun, A.~Boora, R.~Pamula, C.-H. Huang, D.~Peña-Colaiocco, and V.~S. Sathe, ``Model predictive control of an integrated buck converter for digital soc domains in 65nm cmos,'' in \emph{2020 IEEE Symp. on VLSI Circ.}, 2020, pp. 1--2.

\bibitem{Wang2011AdaptivePC}
\BIBentryALTinterwordspacing
X.~Wang, K.~Ma, and Y.~Wang, ``Adaptive power control with online model estimation for chip multiprocessors,'' \emph{IEEE Trans. on Parallel and Distrib. Syst.}, vol.~22, pp. 1681--1696, 2011. [Online]. Available: \url{https://api.semanticscholar.org/CorpusID:11497998}
\BIBentrySTDinterwordspacing

\bibitem{SIPEARL}
\BIBentryALTinterwordspacing
T.~L. Group, ``{SiPearl} {Develops} {ARM} {HPC} {Chip},'' 2020. [Online]. Available: \url{https://www.linleygroup.com/newsletters/newsletter_detail.php?num=6227&year=2020&tag=3}
\BIBentrySTDinterwordspacing

\bibitem{AMDGENOA}
\BIBentryALTinterwordspacing
AMD, ``{E}pyc 7004 {G}enoa,'' 2022. [Online]. Available: \url{https://en.wikichip.org/wiki/amd/cores/genoa}
\BIBentrySTDinterwordspacing

\bibitem{NVIDIAGRACE}
\BIBentryALTinterwordspacing
NVIDIA, ``Nvidia grace cpu superchip,'' 2023. [Online]. Available: \url{https://www.nvidia.com/en-us/data-center/grace-cpu-superchip/}
\BIBentrySTDinterwordspacing

\bibitem{JEREZ_2}
J.~L. Jerez, P.~J. Goulart, S.~Richter, G.~A. Constantinides, E.~C. Kerrigan, and M.~Morari, ``Embedded online optimization for model predictive control at megahertz rates,'' \emph{IEEE Trans. on Autom. Control}, vol.~59, no.~12, pp. 3238--3251, 2014.

\bibitem{Salzmann_2023}
T.~Salzmann, E.~Kaufmann, J.~Arrizabalaga, M.~Pavone, D.~Scaramuzza, and M.~Ryll, ``Real-time neural mpc: Deep learning model predictive control for quadrotors and agile robotic platforms,'' \emph{IEEE Robot. and Automat. Lett.}, vol.~8, no.~4, p. 2397–2404, Apr. 2023.

\bibitem{alavilli2023tinympc}
K.~Nguyen, S.~Schoedel, A.~Alavilli, B.~Plancher, and Z.~Manchester, ``Tinympc: Model-predictive control on resource-constrained microcontrollers,'' in \emph{2024 IEEE Int. Conf. on Robot. and Automat. (ICRA)}, 2024, pp. 1--7.

\bibitem{Scheffler_2023}
P.~Scheffler, F.~Zaruba, F.~Schuiki, T.~Hoefler, and L.~Benini, ``Sparse stream semantic registers: A lightweight isa extension accelerating general sparse linear algebra,'' \emph{IEEE Trans. on Parallel and Distrib. Syst.}, vol.~34, no.~12, p. 3147–3161, Dec. 2023.

\bibitem{BoydAdmm2011}
S.~Boyd, N.~Parikh, E.~Chu, B.~Peleato, and J.~Eckstein, ``Distributed optimization and statistical learning via the alternating direction method of multipliers,'' \emph{Found. Trends Mach. Learn.}, vol.~3, no.~1, p. 1–122, jan 2011.

\bibitem{NoceWrig06}
J.~Nocedal and S.~J. Wright, \emph{Numer. Optim.}, 2nd~ed.\hskip 1em plus 0.5em minus 0.4em\relax New York, NY, USA: Springer, 2006.

\bibitem{Stellato_2020}
B.~Stellato, G.~Banjac, P.~Goulart, A.~Bemporad, and S.~Boyd, ``Osqp: an operator splitting solver for quadratic programs,'' \emph{Math. Program. Comput.}, vol.~12, no.~4, p. 637–672, Feb. 2020.

\bibitem{FerrauABB_1}
D.~Kouzoupis, A.~Zanelli, H.~Peyrl, and H.~J. Ferreau, ``Towards proper assessment of qp algorithms for embedded model predictive control,'' in \emph{2015 Eur. Control Conf. (ECC)}, 2015, pp. 2609--2616.

\bibitem{activeset_intpoint_comp}
M.~S.~K. Lau, S.~P. Yue, K.~V. Ling, and J.~M. Maciejowski, ``A comparison of interior point and active set methods for fpga implementation of model predictive control,'' in \emph{2009 Eur. Control Conf. (ECC)}, 2009, pp. 156--161.

\bibitem{fast_mpc_boyd}
Y.~Wang and S.~Boyd, ``Fast model predictive control using online optimization,'' \emph{IEEE Trans. on Control Syst. Technol.}, vol.~18, no.~2, pp. 267--278, 2010.

\bibitem{operator_splitting_boyd}
B.~O'Donoghue, G.~Stathopoulos, and S.~Boyd, ``A splitting method for optimal control,'' \emph{IEEE Trans. Control Syst. Technol.}, vol.~21, no.~6, pp. 2432--2442, 2013.

\bibitem{operator_splitting_book_1}
G.~Stathopoulos, H.~Shukla, A.~Szucs, Y.~Pu, and C.~N. Jones, \emph{Operator Splitting Methods in Control}.\hskip 1em plus 0.5em minus 0.4em\relax Now Foundations and Trends, 2016.

\bibitem{Stellato_osqp_conference}
G.~Banjac, B.~Stellato, N.~Moehle, P.~Goulart, A.~Bemporad, and S.~Boyd, ``Embedded code generation using the osqp solver,'' in \emph{2017 IEEE 56th Annu. Conf. on Decis. and Control (CDC)}.\hskip 1em plus 0.5em minus 0.4em\relax IEEE Press, 2017, p. 1906–1911.

\bibitem{AMD_SPARSE}
P.~R. Amestoy, T.~A. Davis, and I.~S. Duff, ``Algorithm 837: Amd, an approximate minimum degree ordering algorithm,'' \emph{ACM Trans. Math. Softw.}, vol.~30, no.~3, p. 381–388, sep 2004.

\bibitem{MPC_3D_1}
H.~Najibi, A.~Levisse, G.~Ansaloni, M.~Zapater, M.~Vasic, and D.~Atienza, ``Thermal and voltage-aware performance management of 3-d mpsocs with flow cell arrays and integrated sc converters,'' \emph{IEEE Trans. on Comput.-Aided Des. of Integr. Circuits and Syst.}, vol.~42, no.~1, pp. 2--15, 2023.

\bibitem{HYPERPERIOD}
I.~Ripoll and R.~Ballester-Ripoll, ``Period selection for minimal hyperperiod in periodic task systems,'' \emph{IEEE Trans. on Comput.}, vol.~62, no.~9, pp. 1813--1822, 2013.

\bibitem{bambini2024arxiv}
G.~Bambini, A.~Ottaviano, C.~Conficoni, A.~Tilli, L.~Benini, and A.~Bartolini, ``Modeling and controlling many-core hpc processors: an alternative to pid and moving average algorithms,'' \emph{ACM Trans. Auton. Adapt. Syst.}, Sep. 2024, just Accepted.

\bibitem{thermal_survey}
H.~Sultan, A.~Chauhan, and S.~R. Sarangi, ``A survey of chip-level thermal simulators,'' \emph{ACM Comput. Surv.}, vol.~52, no.~2, apr 2019.

\bibitem{bartolini2019advances}
A.~Bartolini and D.~Rossi, ``Advances in power management of many-core processors,'' in \emph{Many-Core Computing: Hardware and Software}.\hskip 1em plus 0.5em minus 0.4em\relax IET, 2022, ch.~8, pp. 191--213.

\bibitem{SCHUBIGER202055}
M.~Schubiger, G.~Banjac, and J.~Lygeros, ``Gpu acceleration of admm for large-scale quadratic programming,'' \emph{J. of Parallel and Distrib. Comput.}, vol. 144, pp. 55--67, 2020.

\bibitem{TM-Gray-Beneventi}
F.~{Beneventi}, A.~{Bartolini}, A.~{Tilli}, and L.~{Benini}, ``An effective gray-box identification procedure for multicore thermal modeling,'' \emph{IEEE Trans. on Comput.}, vol.~63, no.~5, pp. 1097--1110, 2014.

\bibitem{bambini_phd_2025}
G.~Bambini, ``Mastering power control in hpc cpus: A journey through modeling, algorithms, and hardware insights,'' Ph.D. thesis, University of Bologna, 2025, to appear. Expected to be available at: \url{https://amslaurea.unibo.it/}. Temporary access link: \url{https://iis-nextcloud.ee.ethz.ch/s/Eda4TRnqBdszoj3}.

\bibitem{FORGOTTEN_UNCORE}
V.~Gupta, P.~Brett, D.~Koufaty, D.~Reddy, S.~Hahn, K.~Schwan, and G.~Srinivasa, ``The forgotten 'uncore': on the energy-efficiency of heterogeneous cores,'' in \emph{Proc. of the 2012 USENIX Conf. on Annu. Tech. Conf.}, ser. USENIX ATC'12, USA, 2012, p.~34.

\bibitem{JEREZ_1}
J.~L. Jerez, P.~J. Goulart, S.~Richter, G.~A. Constantinides, E.~C. Kerrigan, and M.~Morari, ``Embedded predictive control on an fpga using the fast gradient method,'' in \emph{2013 Eur. Control Conf. (ECC)}, 2013, pp. 3614--3620.

\bibitem{saad2003iterative}
Y.~Saad, \emph{Iterative Methods for Sparse Linear Systems}, 2nd~ed.\hskip 1em plus 0.5em minus 0.4em\relax Philadelphia, PA: SIAM, 2003.

\bibitem{DIRECT_SPARSE_TRIANG_SOLV}
A.~M. Bradley, ``A hybrid multithreaded direct sparse triangular solver,'' in \emph{Proc. of the 2016 SIAM Workshop on Combinatorial Scientific Comput. (CSC)}, 2016, pp. 13--22.

\bibitem{ALDER_LAKE}
E.~Rotem, A.~Yoaz, L.~Rappoport, S.~J. Robinson, J.~Y. Mandelblat, A.~Gihon, E.~Weissmann, R.~Chabukswar, V.~Basin, R.~Fenger, M.~Gupta, and A.~Yasin, ``Intel alder lake cpu architectures,'' \emph{IEEE Micro}, vol.~42, no.~3, pp. 13--19, 2022.

\bibitem{NDJE202119}
M.~Ndje, L.~Bitjoka, A.~T. Boum, D.~J. {Fotsa Mbogne}, L.~Buşoniu, J.~C. Kamgang, and G.~V. {Tchané Djogdom}, ``Fast constrained nonlinear model predictive control for implementation on microcontrollers,'' \emph{IFAC-PapersOnLine}, vol.~54, no.~4, pp. 19--24, 2021, 4th IFAC Conf. on Embedded Syst., Comput. Intell. and Telematics in Control CESCIT 2021.

\bibitem{8611361}
P.~Chaber and M.~Ławryńczuk, ``Fast analytical model predictive controllers and their implementation for stm32 arm microcontroller,'' \emph{IEEE Trans. on Ind. Inform.}, vol.~15, no.~8, pp. 4580--4590, 2019.

\bibitem{JEREZ_SPARSE}
J.~L. Jerez, G.~A. Constantinides, and E.~C. Kerrigan, ``An fpga implementation of a sparse quadratic programming solver for constrained predictive control,'' in \emph{Proc. of the 19th ACM/SIGDA Int. Symp. on Field Programmable Gate Arrays}, ser. FPGA '11, 2011, p. 209–218.

\bibitem{7348173}
I.~Malouche, A.~K. Abbes, and F.~Bouani, ``Automatic model predictive control implementation in a high-performance microcontroller,'' in \emph{2015 IEEE 12th Int. Multi-Conf. on Syst., Signals \& Devices (SSD15)}, 2015, pp. 1--6.

\bibitem{SABO}
J.~R. Sabo and A.~A. Adegbege, ``A primal-dual architecture for embedded implementation of linear model predictive control,'' in \emph{2018 IEEE Conference on Decision and Control (CDC)}, 2018, pp. 1827--1832.

\bibitem{4814488}
P.~D. Vouzis, L.~G. Bleris, M.~G. Arnold, and M.~V. Kothare, ``A system-on-a-chip implementation for embedded real-time model predictive control,'' \emph{IEEE Trans. Control Syst. Technol.}, vol.~17, no.~5, pp. 1006--1017, 2009.

\bibitem{5692131}
A.~G. Wills, G.~Knagge, and B.~Ninness, ``Fast linear model predictive control via custom integrated circuit architecture,'' \emph{IEEE Trans. Control Syst. Technol.}, vol.~20, no.~1, pp. 59--71, 2012.

\bibitem{FUTURE_AWARE}
S.~Maity, R.~Roy, A.~Majumder, S.~Dey, and A.~R. Hota, ``Future aware dynamic thermal management in cpu-gpu embedded platforms,'' in \emph{2022 IEEE Real-Time Syst. Symp. (RTSS)}, 2022, pp. 396--408.

\bibitem{ZANINI1_TCM}
F.~Zanini, D.~Atienza, C.~N. Jones, L.~Benini, and G.~De~Micheli, ``Online thermal control methods for multiprocessor systems,'' \emph{ACM Trans. Des. Autom. Electron. Syst.}, vol.~18, no.~1, jan 2013.

\end{thebibliography}

\newcommand{\lucaphd}{He is currently pursuing a Ph.D. degree in the Digital Circuits and Systems group of Prof.\ Benini.}
\newcommand{\ethgrad}[2]{received his BSc and MSc degrees in electrical engineering and information technology from ETH Zurich in #1 and #2, respectively.}
\newcommand{\researchinterests}[1]{His research interests include #1.}

\vspace{-0.5cm}

\begin{IEEEbiography}[%
    {\includegraphics[width=1in,height=1.25in,clip,keepaspectratio]%
        {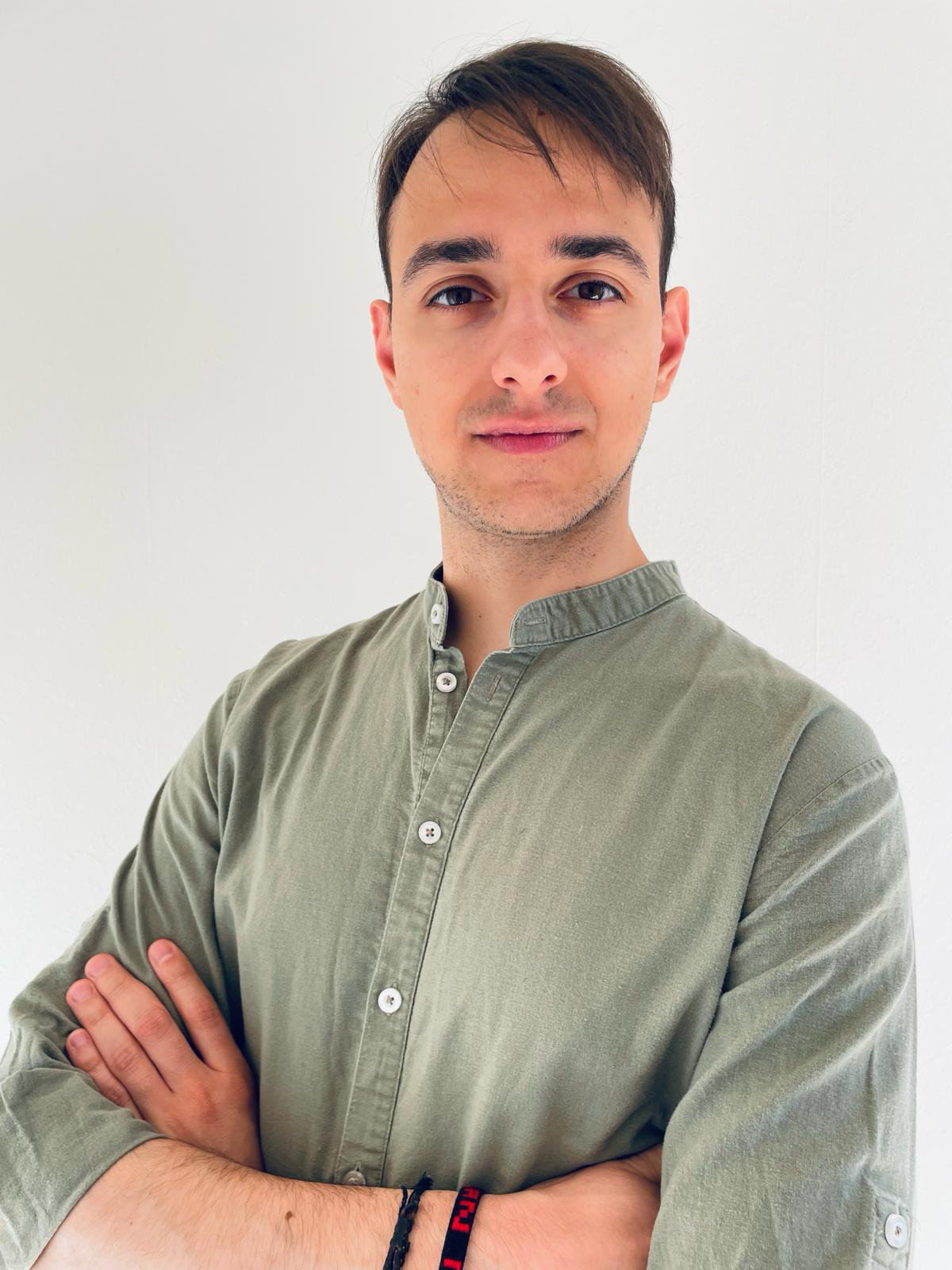}}%
    ]{Alessandro Ottaviano}
    received the B.Sc. in Physical Engineering from Politecnico di Torino, Italy, and the M.Sc. in Electrical Engineering as a joint degree between Politecnico di Torino, Grenoble INP-Phelma and EPFL Lausanne, in 2018 and 2020, respectively. %
    \lucaphd{}
    \researchinterests{%
      embedded system design and energy-efficient computing}
\end{IEEEbiography}

\vspace{-0.5cm}

\begin{IEEEbiography}[%
    {\includegraphics[width=1in,height=1.25in,clip,keepaspectratio]%
        {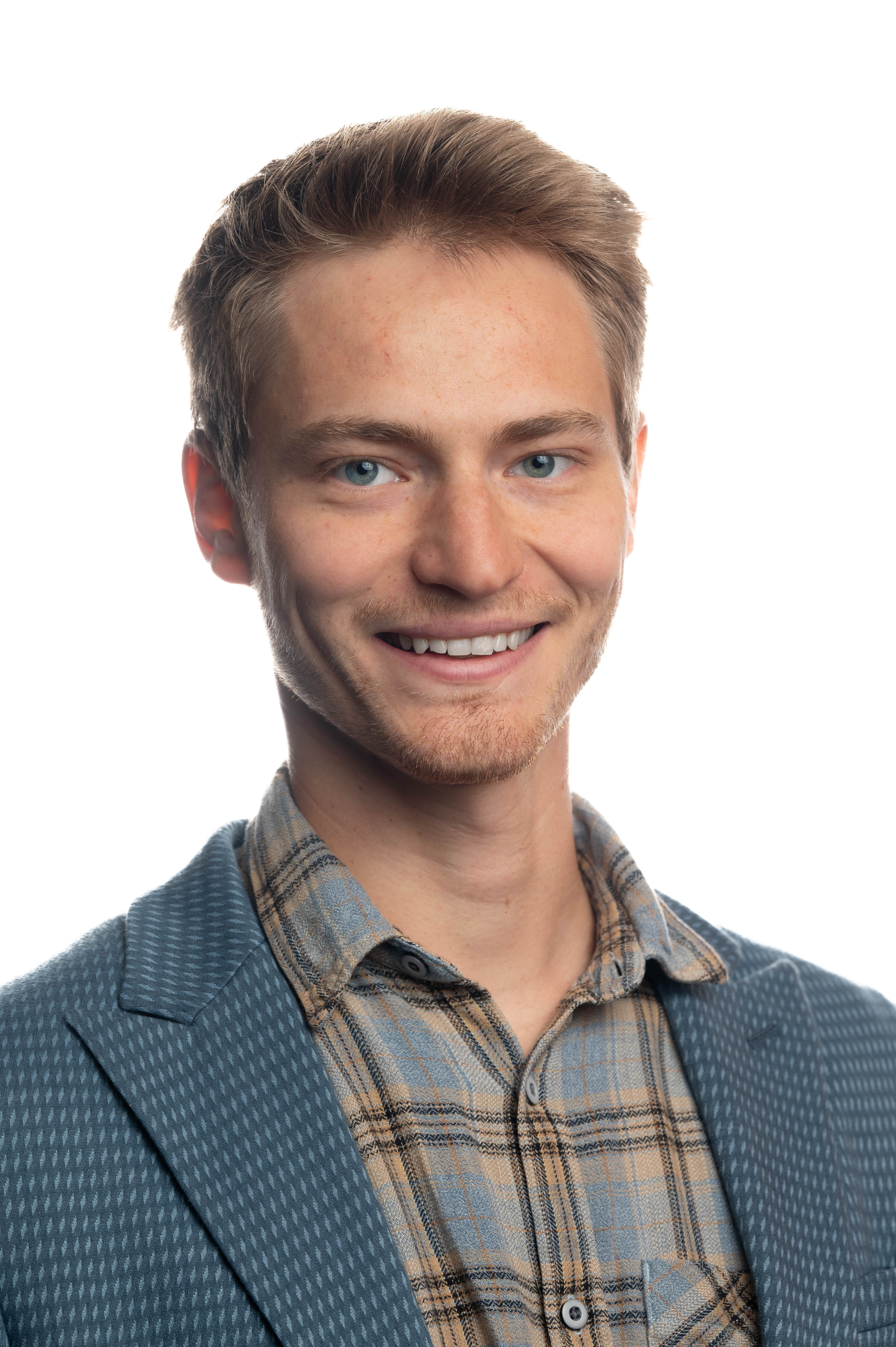}}%
    ]{Andrino Meli}
    \ethgrad{2022}{2024}
    He is currently a research assistant at the Institute for Integrated Systems. %
    \researchinterests{control systems, numerical optimization, parallel algorithm design and linear system solvers}
\end{IEEEbiography}

\vspace{-0.5cm}

\begin{IEEEbiography}[%
    {\includegraphics[width=1in,height=1.25in,clip,keepaspectratio]%
        {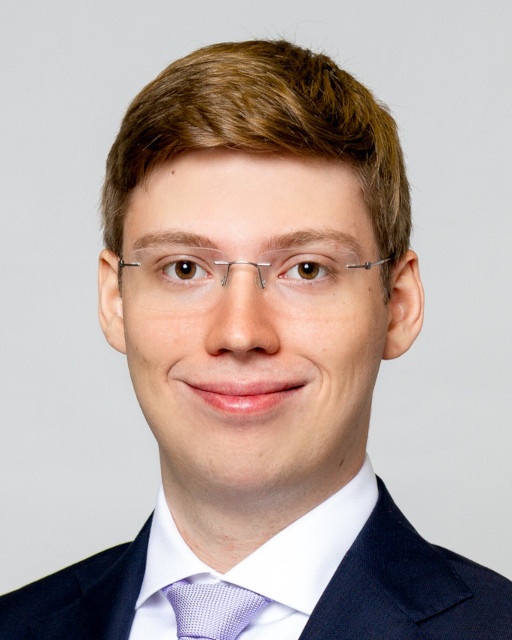}}%
    ]{Paul Scheffler}
    \ethgrad{2018}{2020}
    \lucaphd{}
    \researchinterests{hardware acceleration of sparse and irregular workloads, on-chip interconnects, manycore architectures, and high-performance computing}
\end{IEEEbiography}

\vspace{-0.5cm}

\begin{IEEEbiography}[%
    {\includegraphics[width=1in,height=1.25in,clip,keepaspectratio]%
        {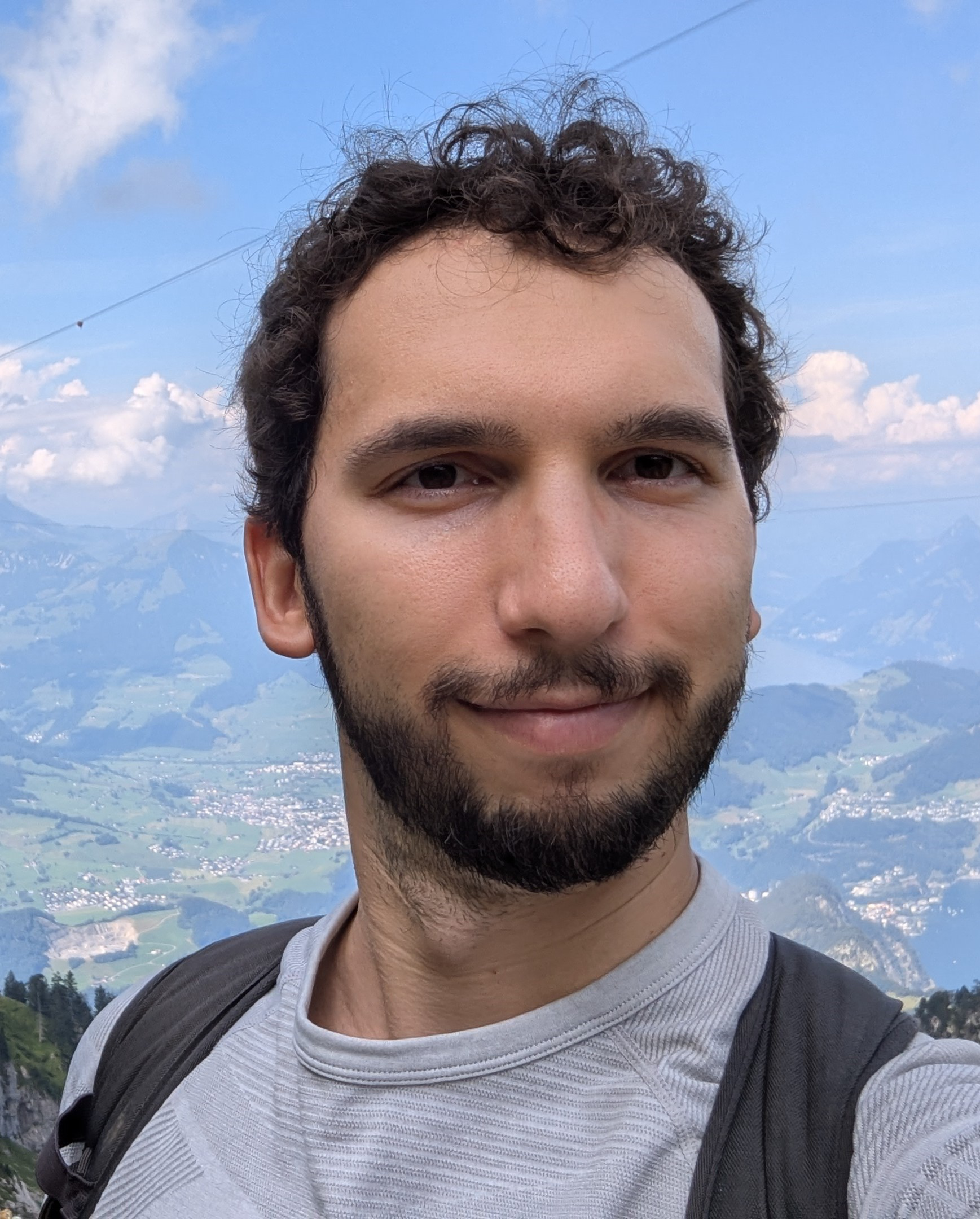}}%
    ]{Giovanni Bambini}
     studied Automation Engineering at the University of Bologna, with a focus on control theory and microcontrollers. He earned a PhD in Electronic Engineering, specializing in modeling and control of HPC CPUs. During his doctoral research, he actively contributed to the European Processor Initiative (EPI), working on thermal and power management control, chip modeling and design. His research interests include advanced control theory, modeling and identification, and real-time firmware development.
\end{IEEEbiography}

\vspace{-0.5cm}

\begin{IEEEbiography}[%
    {\includegraphics[width=1in,height=1.25in,clip,keepaspectratio]%
        {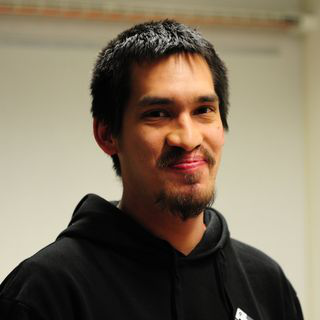}}%
    ]{Robert Balas}
    \ethgrad{2015}{2017}
    \lucaphd{}
    \researchinterests{real-time computing, compilers, and operating-systems}
\end{IEEEbiography}

\vspace{-0.5cm}

\begin{IEEEbiography}[%
    {\includegraphics[width=1in,height=1.25in,clip,keepaspectratio]%
        {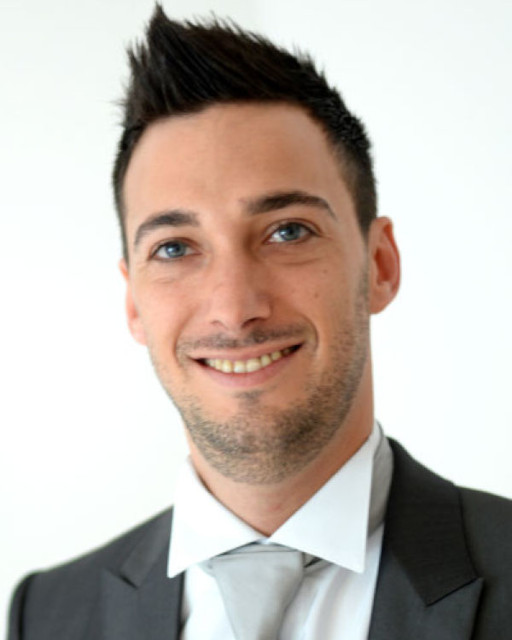}}%
    ]{Davide Rossi}
     received the Ph.D. degree from the University of Bologna, Bologna, Italy, in 2012. He has been a Post-Doctoral Researcher with the Department of Electrical, Electronic and Information Engineering “Guglielmo Marconi,” University of Bologna, since 2015, where he is currently an Associate Professor. His research interests focus on energy-efficient digital architectures. In this field, he has published more than 100 papers in international peer-reviewed conferences and journals.
\end{IEEEbiography}

\vspace{-0.5cm}

\begin{IEEEbiography}[%
    {\includegraphics[width=1in,height=1.25in,clip,keepaspectratio]%
        {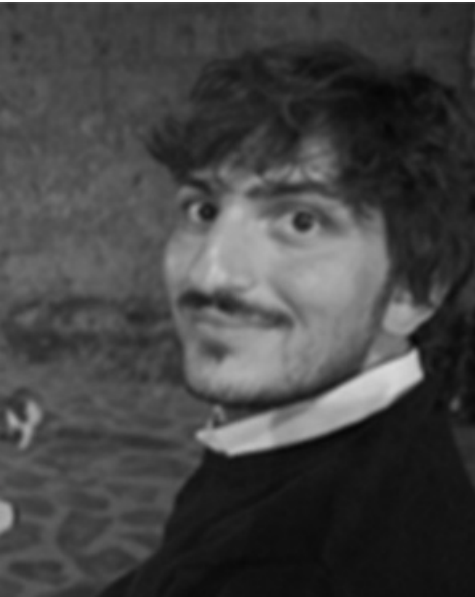}}%
    ]{Andrea Bartolini}
    (Member, IEEE) received the Ph.D. degree from the University of Bologna, Bologna, Italy, in 2013. He is an Associate Professor with the Department of Electrical, Electronic and Information Engineering Guglielmo Marconi, University of Bologna, Bologna, Italy. He has published more than 120 papers in peer-reviewed international journals and conferences and several book chapters with focus on dynamic resource management—ranging from embedded to large-scale HPC systems.
\end{IEEEbiography}

\vspace{-0.5cm}

\begin{IEEEbiography}[%
    {\includegraphics[width=1in,height=1.25in,clip,keepaspectratio]%
        {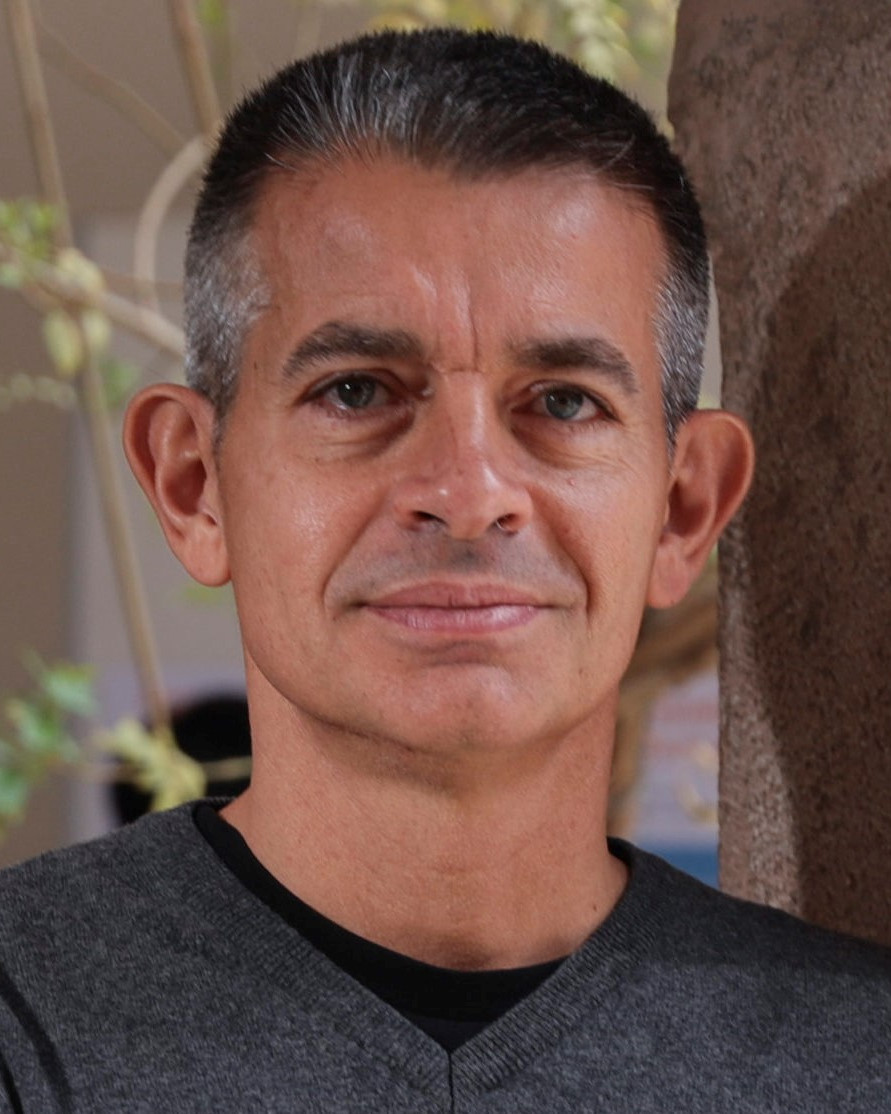}}%
    ]{Luca Benini}
    (F'07) holds the chair of Digital Circuits and Systems at ETH Zurich and is Full Professor at the Università di Bologna.
    Dr.\ Benini’s research interests are in energy-efficient computing systems design, from embedded to high-performance.
    He has published more than 1000 peer-reviewed papers and five books.
    He is a Fellow of the ACM and a member of Academia Europaea.
\end{IEEEbiography}

\vfill
%

\end{document}